\documentclass[preprint,12pt]{elsarticle}
\graphicspath{{./}}

\usepackage[dvipsnames]{xcolor}

\usepackage{amssymb}
\usepackage{subfigure}
\usepackage{amsmath}
\usepackage{color}
\usepackage{hyperref}

\usepackage{amsthm}
\usepackage{enumitem}
\usepackage{pgfplotstable}
\usepackage{pgfplots}
\pgfplotsset{compat=1.14}
\pgfplotsset{
	colormap={my basis colormap}{
		rgb255=(0, 114, 189);
		rgb255=(54, 106, 148);
		rgb255=(108, 98, 107);
		rgb255=(163, 91, 66);
		rgb255=(217, 83, 25);
	}
}
\pgfplotsset{
	colormap={my parula}{
		rgb255=(53.0655, 42.406, 134.9460);
		rgb255=(20.2336, 132.6061, 211.9511);
		rgb255=(55.5463, 184.8859, 157.9118);
		rgb255=(208.7187, 186.8470,  89.1966);
		rgb255=(248.9565, 250.6905, 13.7190);
	}
}
\usepackage{tikz}
\usepgfmodule{oo}
\usetikzlibrary{calc}
\usetikzlibrary{shapes}
\usetikzlibrary{plotmarks}
\usetikzlibrary{backgrounds}
\usetikzlibrary{decorations.pathmorphing}
\usetikzlibrary{external}
\usetikzlibrary{arrows,fit,matrix,positioning,shapes.geometric}

\usepackage{titlesec}

\titleformat{\paragraph}
{\normalfont\normalsize\it}{\theparagraph}{1em}{}
\titlespacing*{\paragraph}
{0pt}{3.25ex plus 1ex minus .2ex}{1.5ex plus .2ex}

\journal{Computer Methods in Applied Mechanics and Engineering}

\def\mathbf#1{\boldsymbol{#1}}

\def\vec#1{\boldsymbol{#1}}

\theoremstyle{plain}

\theoremstyle{remark}

\theoremstyle{definition}

\definecolor{myBlue}{rgb} {0,0.4470,0.7410}
\definecolor{myRed}{rgb} {0.8500,0.3250,0.0980}

\usepackage{marginnote}

\usepackage{siunitx}

\begin{document}

\begin{frontmatter}

  \title{Isogeometric analysis using G-spline surfaces with arbitrary unstructured quadrilateral layout}
 

\author[label1]{Zuowei Wen}
\author[label1]{Md. Sadman Faruque}
\author[label2]{Xin Li}
\author[label3]{Xiaodong Wei}
\author[label1]{Hugo Casquero\corref{cor1}}
\ead{casquero@umich.edu}
\address[label1]{Department of Mechanical Engineering, University of Michigan – Dearborn, 4901 Evergreen Road, Dearborn, MI 48128-1491, U.S.A.}
\address[label2]{School of Mathematical Science, USTC, Hefei, China.}
\address[label3]{University of Michigan-Shanghai Jiao Tong University Joint Institute, Shanghai Jiao Tong University, Shanghai, China.}

\cortext[cor1]{Corresponding author.}


\begin{abstract}

G-splines are a generalization of B-splines that deals with extraordinary points by imposing $G^1$ constraints across their spoke edges, thus obtaining a continuous tangent plane throughout the surface.  Using the isoparametric concept and the Bubnov-Galerkin method to solve partial differential equations with G-splines results in discretizations with global $C^1$ continuity in physical space. Extraordinary points (EPs) are required to represent manifold surfaces with arbitrary topological genus. In this work, we allow both interior and boundary EPs and there are no limitations regarding how close EPs can be from each other. Reaching this level of flexibility is necessary so that splines with EPs can become mainstream in the design-through-analysis cycle of the complex thin-walled structures that appear in engineering applications. To the authors' knowledge, the two EP constructions based on imposing $G^1$ constraints proposed in this work are the first two EP constructions used in isogeometric analysis (IGA) that combine the following distinctive characteristics: (1) Only vertex-based control points are used and they behave as geometric shape handles, (2) any control point of the control net can potentially be an EP, (3) global $C^1$ continuity in physical space is obtained without introducing singularities, (4) faces around EPs are not split into multiple elements, i.e., B\'ezier meshes with uniform element size are obtained, and (5) good surface quality is attained. The studies of convergence and surface quality performed in this paper suggest that G-splines are more suitable for IGA than EP constructions based on the D-patch framework. Finally, we have represented the stiffener, the inner part, and the outer part of a B-pillar with G-spline surfaces and solved eigenvalue problems using both Kirchhoff-Love and Reissner-Mindlin shell theories. The results are compared with bilinear quadrilateral meshes and excellent agreement is found between G-splines and conventional finite elements. In summary, G-splines are a viable alternative to design and analyze thin-walled structures using the same geometric representation so as to streamline the design-through-analysis cycle.



\end{abstract}


\end{frontmatter}


\renewcommand{\thefootnote}{\fnsymbol{footnote}}

\section{Introduction}

In computer-aided-design (CAD) software, the industry standard to represent thin-walled structures is to use trimmed non-uniform rational B-spline (NURBS) patches, namely, hundreds or thousands of trimmed NURBS patches are often used to represent a complex structural part. However, in finite-element-analysis (FEA) software, the industry standard to represent thin-walled structures is to use quad-dominant meshes, i.e., meshes that have predominantly bilinear quadrilaterals as elements, but also have some elements that are linear triangles. The fact that the underlying technologies used to represent thin-walled structures in CAD and FEA programs are completely different results in numerous interoperability issues in the design-through-analysis cycle of the complex thin-walled structures used in engineering applications.

Isogeometric analysis (IGA) \cite{1003.000} aims at using only one geometric representation of thin-walled structures in the whole design-through-analysis cycle so as to significantly cut the overall time spent on this process \cite{cottrell2009isogeometric}. This can be achieved if a type of spline that simultaneously meets the needs of both CAD and FEA programs is found. As a first step, an untrimmed NURBS patch was shown to be not only suitable for analysis \cite{bazilevs2006isogeometric, hughes2008duality, dof, da2011some}, but its higher continuity across element boundaries in comparison with the $C^0$ continuity across element boundaries of Lagrange polynomials resulted in the following benefits: enhanced robustness in handling mesh distortion is obtained \cite{Lipton2010}, enhanced accuracy in spectrum analysis is achieved \cite{Cottrell2006, hughes2014finite, oesterle2022study}, fourth-order partial differential equations can be solved in primal form \cite{Gomez2008, Kiendl2009, nsk, Liu2015a, bueno2018three}, $H^1$-conforming discretizations that are either divergence-conforming or curl-conforming are attained \cite{buffa2011isogeometric, buffa2010isogeometric, john333, kamensky2017immersogeometric, CASQUERO2021109872}, among others. A NURBS patch (or a B-spline patch) is defined on a control net with structured quadrilateral layout. The construction of smooth splines defined on a control net with unstructured quadrilateral layout is a nontrivial task. Extraordinary points (EPs) are the control points around which the connectivity of the control net is unstructured. Developing different strategies to handle EPs has been an active research direction in the communities of CAD and computer animation for more than 40 years. A comprehensive review can be found in \cite{peters2019splines}. This research direction has gained renewed interest since the advent of IGA. To date, the types of splines with EPs that have been used to solve partial differential equations (PDEs) include the following

\begin{itemize} 

\item G-splines \cite{hollig1990g, gregory1990smooth, peters2002geometric} generalize B-splines by enforcing geometric continuity \cite{derose1985intuitive, derose1985geometric, farin2014curves} of each basis function across the edges that emanate from EPs. As shown in \cite{groisser2015matched}, when using the isoparametric concept and the Bubnov-Galerkin method, using basis functions with geometric continuity of order $k$ ($G^k$ continuity) across an element boundary results in discretizations with $C^k$ continuity in physical space across that element boundary. PDEs were solved using G-splines in \cite{Scott2013, nguyen2014comparative, karvciauskas2016generalizing, collin2016analysis, kapl2017dimension, kapl2018construction, kapl2019isogeometric, kapl2021family}. From the convergence studies included in \cite{Scott2013, nguyen2014comparative, karvciauskas2016generalizing, collin2016analysis}, it can be concluded that G-spline constructions do not generally result in optimal asymptotic convergence rates. A subset with optimal asymptotic convergence rates was defined in \cite{collin2016analysis, kapl2017dimension, kapl2018construction, kapl2019isogeometric, kapl2021family}. However, it is not possible to design free-form surfaces within this subset. Asymptotic convergence rates are not the only metrics to measure the approximation power of a functional space. The convergence constant heavily affects the accuracy obtained on coarse meshes. In large-scale industrial applications, relatively coarse meshes are needed to make the problem computable in a reasonable amount of time. Thus, from an engineering point of view, the level of accuracy obtained with relatively coarse meshes is at least as relevant as the level of accuracy obtained with very fine meshes.

\item Catmull-Clark Subdivision surfaces (SubD) \cite{catmull1978recursively, stam1998exact, reif1995unified, peters2008subdivision} collapse the faces around EPs using infinite recursion formulas to reach global $C^1$ continuity in the limit. Since it is a singular construction, the surface quality of SubD is limited \cite{karvciauskas2015improved}. PDEs were solved using Catmull-Clark SubD in \cite{wei2015truncated, pan2016isogeometric, wei2016extended, zhang2018subdivision, BANDARA201862, wei2021tuned}. The main challenge of SubD is to define accurate and efficient rules for their numerical integration \cite{WAWRZINEK201660, juttler2016numerical}.

\item The D-patch framework \cite{reif1997refineable} achieves $C^1$ continuity by collapsing one ring of extraction coefficients for each basis function around EPs. Since it is a singular construction, the surface quality of the D-patch framework is limited \cite{karvciauskas2017improved}. PDEs were solved using the D-patch framework in \cite{nguyen2016refinable, toshniwal2017smooth, casquero2020seamless, wei2022analysis, yang2023non}.

\item Manifold splines \cite{grimm1995modeling, navau2000modeling, ying2004simple, gu2005manifold, tosun2011manifold} use the partition of unity method to smoothly blend splines. The coefficients that multiply the basis functions in manifold splines do not behave as geometric shape handles. PDEs were solved using manifold splines in \cite{majeed2017isogeometric, zhang2020manifold, koh2022optimally}.

\item Different types of splines with EPs that do not reach global $C^1$ continuity in physical space, but have optimal asymptotic convergence rates \cite{Wei2018, toshniwal2022quadratic, takacs2023almost}.

\end{itemize}

In this work, we develop two G-spline constructions that simultaneously satisfy the following distinguishing features

\begin{itemize}

\item The control points behave as geometric shape handles. Any EP construction that aspires to meet the demands of CAD programs must satisfy this.

\item Unlike in the preceding G-spline constructions used in IGA thus far \cite{Scott2013, nguyen2014comparative, karvciauskas2016generalizing, collin2016analysis, kapl2017dimension, kapl2018construction, kapl2019isogeometric, kapl2021family}, three-dimensional control nets with arbitrary unstructured quadrilateral layout are supported, i.e., both interior and boundary EPs are allowed and there are no restrictions with respect to how close EPs can be from each other. Allowing multiple EPs per face is needed to capture small features such as holes without having to introduce very small elements, which would in turn decrease the size of the maximum stable time step.

\item B\'ezier meshes with uniform element size are obtained, i.e., faces around EPs are not split into multiple elements.

\item Global $C^1$ continuity in physical space is obtained without introducing any singularity.

\end{itemize}

In order to evaluate the behavior of the two proposed EP constructions at length, we performed convergence studies, study the surface quality, and solve eigenvalue problems on complex thin-walled structures. In addition, we included detailed comparisons with respect to the D-patch framework and conventional finite elements.


The paper is outlined as follows. Section 2 describes in detail the two types of G-splines introduced in this work. Section 3 studies the convergence of G-splines when solving second-order linear elliptic problems and their accuracy is compared with the D-patch framework. Section 4 compares the surface quality of G-splines and the D-patch framework. Section 5 solves eigenvalue problems on the stiffener, the inner part, and the outer part of an automotive B-pillar using both G-splines and bilinear quadrilaterals. Conclusions are drawn in Section 6.

\section{G-splines}

In this section, we explain how to construct and refine G-spline surfaces defined on control nets with arbitrary unstructured quadrilateral layout.

\subsection{C-net}

\begin{figure} [t!] 
 \centering
 \subfigure[C-net]{\includegraphics[scale=.74]{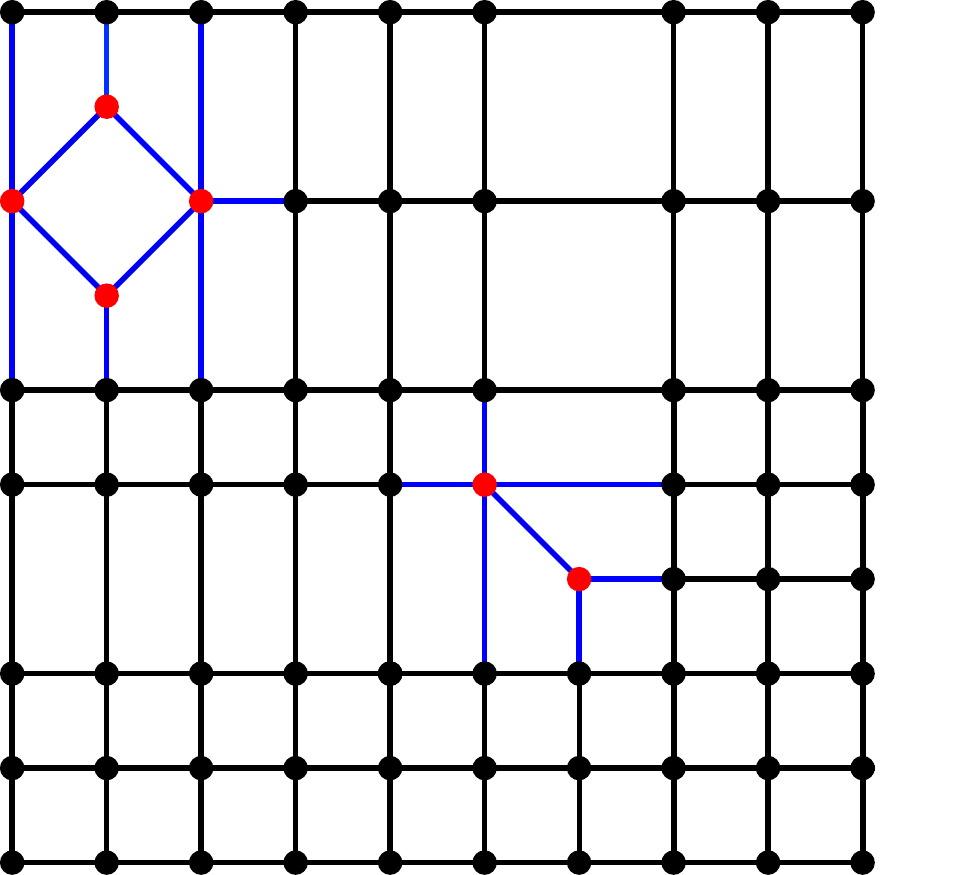}} 
 \subfigure[Control net]{\includegraphics[scale=.33]{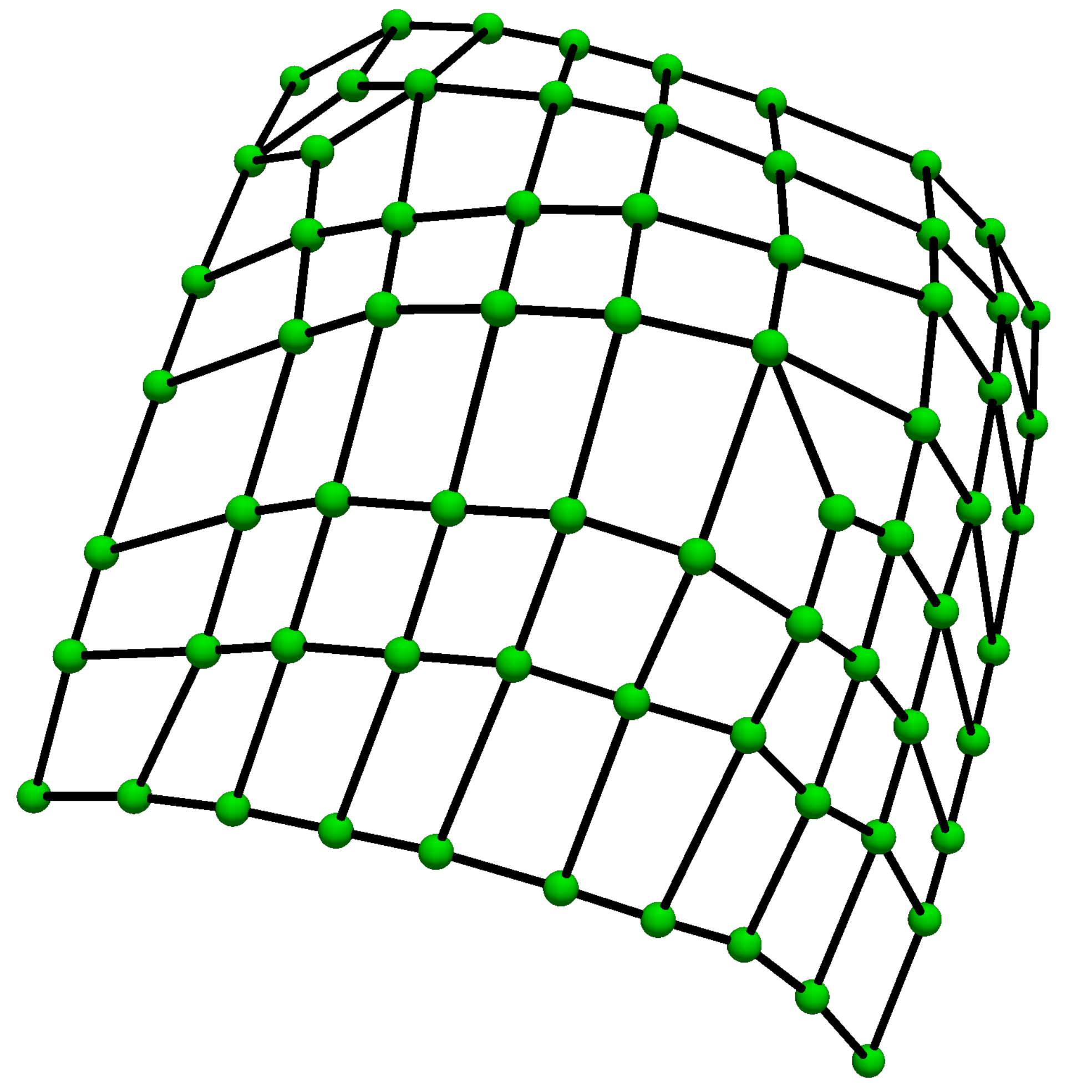}} \\
\caption{(Color online) (a) C-net with boundary and interior extraordinary vertices. The extraordinary vertices and the spoke edges are marked with red circles and blue lines, respectively. (b) A possible control net associated with the C-net given in (a).}
\label{tnet}
\end{figure}

A \textit{C-net} is an unstructured quadrilateral layout that specifies the way in which the different constituent parts of a G-spline surface are interrelated or arranged (the C of C-net stands for connectivity). Fig. \ref{tnet} a) shows an example of a C-net. The vertices, edges, and faces of the C-net are represented by circles, solid lines, and white regions in Fig. \ref{tnet} a), respectively. A control point is assigned to each vertex of the C-net. Control points are points in $\mathbb{R}^3$. In CAD programs, the control points are used as geometric shape handles to construct the surface. The total number of control points is denoted by $n_{cp}$. The control net is the unstructured quadrilateral net obtained by performing bilinear interpolations of the control points. A possible control net associated with the C-net given in Fig. \ref{tnet} a) is plotted in Fig. \ref{tnet} (b). The valence of a vertex is the number of faces that share that vertex. Extraordinary vertices are either interior vertices with a valence different than four or boundary vertices with a valence greater than two. Extraordinary vertices are marked with red circles in Fig. \ref{tnet} a). Extraordinary points (EPs) are control points associated with extraordinary vertices. A knot span of unit length is assigned to each edge of the C-net. An element is assigned to each face of the C-net. Elements are the regions in which the basis functions are either polynomial functions or rational functions. Thus, elements are the regions in which each basis function can be written in terms of Bernstein polynomials. The total number of elements is denoted by $n_{el}$. Spoke edges are the edges that emanate from an extraordinary vertex. Spoke edges are marked with blue lines in Fig. \ref{tnet} a). The \textit{1-ring faces} of an EP are the faces that are in contact with the EP. For $m > 1$, the \textit{m-ring faces} of an EP are all faces that touch the ($m$-1)-ring faces and are not a part of the ($m$-2)-ring faces. The set of \textit{0-ring vertices} of an EP contains only the EP itself. For $m > 0$, the
\textit{m-ring vertices} of an EP contain all the vertices that lie on the $m$-ring faces and are not a part of the ($m$-1)-ring vertices. 

Unlike preceding works \cite{Scott2013, nguyen2014comparative, karvciauskas2016generalizing, collin2016analysis, kapl2017dimension, kapl2018construction, kapl2019isogeometric, kapl2021family, nguyen2016refinable, toshniwal2017smooth, casquero2020seamless, wei2022analysis, yang2023non}, we allow any control point of a three-dimensional control net to potentially be an extraordinary point. In other words, there are no limitations regarding how close EPs can be from each other (potentially all four control points of a face can be EPs) and both interior and boundary EPs are allowed. Developing EP constructions with this level of flexibility is necessary to properly meet the design and analysis demands of complex engineering applications involving thin-walled structures.



\subsection{B\'ezier extraction}

A piecewise polynomial basis function $N$ of degree $p$ in each parametric direction restricted to an element $e$ can be represented as a linear combination of $(p+1)^2$ bivariate Bernstein polynomials, viz.,
\begin{equation}    
N^e \left( \xi , \eta \right) = \sum_{i=1}^{p+1} \sum_{j=1}^{p+1} c^e_{i,j} b_{i,j} \left( \xi , \eta \right),  \quad   \left( \xi , \eta \right)  \in   \left[ 0,1 \right]^2  \text{,}
\end{equation}
with
\begin{equation}    
b_{i,j} \left( \xi , \eta \right) = \mathfrak{B}_i \left( \xi \right) \mathfrak{B}_j \left( \eta \right)  \text{,}
\end{equation}
\begin{equation}    
\mathfrak{B}_{i} \left( \xi \right) =  \frac{p!}{(i-1)!(p-i+1)!} {\xi}^{i-1} \left( 1 - \xi \right)^{p-i+1}  \text{,}
\end{equation}
\begin{equation}    
\mathfrak{B}_{j} \left( \eta \right) =  \frac{p!}{(j-1)!(p-j+1)!} {\eta}^{j-1} \left( 1 - \eta \right)^{p-j+1}  \text{,}
\end{equation}
where $\left[ 0,1 \right]^2$ is the parent element domain, $\mathfrak{B}_i$ and $\mathfrak{B}_j$ are univariate Bernstein polynomials, $b_{i,j}$ is a bivariate Bernstein polynomial, and $c^e_{i,j}$ is an extraction coefficient.

Let us collect all the basis functions with support on element $e$ in a column vector, namely, $\mathbf{N}^e = ( N_1^e, N_2^e, ..., N_{n^e}^e )^T$, where $n^e$ is the number of basis functions with support on element $e$. Let us also collect the bivariate Bernstein polynomials in a column vector, namely, $\mathbf{b}= ( b_1, b_2, ..., b_{(p+1)^2})^T$, where $b_k = b_{i,j}$ with $k = (p+1)(j-1) + i$. For element $e$, the \textit{spline extraction operator} $\mathbf{C}^e$ is a matrix of dimension $n^e \times (p+1)^2$ that relates the basis functions with the bivariate Bernstein polynomials as follows
\begin{equation}
\mathbf{N}^e \left(  \xi , \eta \right)=\mathbf{C}^e \mathbf{b} \left(  \xi , \eta \right),\quad \left( \xi , \eta \right) \in  \left[ 0,1 \right]^2  \text{.}
\end{equation}

Analogously, for each element, spline control points $\mathbf{P}^e=( \mathbf{P}_1^e, \mathbf{P}_2^e, ..., \mathbf{P}_{n^e}^e )^T$ can be related to B\'ezier control points $\mathbf{B}^e= ( \mathbf{B}^e_1, \mathbf{B}^e_2, ..., \mathbf{B}^e_{(p+1)^2})^T$, where $\mathbf{B}^e_k = \mathbf{B}^e_{i,j}$ with $k = (p+1)(j-1) + i$, as follows
\begin{equation}\label{bextractioncoef}
\mathbf{B}^e  =  \left( \mathbf{C}^e \right)^T   \mathbf{P}^e  \text{,}
\end{equation}
where $\mathbf{P}^e$ and $\mathbf{B}^e$ are matrices of dimension $n^e \times 3$ and $(p+1)^2 \times 3$, respectively. $\mathbf{E}^e = (\mathbf{C}^e)^T$ is the \textit{B\'ezier extraction operator}. A common manner to implicitly define extraction operators is to establish relations among B\'ezier control points and spline control points.

When applying the Bubnov-Galerkin method and the isoparametric concept, the polynomial basis functions $\mathbf{N}^e$ can be used to define the weighting space, the trial space, and the geometry if the property of partition of unity is satisfied, i.e., if
\begin{equation}    
\sum_{a=1}^{n^e} N^e_a \left( \xi , \eta \right) = 1,  \quad   \forall \left( \xi , \eta \right)  \in   \left[ 0,1 \right]^2  \text{,}
\end{equation}
If the polynomial basis functions $\mathbf{N}^e$ do not satisfy the property of partition of unity, the following rational basis functions can be defined
\begin{equation}  \label{rationalbasis}  
R^e_a \left( \xi , \eta \right) = \frac{N^e_a \left( \xi , \eta \right)}{\sum_{b=1}^{n^e} N^e_b \left( \xi , \eta \right)} ,  \quad   \left( \xi , \eta \right)  \in   \left[ 0,1 \right]^2, \quad \forall a\in\{1,2,...,n^{e}\}  \text{.}
\end{equation}
These rational basis functions satisfy the property of partition of unity and can be used to define the weighting space, the trial space, and the geometry when applying the Bubnov-Galerkin method and the isoparametric concept.

\subsection{Basis functions}


%
%
%

A basis function is assigned to each vertex. As an aid to define basis functions, we classify the elements and the basis functions as follows:

\begin{itemize}
\item \textit{Irregular elements} are the elements associated with 1-ring faces of EPs. \textit{Transition elements} are the elements associated  with 2-ring faces of EPs. The remaining elements are \textit{regular elements}.

\item \textit{Irregular basis functions} are the basis functions associated with 0-, 1-, and 2-ring vertices of EPs. The remaining basis functions are \textit{regular basis functions}. 
\end{itemize}

 The regular basis functions are globally $C^2$-continuous while the irregular basis functions are globally $G^1$-continuous. The continuity across the shared boundary of two irregular elements is $G^1$. The continuity across the shared boundary of an irregular element and a transition element is $C^1$. The continuity across the remaining element boundaries is $C^2$. In the following, the basis functions are specified by the extraction operators defined in each element.

\subsubsection{Preliminary $C^0$ construction}

	\begin{figure} [t!] 
\centering
\includegraphics[width=5.5cm]{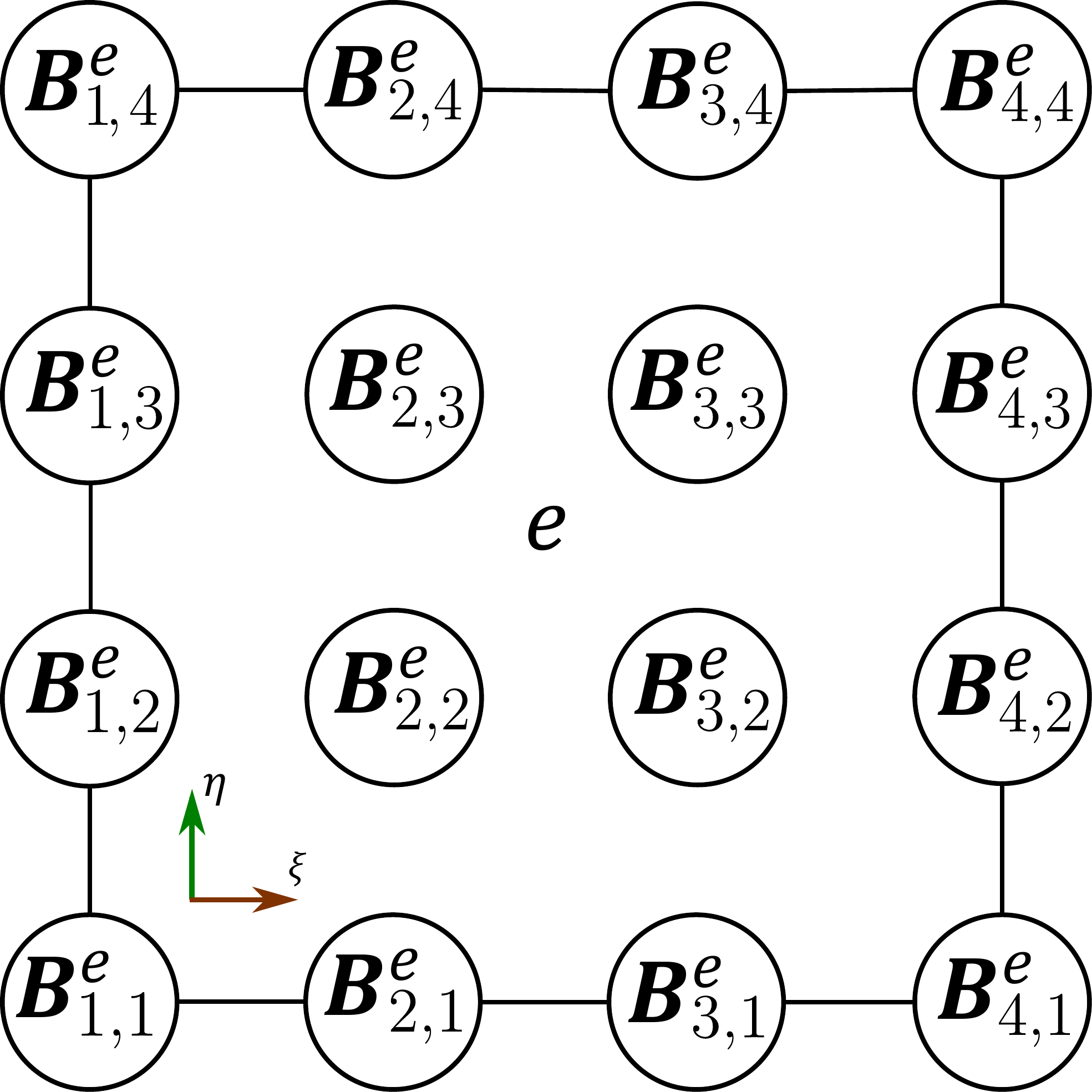}
\caption{B\'ezier control points are classified in face, edge, and vertex B\'ezier control points. $\mathbf{B}_{2,2}^e$, $\mathbf{B}_{3,2}^e$, $\mathbf{B}_{2,3}^e$, and $\mathbf{B}_{3,3}^e$ are face B\'ezier control points. $\mathbf{B}_{2,1}^e$, $\mathbf{B}_{3,1}^e$, $\mathbf{B}_{1,2}^e$, $\mathbf{B}_{4,2}^e$, $\mathbf{B}_{1,3}^e$, $\mathbf{B}_{4,3}^e$, $\mathbf{B}_{2,4}^e$, and $\mathbf{B}_{3,4}^e$ are edge B\'ezier control points. $\mathbf{B}_{1,1}^e$, $\mathbf{B}_{4,1}^e$, $\mathbf{B}_{1,4}^e$, and $\mathbf{B}_{4,4}^e$ are vertex B\'ezier control points.} 
\label{beziercp}
\end{figure}

\begin{figure} [t!] 
 \centering
 \subfigure[]{\includegraphics[scale=0.3]{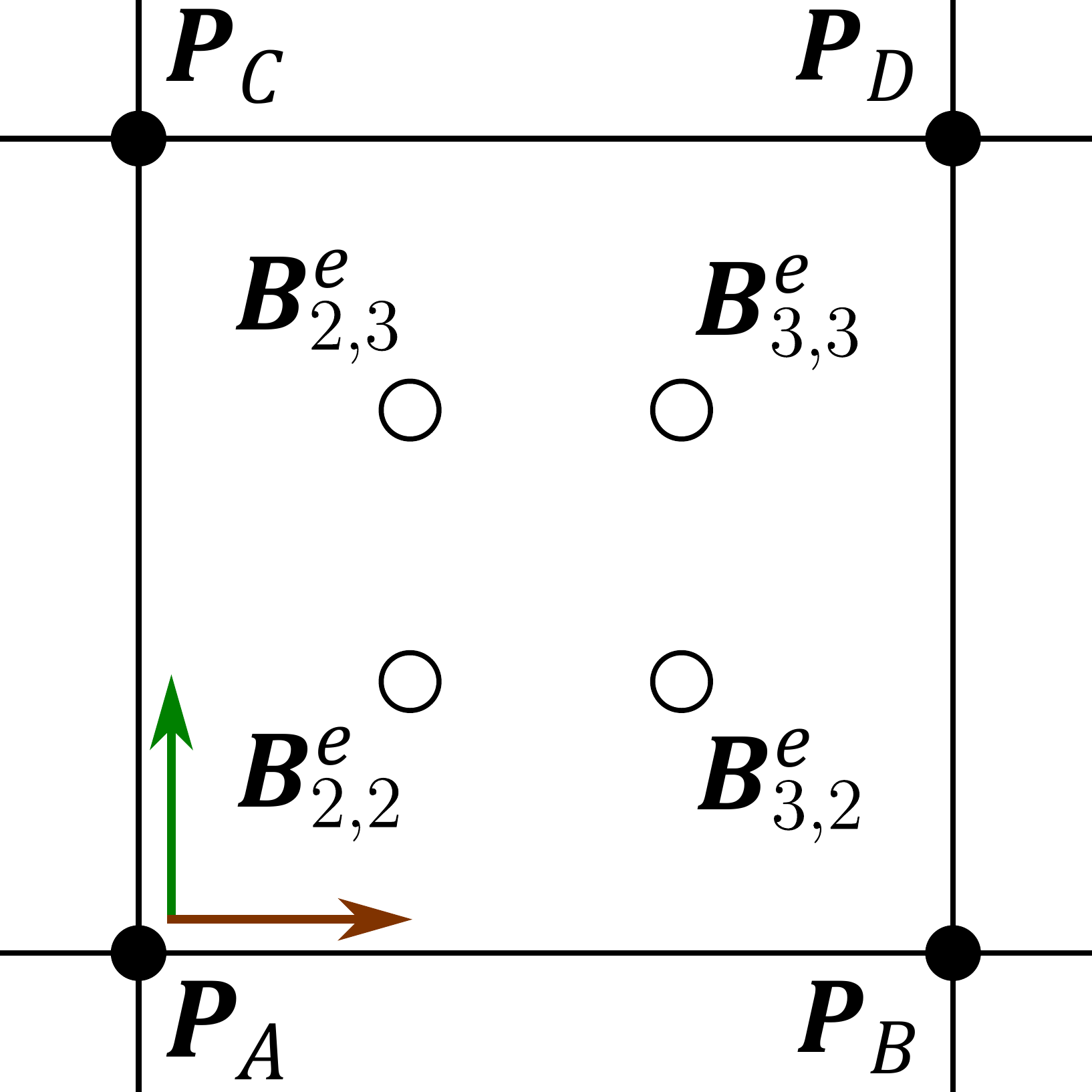}} \hspace*{+1.0mm}
 \subfigure[]{\includegraphics[scale=0.3]{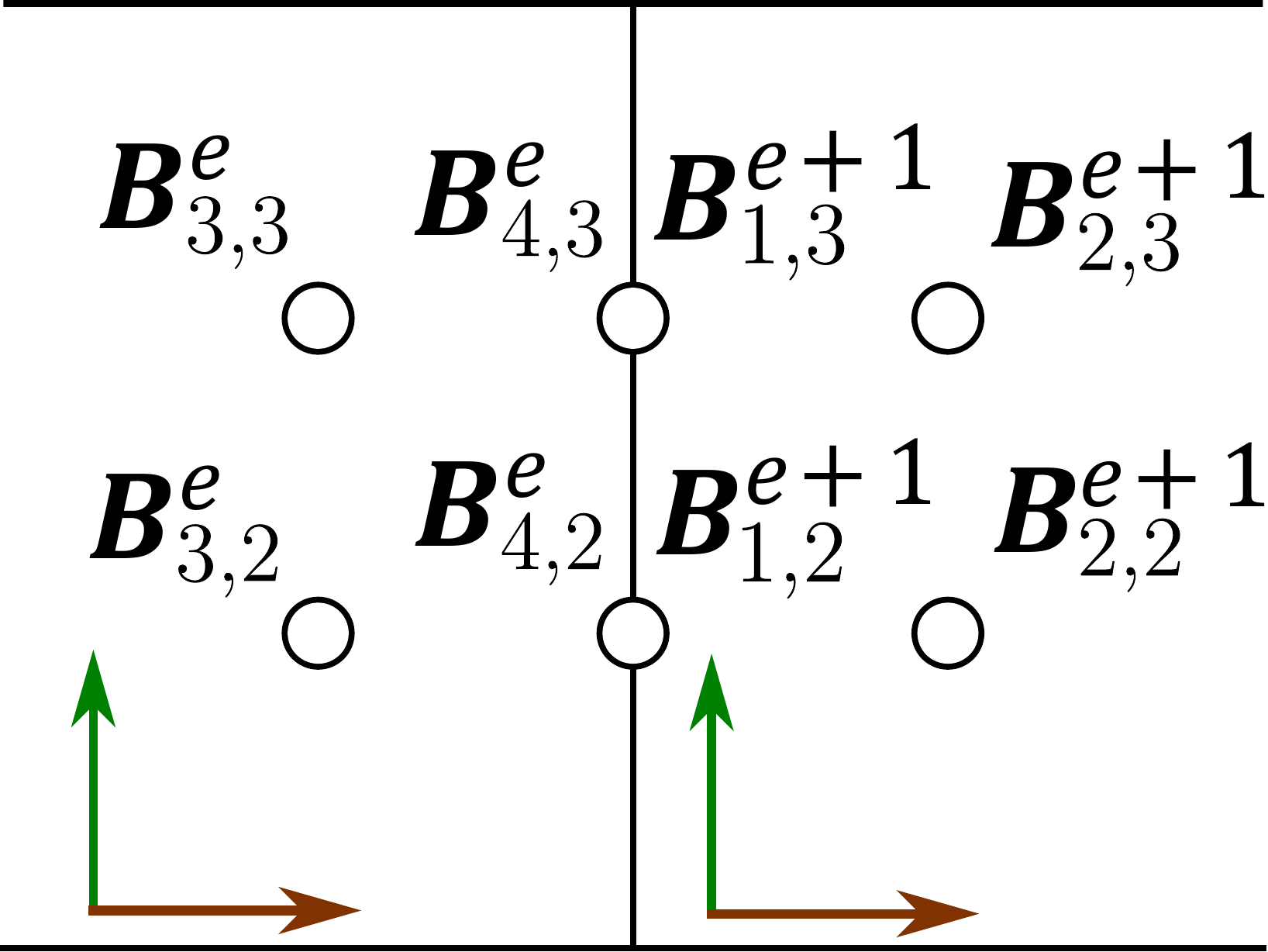}}  \hspace*{+1.0mm}
  \subfigure[]{\includegraphics[scale=0.3]{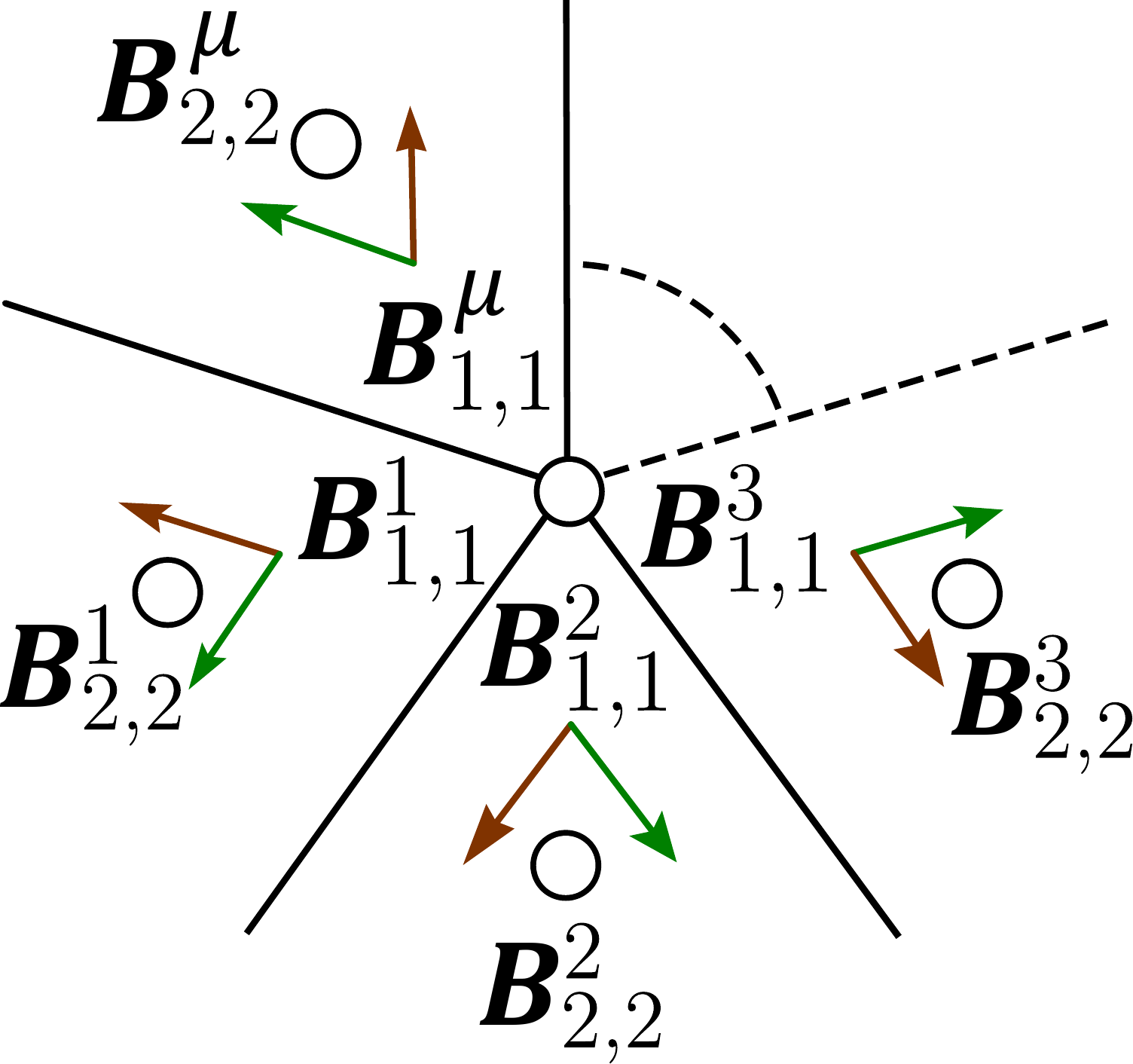}} \\
\caption{(a) Face B\'ezier control points are defined in terms of spline control points using Eqs. \eqref{firsteq}-\eqref{4theq}. (b) Edge B\'ezier control points that are not located at the boundary are defined in terms of adjacent face B\'ezier control points using Eqs. \eqref{5theq}-\eqref{6theq}. (c) Vertex B\'ezier control points that are not located at the boundary are defined in terms of adjacent face B\'ezier control points using Eq. \eqref{7theq}.}
\label{fevbcp}
\end{figure}

\begin{figure} [t!] 
 \centering
 \subfigure[]{\includegraphics[scale=0.3]{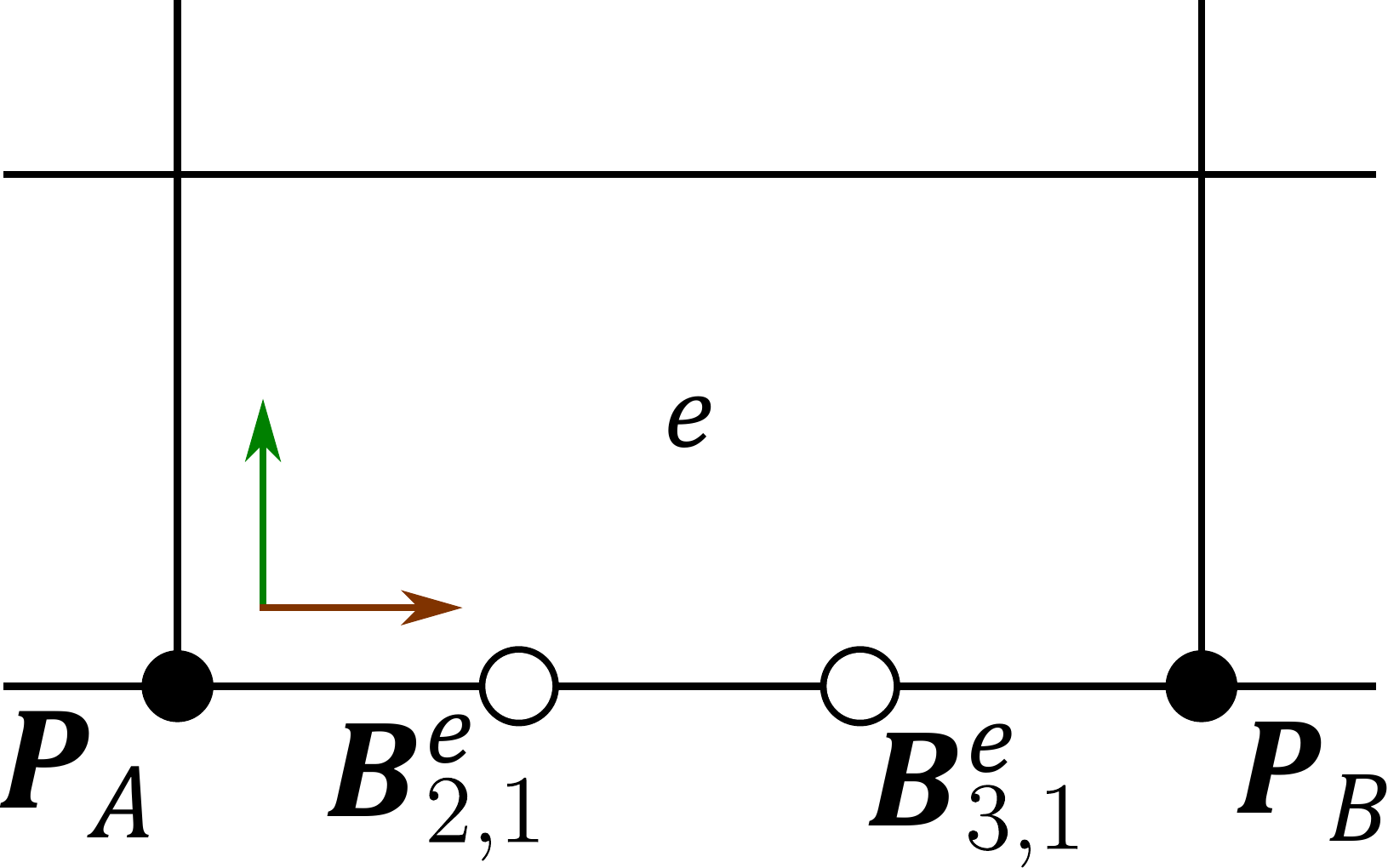}} 
 \subfigure[]{\includegraphics[scale=0.3]{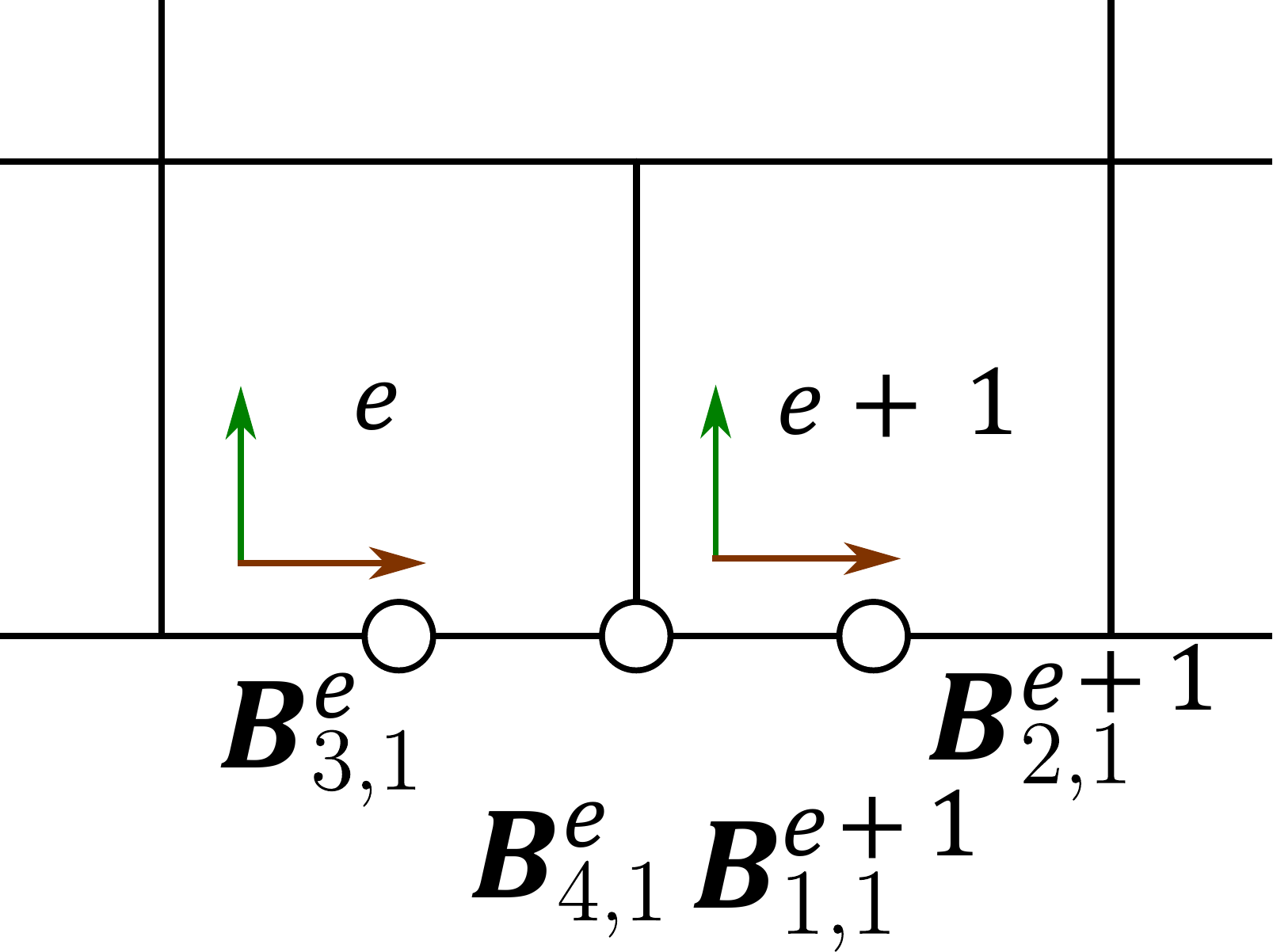}}
  \subfigure[]{\includegraphics[scale=0.3]{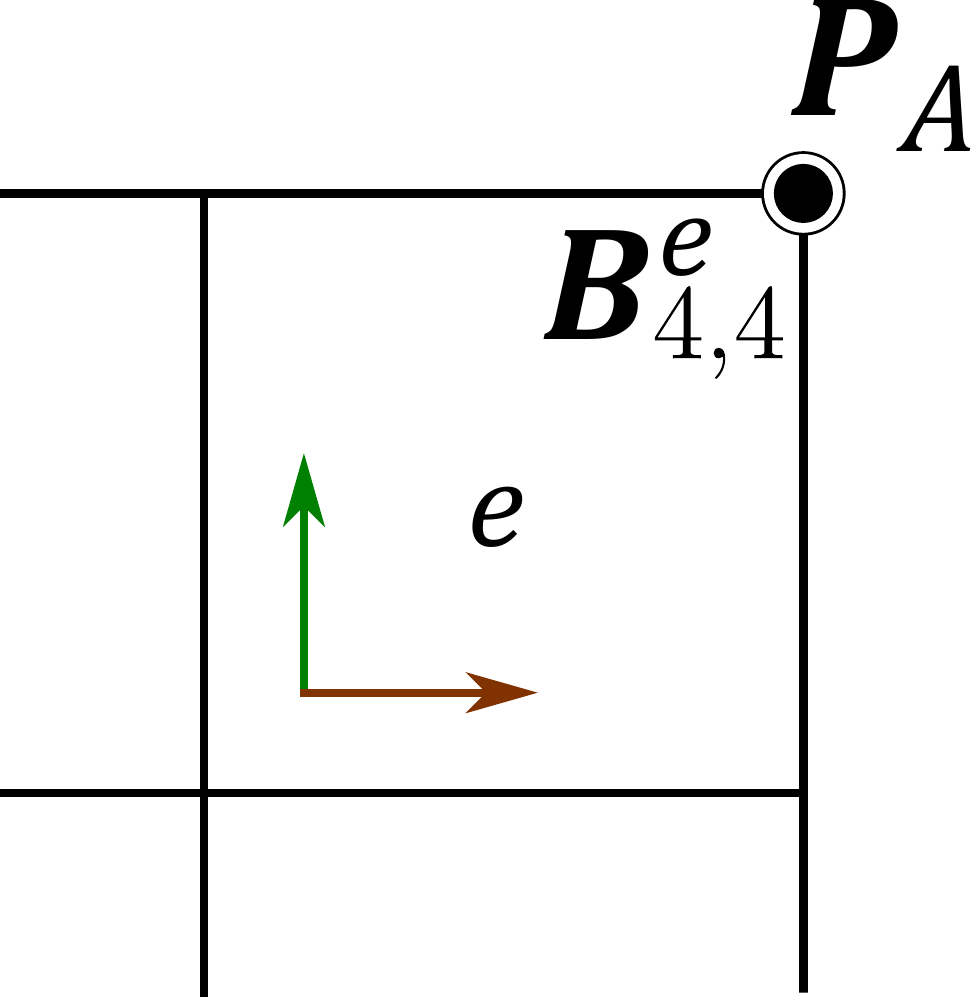}}
\caption{(a) Edge B\'ezier control points that are located at the boundary are defined in terms of adjacent spline control points using Eqs. \eqref{8theq}-\eqref{9theq}. (b) Vertex B\'ezier control points that are located at the boundary and not placed at a corner are defined in terms of adjacent edge B\'ezier control points using Eq. \eqref{10theq}. (c) Vertex B\'ezier control points that are located at the boundary and placed at a corner are equal to the spline control point placed at that corner.}
\label{bevbcp}
\end{figure}

We begin using bi-cubic polynomials for all elements. The 16 B\'ezier control points are classified in face, edge, and vertex B\'ezier control points as shown in Fig. \ref{beziercp}. Face B\'ezier control points are defined in terms of the spline control points using the expressions of a uniform bi-cubic B-spline surface, viz.,
\begin{equation}   \label{firsteq} 
\mathbf{B}_{2,2}^e = \frac{4}{9} \mathbf{P}_A + \frac{2}{9} \mathbf{P}_B + \frac{2}{9} \mathbf{P}_C + \frac{1}{9} \mathbf{P}_D \text{,} 
\end{equation}
\begin{equation}    
\mathbf{B}_{3,2}^e = \frac{2}{9} \mathbf{P}_A + \frac{4}{9} \mathbf{P}_B + \frac{1}{9} \mathbf{P}_C + \frac{2}{9} \mathbf{P}_D \text{,} 
\end{equation}
\begin{equation}    
\mathbf{B}_{2,3}^e = \frac{2}{9} \mathbf{P}_A + \frac{1}{9} \mathbf{P}_B + \frac{4}{9} \mathbf{P}_C + \frac{2}{9} \mathbf{P}_D \text{,} 
\end{equation}
\begin{equation}    \label{4theq} 
\mathbf{B}_{3,3}^e = \frac{1}{9} \mathbf{P}_A + \frac{2}{9} \mathbf{P}_B + \frac{2}{9} \mathbf{P}_C + \frac{4}{9} \mathbf{P}_D \text{.} 
\end{equation}
The labels used in Eqs. \eqref{firsteq}-\eqref{4theq} are depicted in Fig. \ref{fevbcp} a). Interior edge B\'ezier control points are defined in terms of face B\'ezier control points  using the expressions of a uniform bi-cubic B-spline surface, viz.,
\begin{equation}    \label{5theq}
\mathbf{B}_{4,2}^e = \mathbf{B}_{1,2}^{e+1} = \frac{1}{2} \mathbf{B}_{3,2}^e + \frac{1}{2} \mathbf{B}_{2,2}^{e+1} \text{,} 
\end{equation}
\begin{equation}    \label{6theq}
\mathbf{B}_{4,3}^e = \mathbf{B}_{1,3}^{e+1} = \frac{1}{2} \mathbf{B}_{3,3}^e + \frac{1}{2} \mathbf{B}_{2,3}^{e+1} \text{.} 
\end{equation}
The labels used in Eqs. \eqref{5theq}-\eqref{6theq} are depicted in Fig. \ref{fevbcp} b). Interior vertex B\'ezier control points are defined in terms of face B\'ezier control points generalizing the expression for a uniform bi-cubic B-spline surface so that partition of unity is satisfied, viz.,
\begin{equation}   \label{7theq} 
\mathbf{B}_{1,1}^1 = \mathbf{B}_{1,1}^2 = ... = \mathbf{B}_{1,1}^{\mu} = \frac{1}{\mu} \sum_{j=1}^{\mu} \mathbf{B}_{2,2}^{j}  \text{,} 
\end{equation}
where $\mu$ is the valence of the vertex. The labels used in Eq. \eqref{7theq} are depicted in Fig. \ref{fevbcp} c). Boundary edge B\'ezier control points are defined in terms of the spline control points using the expressions of a uniform cubic B-spline curve, viz.,
\begin{equation}    \label{8theq} 
\mathbf{B}_{2,1}^e = \frac{2}{3} \mathbf{P}_A + \frac{1}{3} \mathbf{P}_B \text{,} 
\end{equation}
\begin{equation}   \label{9theq} 
\mathbf{B}_{3,1}^e =  \frac{1}{3} \mathbf{P}_A + \frac{2}{3} \mathbf{P}_B \text{.} 
\end{equation} 
The labels used in Eqs. \eqref{8theq}-\eqref{9theq} are depicted in Fig. \ref{bevbcp} a). Boundary vertex B\'ezier control points that are not placed at a corner are defined in terms of the boundary edge B\'ezier control points using the expression of a uniform cubic B-spline curve, viz.,
\begin{equation}   \label{10theq} 
\mathbf{B}_{4,1}^e = \mathbf{B}_{1,1}^{e+1} = \frac{1}{2} \mathbf{B}_{3,1}^e + \frac{1}{2} \mathbf{B}_{2,1}^{e+1}  \text{.} 
\end{equation}
The labels used in Eq. \eqref{10theq} are depicted in Fig. \ref{bevbcp} b). Boundary vertex B\'ezier control points that are placed at a corner are equal to the spline control point at that corner, viz.,
\begin{equation}   \label{11theq} 
\mathbf{B}_{4,4}^e = \mathbf{P}_{A}  \text{.} 
\end{equation}
The labels used in Eq. \eqref{11theq} are depicted in Fig. \ref{bevbcp} c).

The above relations between  B\'ezier control points and spline control points define preliminary expressions for the spline extraction operator of each element. Using these spline extraction operators, spline basis functions are only $C^0$-continuous across spoke edges, but $C^2$-continuous across all the other edges. In the following, we will modify the basis functions with support on irregular elements one by one in order to obtain $G^1$ continuity across spoke edges.

\subsubsection{$G^1$ constraints}

\begin{figure} [t!] 
 \centering
 \subfigure[$G^1$ constraints]{\includegraphics[scale=.70]{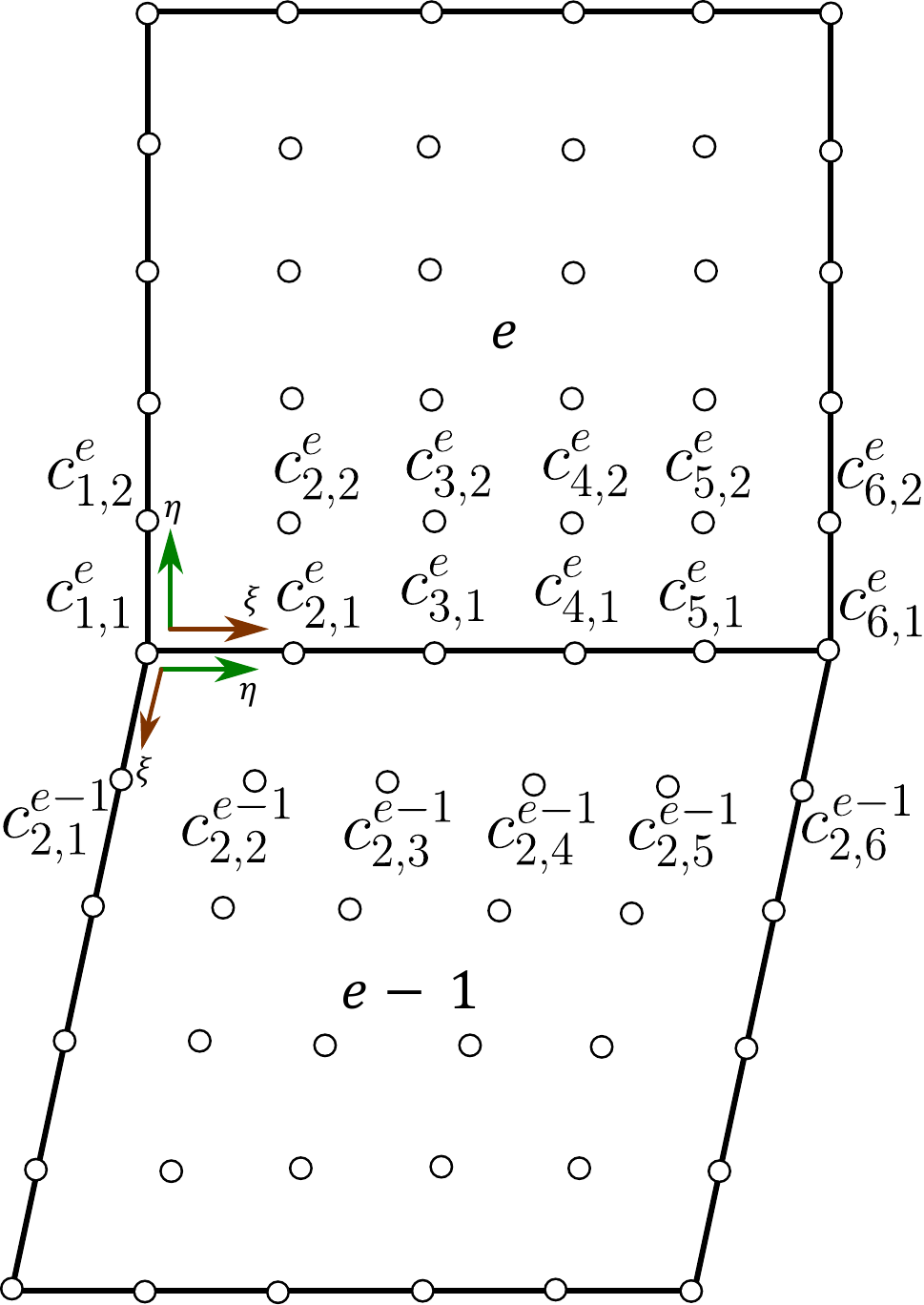}} \hspace*{+3.0mm}
 \subfigure[$C^1$ continuity]{\includegraphics[scale=.70]{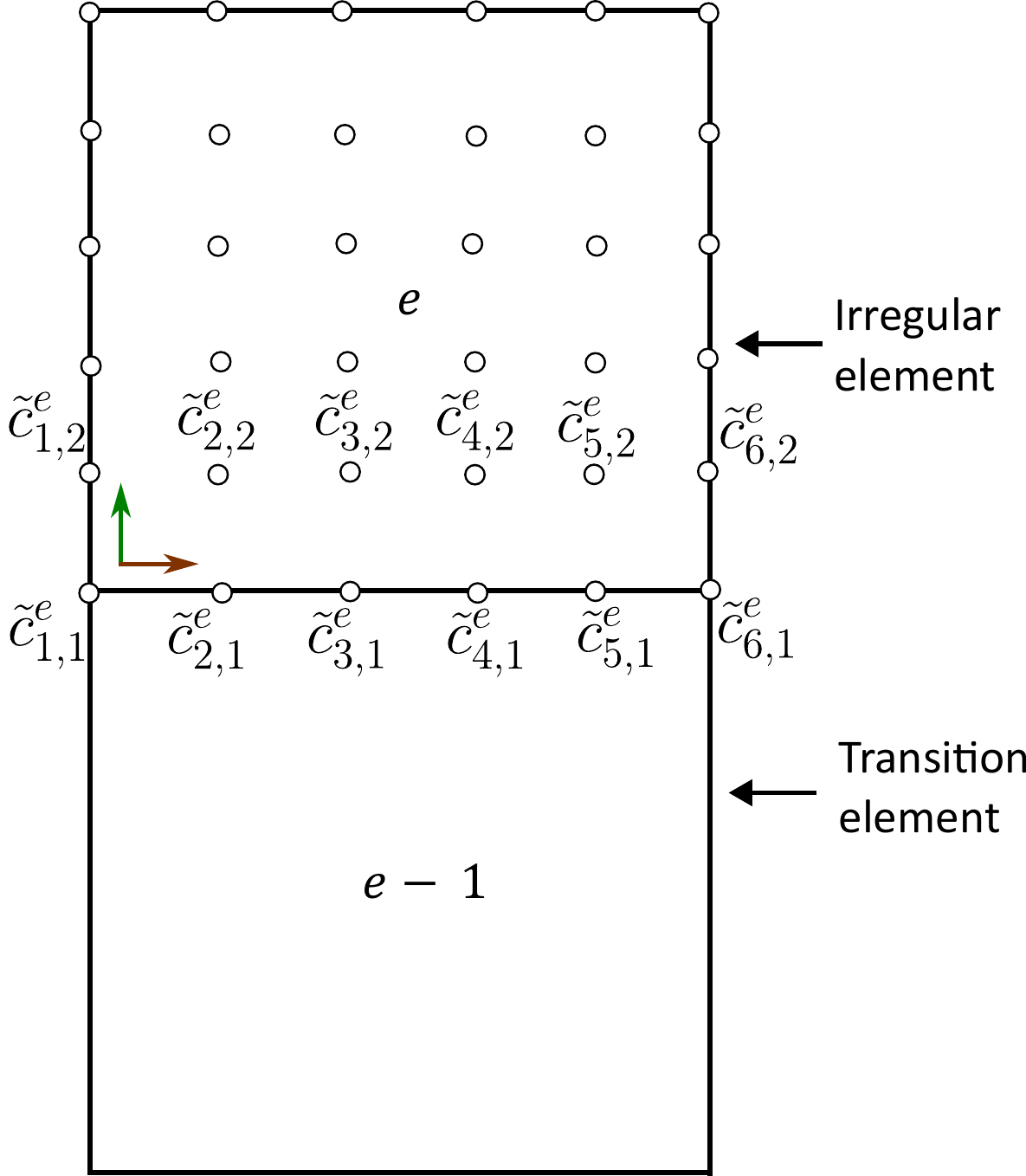}} \\
\caption{Imposing $G^1$ constraints at spoke edges and maintaining $C^1$ continuity at element boundaries between irregular and transition elements.}
\label{g1constraints}
\end{figure}

When imposing  the equations that enforce $G^1$ continuity of each basis function across spoke edges, the extraction coefficients of the basis function in irregular elements are the degrees of freedom whose values will be changed to achieve $G^1$ continuity. In order to provide additional degrees of freedom to satisfy the $G^1$ constraints, each basis function is degree elevated twice in each irregular element to be bi-quintic. After the algorithm for degree elevation of a B\'ezier surface is applied twice, a given basis function $N$ on an irregular element $e$ is represented as
\begin{equation}    
N^e \left( \xi , \eta \right) = \sum_{i=1}^{6} \sum_{j=1}^{6} \tilde{c}^e_{i,j} b_{i,j} \left( \xi , \eta \right),  \quad   \left( \xi , \eta \right)   \in   \left[ 0,1 \right]^2  \text{.}
\end{equation}
Given a basis function $N$ with $C^0$ continuity across spoke edges, as shown in \cite{gregory1990smooth}, $N$ has $G^1$ continuity across the shared boundary of two irregular elements with indices $e$ and $e-1$ if
\begin{equation} \label{g1c}   
\frac{ \partial N^{e-1}}{ \partial \xi}  (0,v) + b(v) \frac{ \partial N^e}{ \partial \xi} (v,0) + \frac{ \partial N^e}{ \partial \eta} (v,0)= 0,  \quad   v   \in   [0,1]  \text{,}
\end{equation}
where the axes of the parametric coordinates used for $N^{e-1}$ and $N^e$ are plotted in Fig. \ref{g1constraints} a). We choose $b(v)$ to be a quadratic polynomial, viz.,
\begin{equation} \label{bofu}   
b(v)= - 2 \omega_1 (1-v)^2 + 2 \omega_2 v^2,  \quad   v   \in   [0,1]  \text{,}
\end{equation}
with
\begin{equation} \label{omega}   
\omega_1 = \cos \left( \frac{a_1 \pi}{\mu_1}  \right), \; \omega_2 = \cos \left( \frac{a_2 \pi}{\mu_2}  \right)  \text{,}
\end{equation}
where $a_1$ is equal to 2 if vertex 1 is an interior vertex and equal to 1 if vertex 1 is a boundary vertex, $a_2$ is equal to 2 if vertex 2 is an interior vertex and equal to 1 if vertex 2 is a boundary vertex, $\mu_1$ and $\mu_2$ are the valences of vertices 1 and 2, respectively. Using the local axes of element $e$ specified in Fig. \ref{g1constraints} a), the coordinates of vertices 1 and 2 are $(\xi, \eta) = (0,0)$ and $(\xi, \eta) = (1,0)$, respectively. Eq. \ref{bofu} results in spoke edges that are evenly distributed around each EP. As shown in \cite{gregory1990smooth}, choosing $b(v)$ to be a linear polynomial results in a singular parameterization when vertex 1 is an extraordinary vertex and vertex 2 is not or vice versa.

Using the notation $ \langle c^e_1, c^e_2, ..., c^e_{p+1} \rangle^p  (v) = \sum_{i=1}^{p+1} c^e_i \mathfrak{B}_i (v)$ where $\mathfrak{B}_i (v)$ is the $i$th univariate Bernstein polynomial of degree $p$ in $v$, the terms from Eq. \eqref{g1c} can be written as follows:
\begin{equation}   
\frac{ \partial N^{e-1}}{ \partial \xi}  (0,v) = 5 \langle c^{e-1}_{2,1} - c^e_{1,1}, c^{e-1}_{2,2} - c^e_{2,1}, c^{e-1}_{2,3} - c^e_{3,1}, c^{e-1}_{2,4} - c^e_{4,1}, c^{e-1}_{2,5} - c^e_{5,1}, c^{e-1}_{2,6} - c^e_{6,1} \rangle^5  (v) \text{,}
\end{equation}
\begin{align} \label{bu}
b(v) \frac{ \partial N^e}{ \partial \xi} (v,0) & =  \langle - 2 \omega_1, 0, 2 \omega_2 \rangle^2  (v) \langle 5 (c^e_{2,1} - c^e_{1,1}), 5/3 (4c^e_{3,1} - 5c^e_{2,1} + c^e_{1,1}),  \nonumber \\
 & 5/3 (5 c^e_{5,1} - 4c^e_{4,1} - c^e_{6,1}), 5 (c^e_{6,1} - c^e_{5,1}) \rangle^3  (v) \nonumber \\
 & = \langle - 10 \omega_1 ( c^{e}_{2,1} - c^{e}_{1,1} ), - \omega_1 (8 c^{e}_{3,1} - 10 c^{e}_{2,1} + 2 c^{e}_{1,1} ),  \nonumber \\
 & - \omega_1 (5 c^{e}_{5,1} - 4 c^{e}_{4,1} - c^{e}_{6,1} ) + \omega_2 ( c^{e}_{2,1} - c^{e}_{1,1} ) ,  \nonumber \\
 & - \omega_1 ( c^{e}_{6,1} - c^{e}_{5,1} ) + \omega_2 (4 c^{e}_{3,1} - 5 c^{e}_{2,1} + c^{e}_{1,1} ),  \nonumber \\
 & \omega_2 (10 c^{e}_{5,1} - 8 c^{e}_{4,1} - 2 c^{e}_{6,1} ), 10 \omega_2 ( c^{e}_{6,1} - c^{e}_{5,1} ) \rangle^5  (v)   \text{,}
\end{align}
\begin{equation}   
\frac{ \partial N^{e}}{ \partial \eta}  (v,0) = 5 \langle c^{e}_{1,2} - c^e_{1,1}, c^{e}_{2,2} - c^e_{2,1}, c^{e}_{3,2} - c^e_{3,1}, c^{e}_{4,2} - c^e_{4,1}, c^{e}_{5,2} - c^e_{5,1}, c^{e}_{6,2} - c^e_{6,1} \rangle^5  (v) \text{.}
\end{equation}
The left hand side of Eq. \eqref{g1c} is a polynomial of degree five in $v$ which is zero only if the following six equations are satisfied
\begin{align} \label{feqq} 
   5 ( c^{e-1}_{2,1} - c^{e}_{1,1} ) - 10 \omega_1 ( c^{e}_{2,1} - c^{e}_{1,1} )  + 5 ( c^{e}_{1,2} - c^{e}_{1,1} ) &=  0  \text{,} \\
   5 ( c^{e-1}_{2,2} - c^{e}_{2,1} ) - \omega_1 (8 c^{e}_{3,1} - 10 c^{e}_{2,1} + 2 c^{e}_{1,1} ) + 5 ( c^{e}_{2,2} - c^{e}_{2,1} ) &=  0  \text{,} \\
   5 ( c^{e-1}_{2,3} - c^{e}_{3,1} ) - \omega_1 (5 c^{e}_{5,1} - 4 c^{e}_{4,1} - c^{e}_{6,1} ) + \omega_2 ( c^{e}_{2,1} - c^{e}_{1,1} ) + 5 ( c^{e}_{3,2} - c^{e}_{3,1} ) &=  0  \text{,} \\
   5 ( c^{e-1}_{2,4} - c^{e}_{4,1} ) - \omega_1 ( c^{e}_{6,1} - c^{e}_{5,1} ) + \omega_2 (4 c^{e}_{3,1} - 5 c^{e}_{2,1} + c^{e}_{1,1} )  + 5 ( c^{e}_{4,2} - c^{e}_{4,1} ) &=  0  \text{,} \\
   5 ( c^{e-1}_{2,5} - c^{e}_{5,1} ) + \omega_2 (10 c^{e}_{5,1} - 8 c^{e}_{4,1} - 2 c^{e}_{6,1} )  + 5 ( c^{e}_{5,2} - c^{e}_{5,1} ) &=  0  \text{,} \\
   5 ( c^{e-1}_{2,6} - c^{e}_{6,1} ) + 10 \omega_2 ( c^{e}_{6,1} - c^{e}_{5,1} ) + 5 ( c^{e}_{6,2} - c^{e}_{6,1} ) &=  0  \text{.}
\end{align}

Eq. \eqref{bu} assumes that the shared boundary curve is a quartic polynomial. This is imposed by enforcing that the fifth derivative of the shared boundary curve vanishes, viz.,
\begin{equation}  \label{leqq} 
   - c^{e}_{1,1} + 5 c^{e}_{2,1} - 10 c^{e}_{3,1} + 10 c^{e}_{4,1} - 5 c^{e}_{5,1} + c^{e}_{6,1}   =  0  \text{.}
\end{equation}
For a given basis function, Eqs. \eqref{feqq}-\eqref{leqq} are applied to all the spoke edges of the extraordinary vertices in which the basis function (as defined in Section 2.3.1) has support. If these spoke edges reach a new extraordinary vertex, Eqs. \eqref{feqq}-\eqref{leqq} are applied to all the spoke edges of that extraordinary vertex as well. This operation is repeated recursively. As a result, the support of a basis function can increase after enforcing $G^1$ continuity.

Applying Eqs. \eqref{feqq}-\eqref{leqq} may result in loss continuity across element boundaries between irregular and transition elements. In order to maintain $C^1$ continuity between irregular and transition elements, the extraction coefficients of irregular elements located at element boundaries between irregular and transition elements and one layer inwards from these boundaries are kept unchanged, viz.,
\begin{equation}  \label{c1constraints} 
   c^{e}_{1,1} =  \tilde{c}^{e}_{1,1}, \quad c^{e}_{2,1} =  \tilde{c}^{e}_{2,1}, \quad c^{e}_{3,1} =  \tilde{c}^{e}_{3,1}, \quad c^{e}_{4,1} =  \tilde{c}^{e}_{4,1} , \quad c^{e}_{5,1} =  \tilde{c}^{e}_{5,1},  \quad c^{e}_{6,1} =  \tilde{c}^{e}_{6,1}     \text{.}
\end{equation}
\begin{equation}  \label{c1constraints2} 
   c^{e}_{1,2} =  \tilde{c}^{e}_{1,2}, \quad c^{e}_{2,2} =  \tilde{c}^{e}_{2,2}, \quad c^{e}_{3,2} =  \tilde{c}^{e}_{3,2}, \quad c^{e}_{4,2} =  \tilde{c}^{e}_{4,2} , \quad c^{e}_{5,2} =  \tilde{c}^{e}_{5,2},  \quad c^{e}_{6,2} =  \tilde{c}^{e}_{6,2}     \text{.}
\end{equation}
The labels used in Eqs. \eqref{c1constraints}-\eqref{c1constraints2} are depicted in Fig. \ref{g1constraints} b).

\begin{figure} [t!] 
 \centering
 \subfigure[Before applying $G^1$ constraints]{\includegraphics[scale=.82]{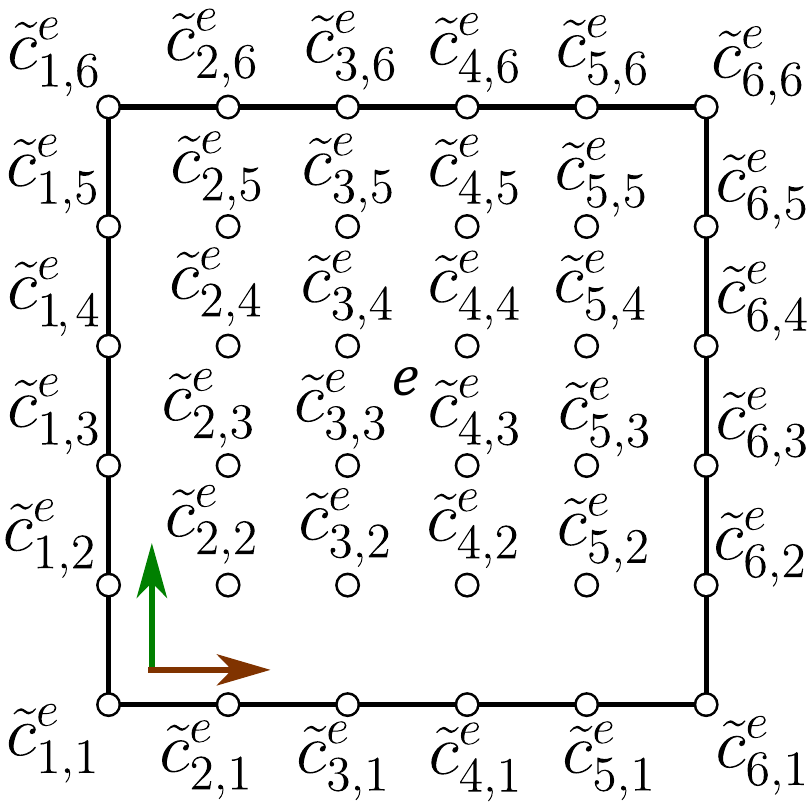}} \hspace*{+5.0mm}
 \subfigure[After applying $G^1$ constraints]{\includegraphics[scale=.82]{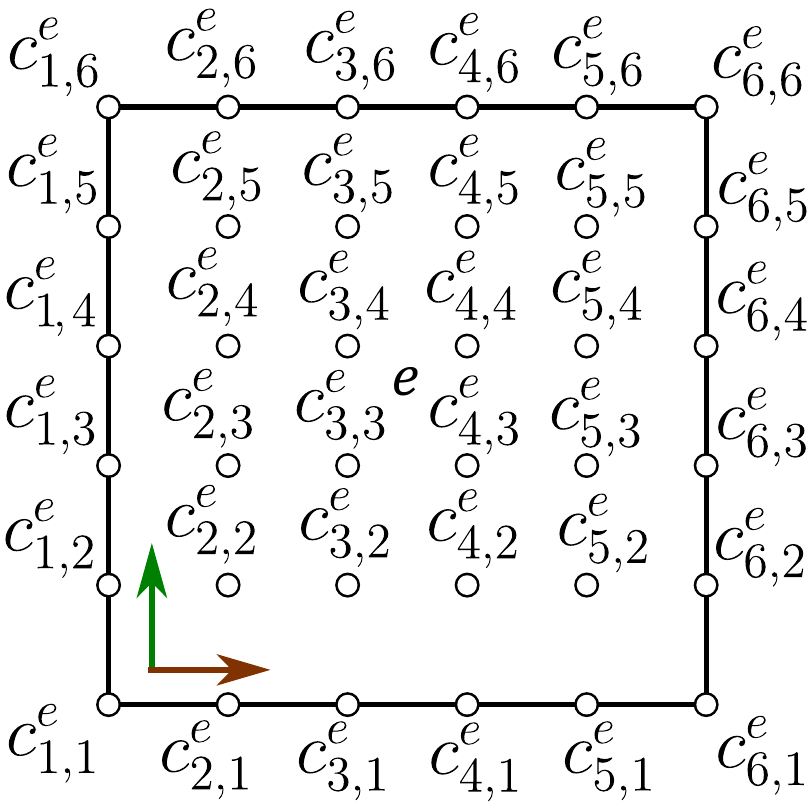}} \\
\caption{Fairing equations.}
\label{fairing}
\end{figure}

For each basis function, all the equations that enforce $G^1$ continuity at spoke edges (Eqs. \eqref{feqq}-\eqref{leqq}) and all the equations that enforce $C^1$ continuity across element boundaries between irregular and transition elements (Eqs. \eqref{c1constraints}-\eqref{c1constraints2}) result in a system of linear algebraic equations with matrix of coefficients $\mathbf{G}$ and right-hand-side vector $\mathbf{g}$. This system of linear algebraic equations has more unknowns than equations and some of the equations are linearly dependent. In order to obtain a high-quality surface, fairing equations are needed. Since the surface quality in the interior of irregular elements is already high before imposing the $G^1$ constraints, a simple and efficient option is to pick the fairing equations so as to minimize the vertical and horizontal differences of the extraction coefficients before and after applying the $G^1$ constraints, viz.,
\begin{align}  
    c^{e}_{i,j} - c^{e}_{i+1,j} &=  \tilde{c}^{e}_{i,j} - \tilde{c}^{e}_{i+1,j}, \quad  1 \le i \le 5, \quad 1\le j \le 6  \label{fairing1} \text{,} \\
   c^{e}_{i,j} - c^{e}_{i,j+1} &=  \tilde{c}^{e}_{i,j} - \tilde{c}^{e}_{i,j+1}, \quad  1 \le i \le 6, \quad 1\le j \le 5 \label{fairing2}  \text{.} 
\end{align}
The labels used in Eqs. \eqref{fairing1}-\eqref{fairing2} are depicted in Fig. \ref{fairing}. For each basis function, all the fairing equations for each irregular element result in a system of linear algebraic equations with matrix of coefficients $\mathbf{F}$ and right-hand-side vector $\mathbf{f}$.

For each basis function, a constrained least square problem needs to be solved in order to obtain its extraction coefficients, viz.,
\begin{equation}  \label{cleasts} 
   \min_{\mathbf{c}}  \vert\vert \mathbf{F} \mathbf{c}  - \mathbf{f} \vert\vert_2 \quad \text{subject to} \quad  \mathbf{G} \mathbf{c}  = \mathbf{g} \text{.}
\end{equation}
From here on out, the EP construction described above is referred to as construction $G^1$P. The polynomial basis functions of the construction $G^1$P satisfy partition of unity. This can be shown in three steps. First, the preliminary expressions for the basis functions defined in Section 2.3.1 satisfy partition of unity since the coefficients of the right-hand sides of Eqs. \eqref{firsteq}-\eqref{11theq} sum to one. Second, the algorithm of degree elevation preserves partition of unity. Third, the constrained least square problem also preserves partition of unity. Since we always apply either Eqs. \eqref{feqq}-\eqref{leqq} or Eqs. \eqref{c1constraints}-\eqref{c1constraints2} to all basis functions for a given edge, this third step can be proven by showing that if all $\tilde{c}^{e}_{i,j}$ are one, then all $c^{e}_{i,j}$ are also one. This is the case as Eqs. \eqref{feqq}-\eqref{fairing2} are satisfied if all $\tilde{c}^{e}_{i,j}$ and all $c^{e}_{i,j}$ are one. Since partition of unity is satisfied if each column of the spline extraction operator sums to one, we have also numerically verified this is the case for all the meshes that we have built using this EP construction.

It is possible to enforce $G^1$ continuity of each basis function at the spoke edges without increasing the support of any basis function. This is accomplished by applying Eqs. \eqref{feqq}-\eqref{leqq} to only the spoke edges that are in the interior of the basis function's support (as defined in Section 2.3.1) and then applying Eqs. \eqref{c1constraints}-\eqref{c1constraints2} not only to element boundaries between irregular and transition elements, but also to the spoke edges that are at the boundary of the basis function's support (as defined in Section 2.3.1). All these equations result in a system of linear algebraic equations with matrix of coefficients $\mathbf{G}$ and right-hand-side vector $\mathbf{g}$. The fairing equations are kept unchanged and a constrained least square problem needs to be solved as in Eq. \eqref{cleasts}. From here on out, this EP construction is referred to as construction $G^1$R. The polynomial basis functions of the construction $G^1$R do not exactly satisfy partition of unity in irregular elements. This is caused by the fact that for a given spoke edge, Eqs. \eqref{feqq}-\eqref{leqq} are applied to certain basis functions, but Eqs. \eqref{c1constraints}-\eqref{c1constraints2} are  applied to other basis functions. In order to recover partition of unity, the basis functions are rationalized in this construction as shown in Eq. \eqref{rationalbasis}.

A mathematical proof of the linear independence of the basis functions obtained with constructions $G^1$P and $G^1$R is beyond the scope of the current work. However, we have extensively tested both constructions solving boundary-value problems and eigenvalue problems. All the numerical results suggest that linear independence is satisfied for both constructions.

\subsection{Geometric map}

\begin{figure} [t!] 
 \centering
 \subfigure[Construction $G^1$P]{\includegraphics[scale=.33]{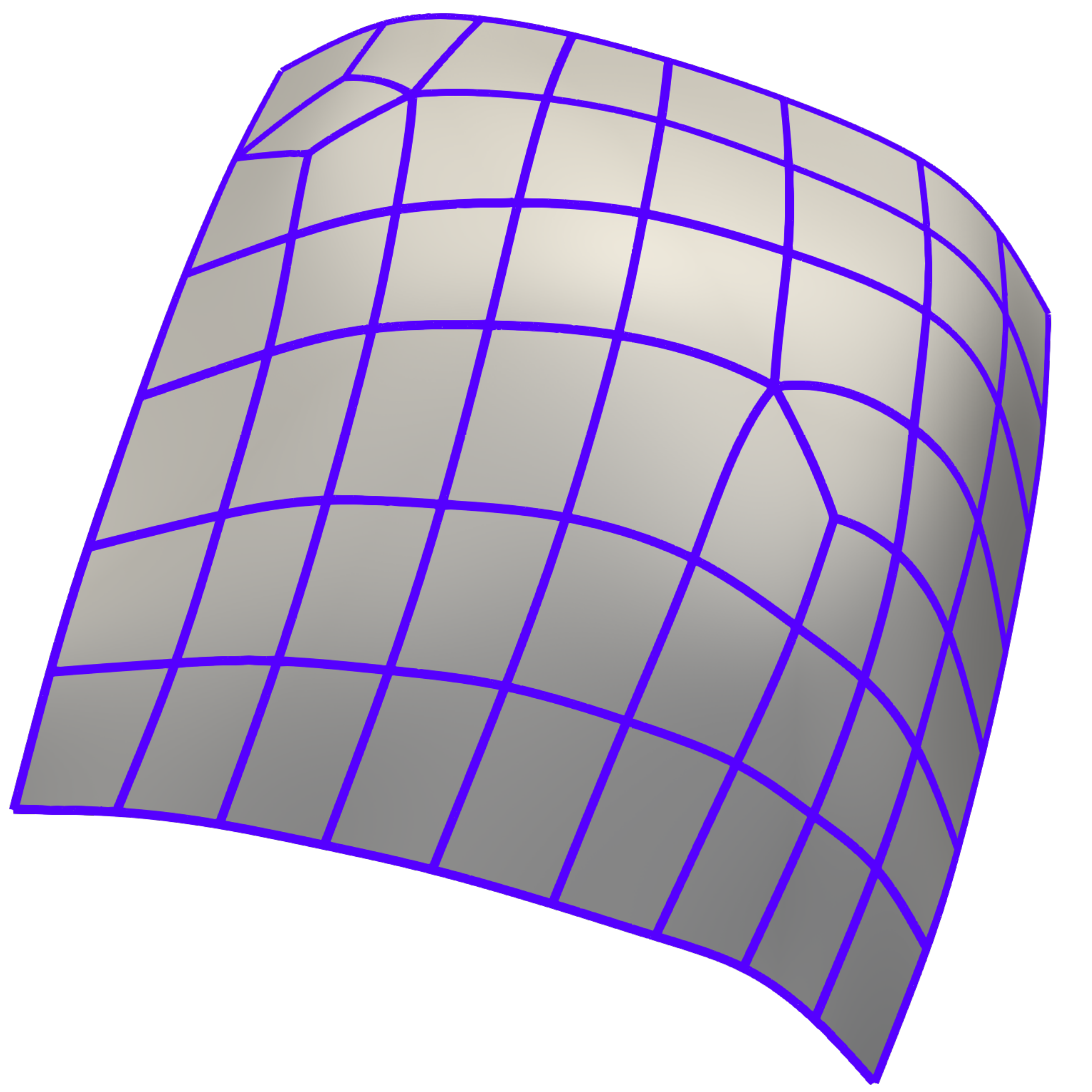}} \hspace*{+5.0mm}
 \subfigure[Construction $G^1$R]{\includegraphics[scale=.32]{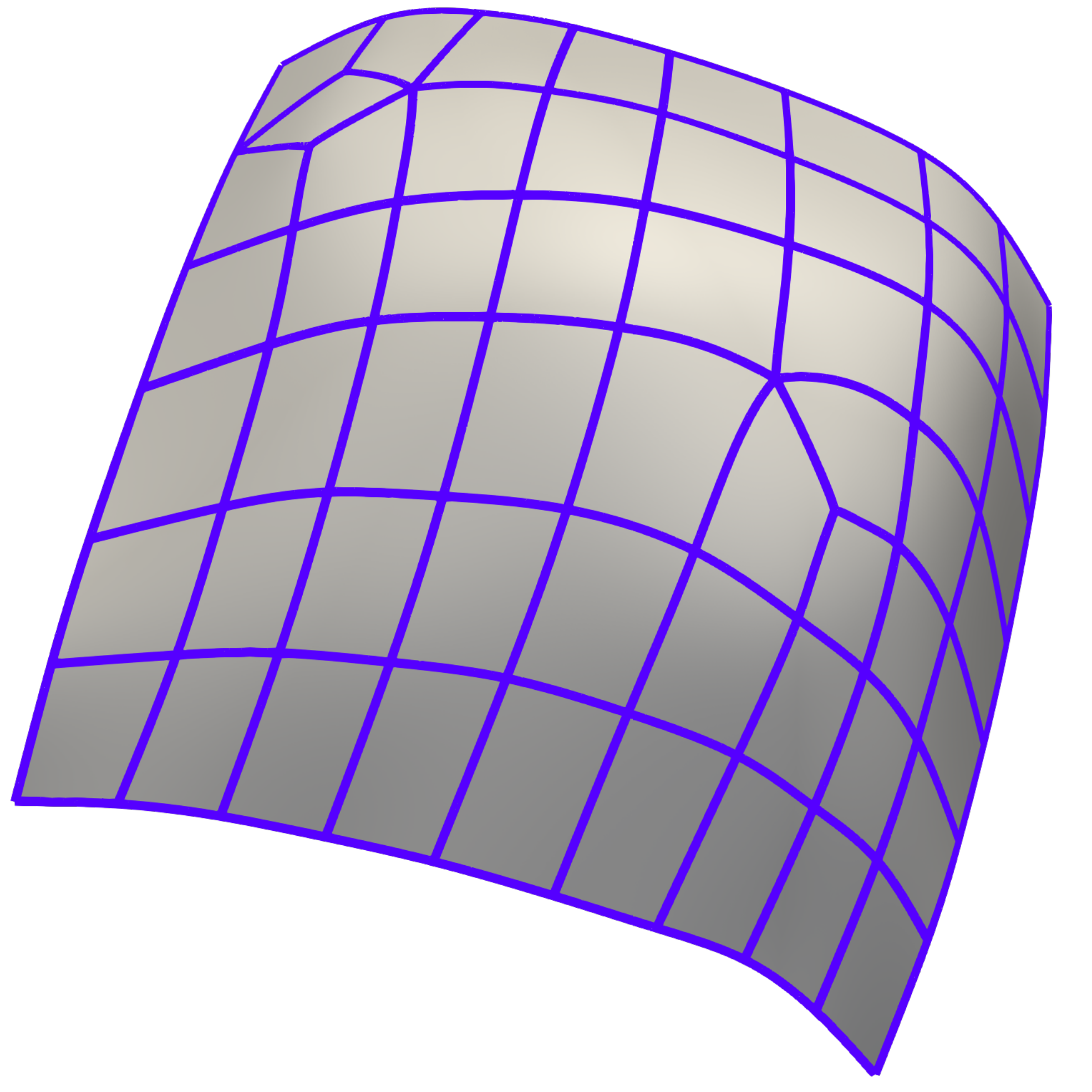}} \\
\caption{(Color online) G-spline surface and B\'ezier mesh.}
\label{beziermesh}
\end{figure}

Using the construction $G^1$P, the G-spline surface is obtained applying the following mapping
\begin{equation}
\vec x^e \left( \xi, \eta \right) = \sum_{a=1}^{n^e} \vec{P}^e_a N^e_a \left( \xi, \eta \right),    \quad \forall e\in\{1,2,...,n_{el}\}, \quad \left( \xi, \eta \right) \in \left[ 0,1 \right]^2 \text{,}
\end{equation}
while using the construction $G^1$R, the G-spline surface is obtained applying the following mapping
\begin{equation}
\vec x^e \left( \xi, \eta \right) = \sum_{a=1}^{n^e} \vec{P}^e_a R^e_a \left( \xi, \eta \right),    \quad \forall e\in\{1,2,...,n_{el}\}, \quad \left( \xi, \eta \right) \in \left[ 0,1 \right]^2 \text{,}
\end{equation}
where $R^e_a = N^e_a$ in transition and regular elements. The B\'ezier mesh is obtained by plotting the element boundaries over the G-spline surface. Fig. \ref{beziermesh} a) and b) plot the B\'ezier mesh associated with the control net shown in Fig. \ref{tnet} (b) using the constructions $G^1$P and $G^1$R, respectively. The G-spline surface and B\'ezier meshes obtained with the constructions $G^1$P and $G^1$R are indistinguishable at the scale of the plot. Thus, for brevity, we will only plot G-spline surfaces and B\'ezier meshes using the construction $G^1$P in the remainder of this paper.

In order to solve partial differential equations (PDEs) using G-splines, the Bubnov-Galerkin method and the isoparametric concept are used, i.e., the same basis functions are used for the geometry, the weighting space, and the trial space.

\subsection{Refinement}

	\begin{figure} [t!] 
\centering
\includegraphics[width=8cm]{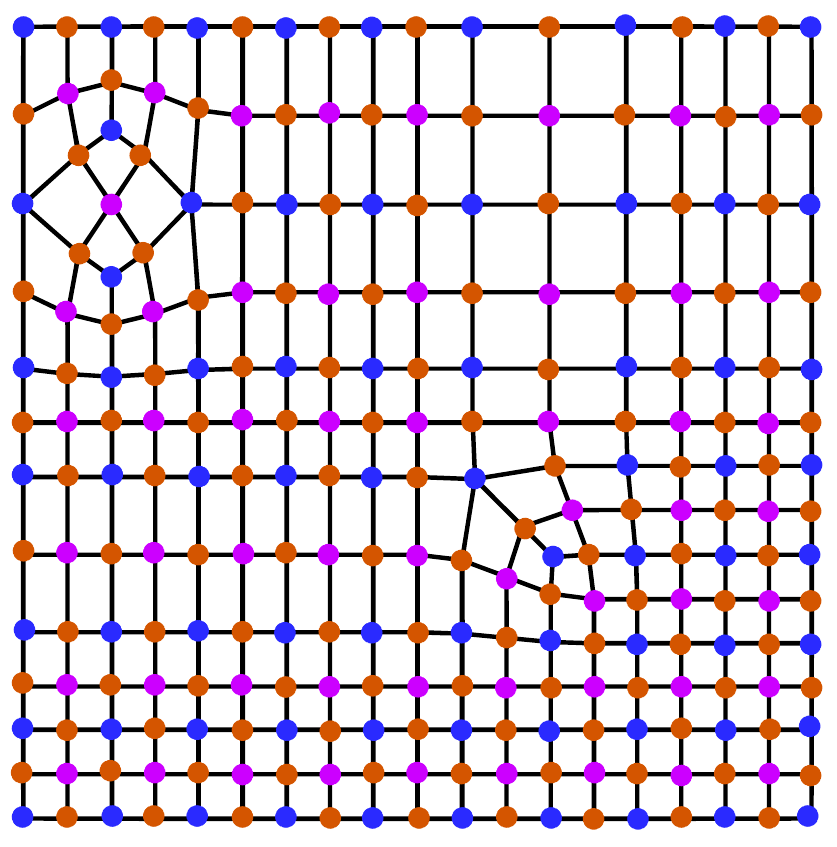}
\caption{(Color online)  C-net obtained after applying one level of global uniform refinement to the C-net given in Fig. \ref{tnet} a). Vertices associated with new face control points, new edge control points, and new vertex control points are marked with purple, brown, and blue circles, respectively.} 
\label{refined}
\end{figure}

Book-keeping means it is needed to keep track of how many refinement levels have already been applied since the algorithm used to perform the next refinement level depends on how many refinement levels have been previously applied. Book-keeping is a pathway to build refinement algorithms for EP constructions that result in either nested spaces \cite{toshniwal2017smooth} or just constant parameterization through refinement \cite{ casquero2020seamless, wei2022analysis}. In turn, either nestedness or constant parameterization through refinement are known pathways to optimal asymptotic convergence rates. When performing a convergence study on a simple geometry, book-keeping does not pose any shortcoming since the geometry can be exactly represented with a very coarse mesh and multiple refinement levels can be performed afterwards. However, in real-world engineering applications that involve thin-walled structures, the engineer is usually interested in keeping all the small features (for example, holes) whose size is above the element size that the engineer has decided to use in the simulations (usually 5 mm in the case of the automotive industry). In this scenario, book-keeping cannot be deployed since the geometry cannot be represented with a coarse mesh that can be refined multiple times afterwards. Thus, we have decided to not pursue a refinement strategy that relies on book-keeping to refine G-splines. Instead, we have chosen a simple refinement strategy with no book-keeping that only involves vertex-based basis functions. This strategy is explained below.

\begin{figure} [t!] 
 \centering
 \subfigure[]{\includegraphics[scale=0.3]{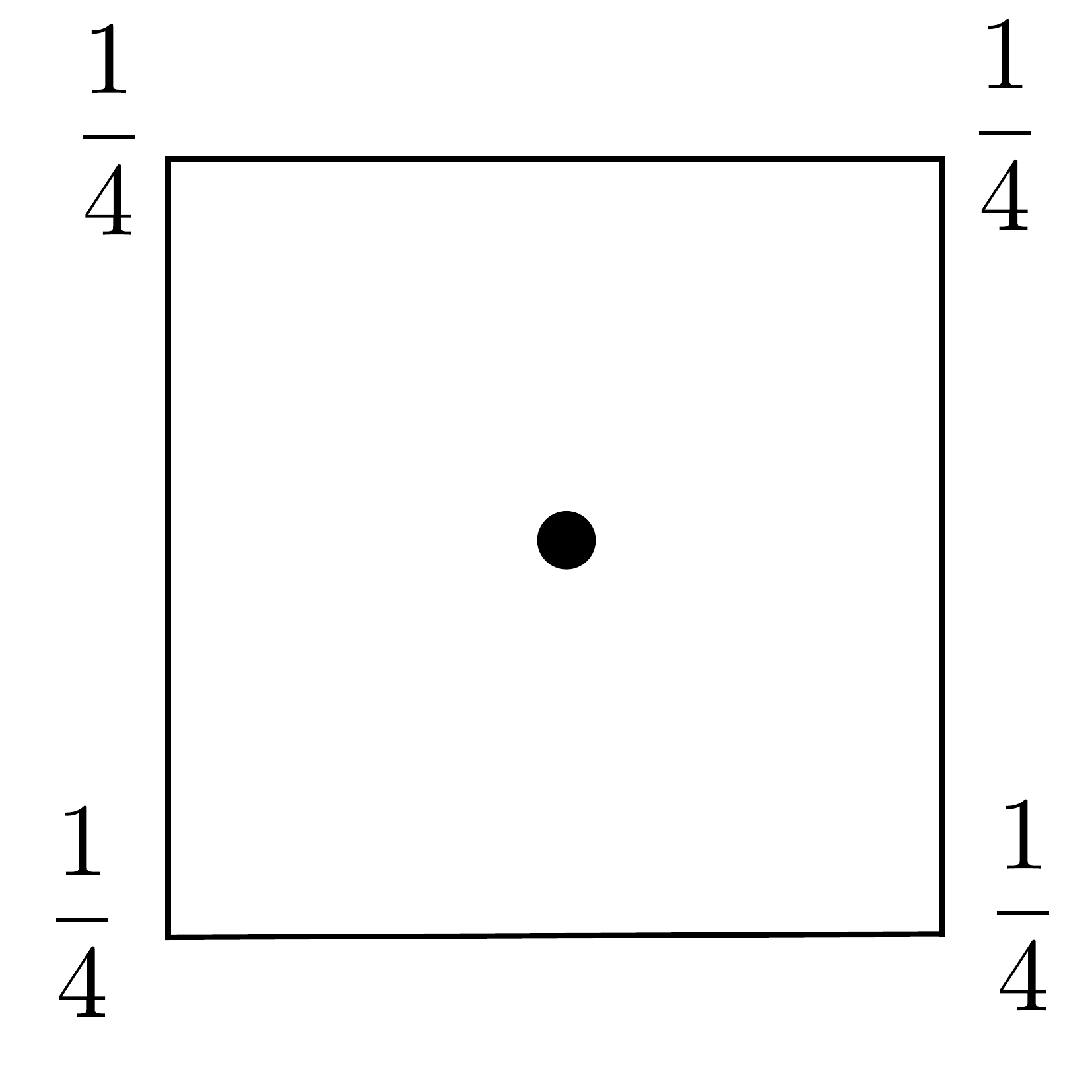}} \hspace*{+1.0mm}\
 \subfigure[]{\includegraphics[scale=0.3]{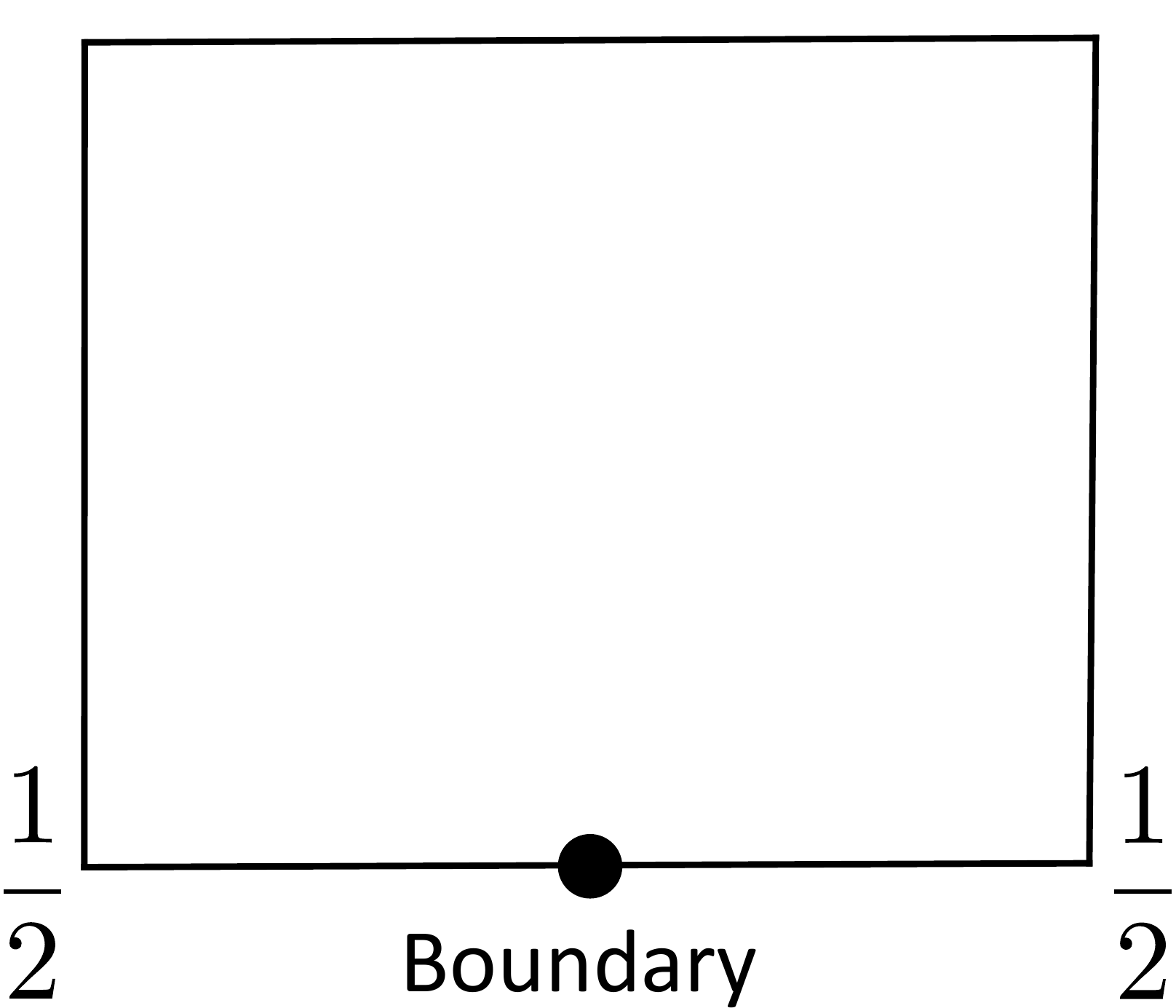}} \hspace*{+1.0mm} \
 \subfigure[]{\includegraphics[scale=0.3]{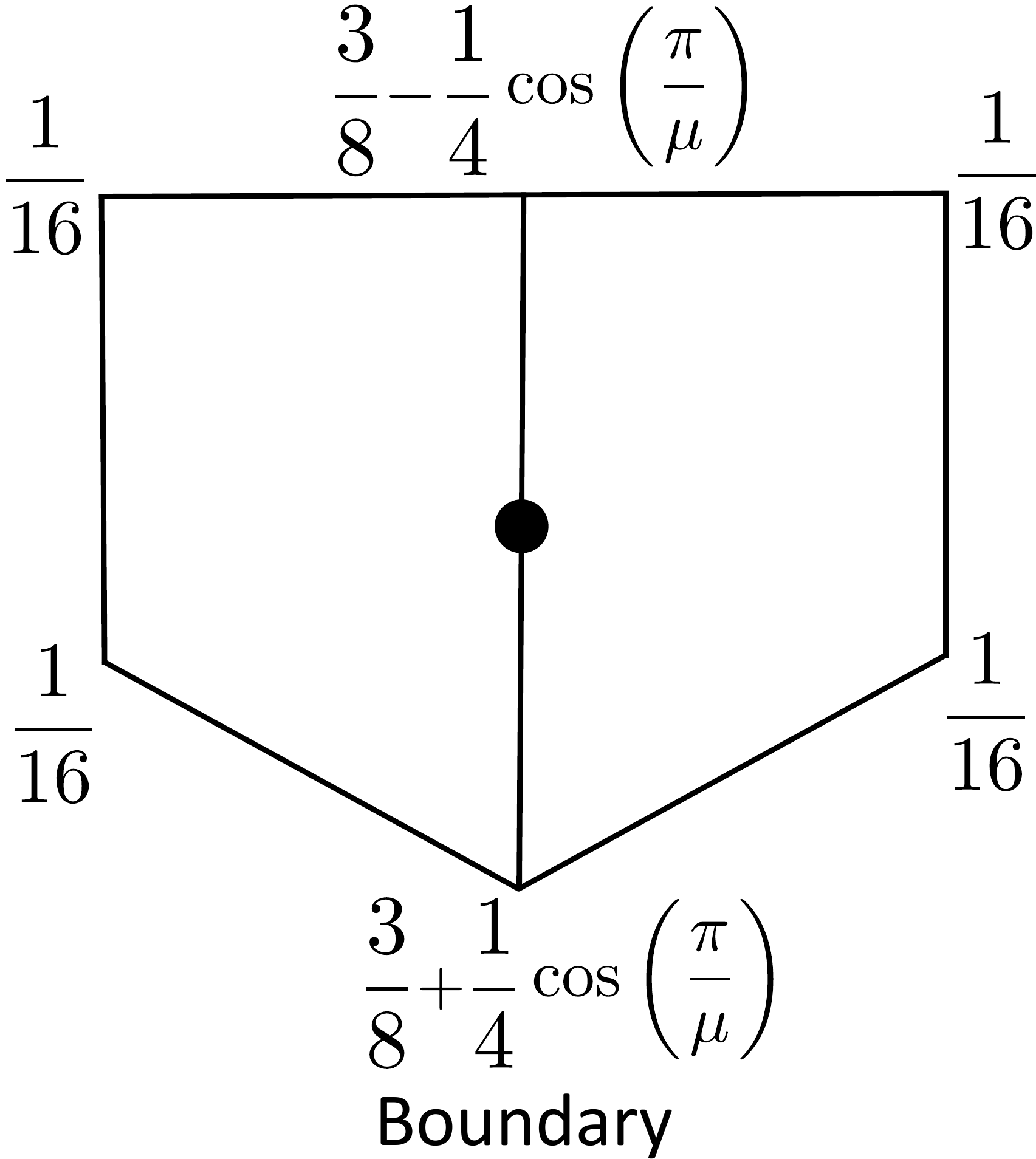}} \
 \subfigure[]{\includegraphics[scale=0.3]{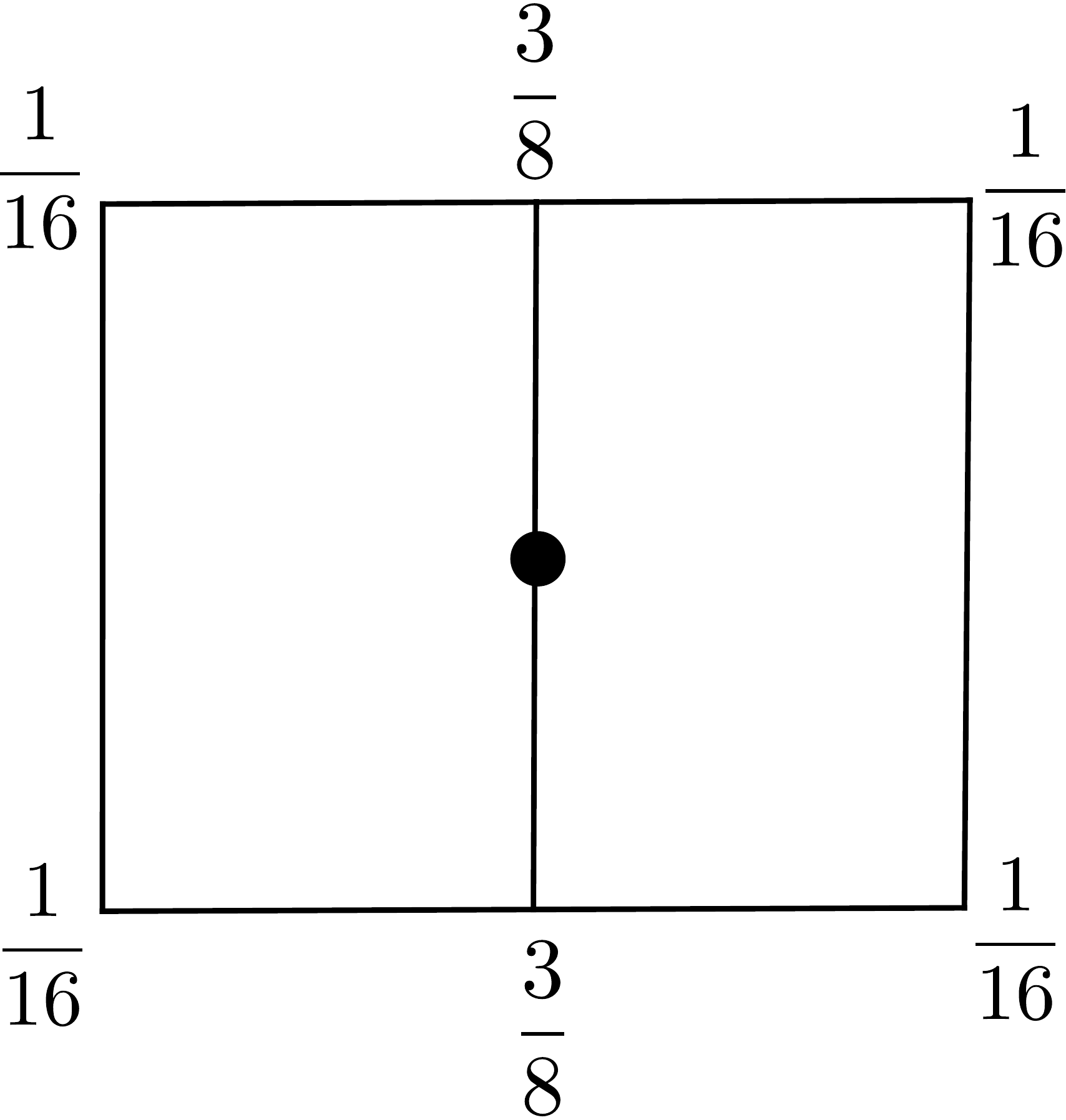}}  \hspace*{+1.0mm}\
 \subfigure[]{\includegraphics[scale=0.3]{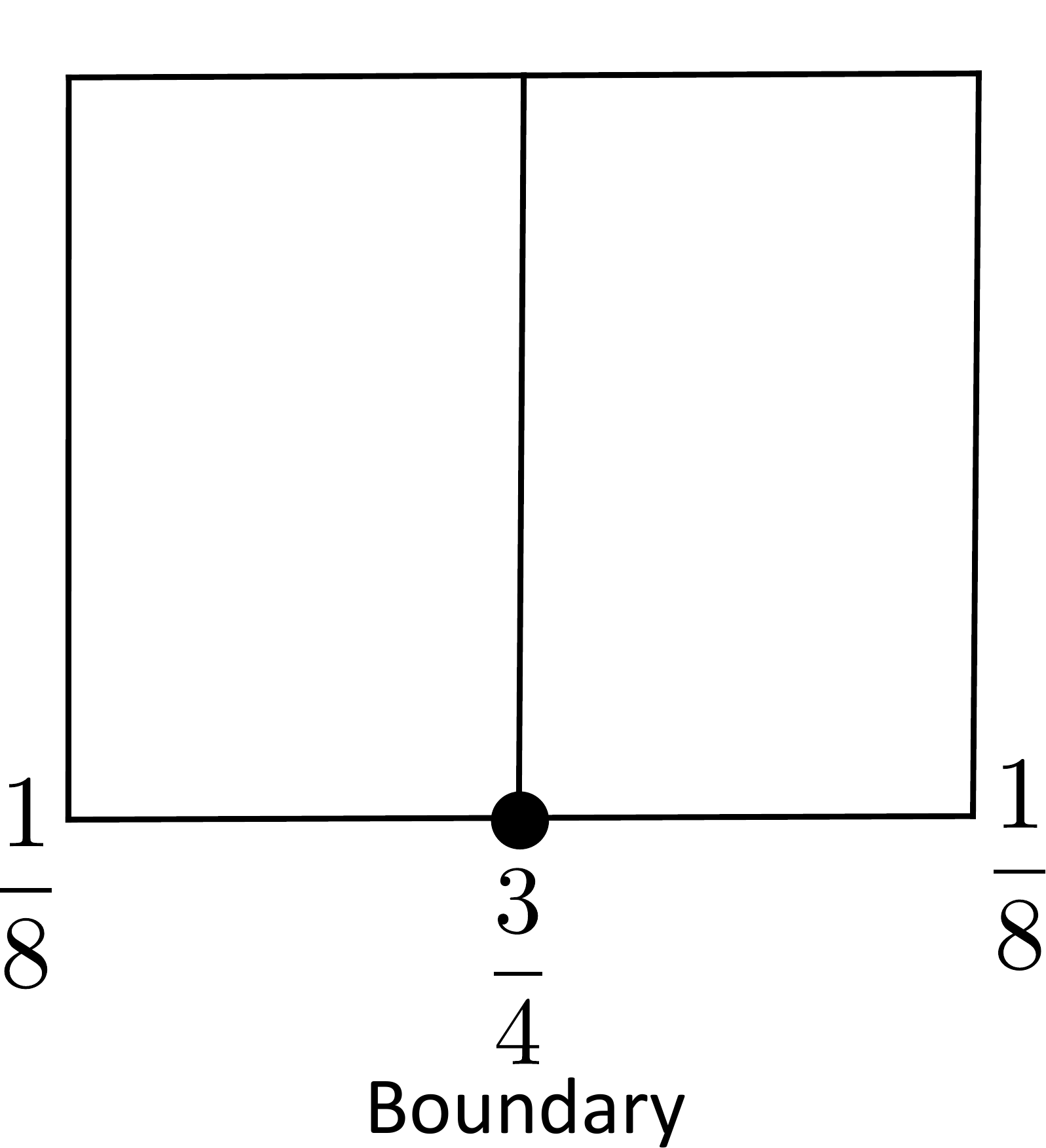}} \hspace*{+1.0mm}\
 \subfigure[]{\includegraphics[scale=0.3]{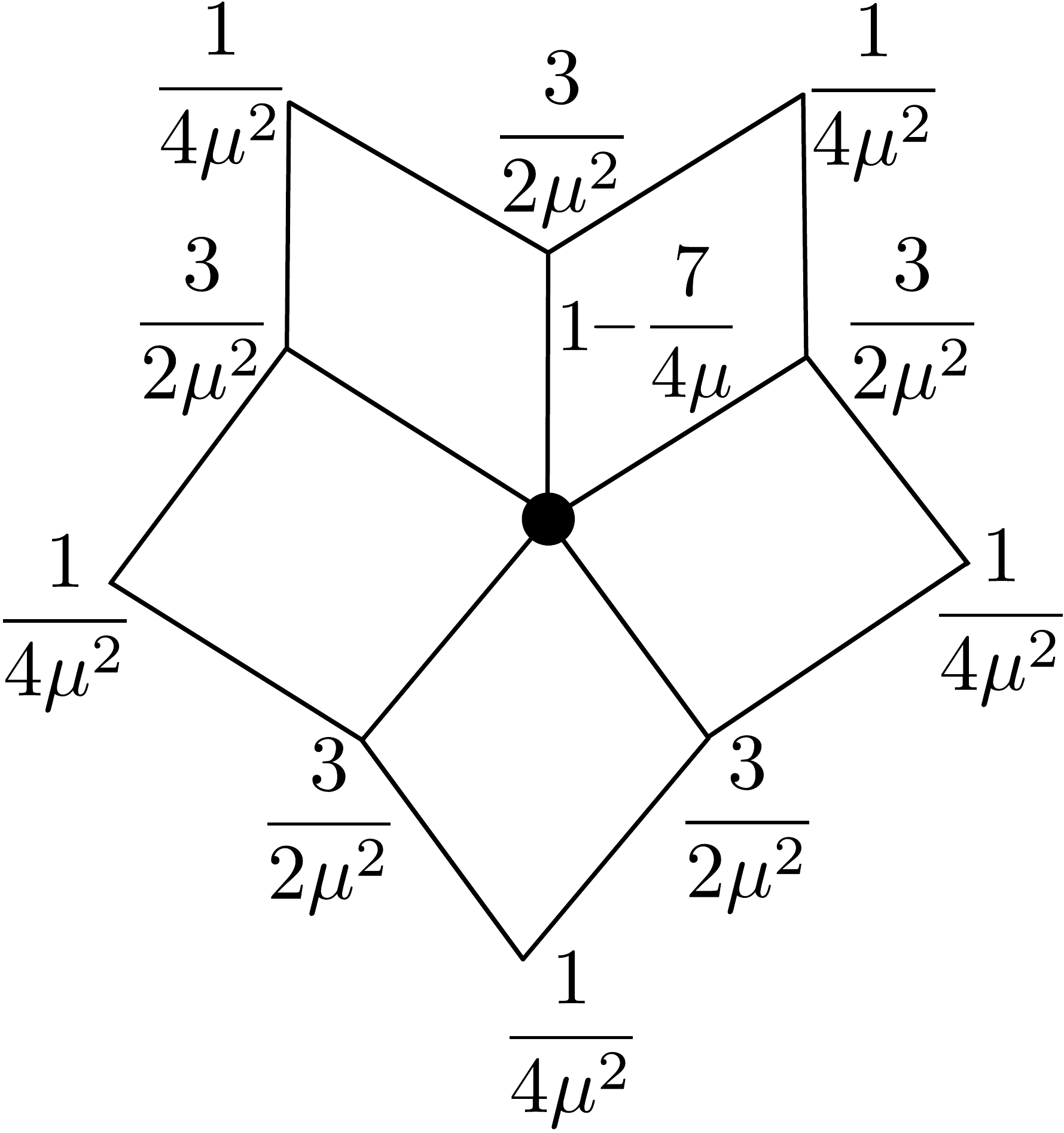}} \\
\caption{Refinement rules. (a) New face control point.  (b) New edge control point at a boundary edge. (c) New edge control point at an edge that emanates from a boundary vertex and is not a boundary edge.  (d) New edge control point at an edge that does not have boundary vertices. (e) New vertex control point at a boundary vertex. (f) New vertex control point at an interior vertex.}
\label{refrules}
\end{figure}

In each level of global uniform refinement, each face of the C-net is split into four faces. Fig. \ref{refined} plots the C-net obtained after applying one level of global uniform refinement to the C-net given in Fig. \ref{tnet} a). The total number of extraordinary vertices remains unchanged through refinement. Since no book-keeping is used, the basis functions associated with the vertices of the refined C-net are obtained as explained in Section 2.3, that is, the steps to follow to compute the basis functions are independent of whether or not the C-net has been refined. The coordinates of the control points associated with the vertices of the refined C-net are obtained using the refinement rules of extended Catmull-Clark subdivision \cite{catmull1978recursively, biermann2000piecewise}. In order to define these refinement rules, the new control points are classified in three groups, namely, new face control points, new edge control points, and new vertex control points. In Fig. \ref{refined}, vertices associated with new face control points, new edge control points, and new vertex control points are marked with purple, brown, and blue circles, respectively. The refinement rules are
\begin{itemize}
\item New face control points. The newly added control point on an old face is the average of the four old control points defining that face. The weight of each old control point is shown in Fig. \ref{refrules} a).
\item New edge control points. Three cases are distinguished
  \begin{itemize}
  \item Boundary edges. The newly added control point on an old boundary edge is the average of the two old control points defining that edge. The weight of each old control point is shown in Fig. \ref{refrules} b).
  \item Edges that emanate from a boundary vertex and are not boundary edges. The six old control points defining the two old faces that share the old edge intervene in the computation of the newly added control point. The weight of the old boundary control point is $3/8 + (1/4) \cos(\pi/\mu)$, where $\mu$ is the valence of the old boundary control point. The weight of the old control point on the opposite end of the old edge is $3/8 - (1/4) \cos(\pi/\mu)$. The weight of the other four old control points is 1/16. The weight of each old control point is shown in Fig. \ref{refrules} c).
  \item Remaining edges. The six old control points defining the two old faces that share the old edge intervene in the computation of the newly added control point. The weight of the two old control points defining the old edge is 3/8. The weight of the other four old control points is 1/16. The weight of each old control point is shown in Fig. \ref{refrules} d).
  \end{itemize}
\item New vertex control points. Two cases are distinguished
  \begin{itemize}
  \item Boundary vertices. The new position of a control point on an old vertex is the sum of its own previous position with weight 3/4 and the two old adjacent boundary control points with weight 1/8. The weight of each old control point is shown in Fig. \ref{refrules} e).
  \item Interior vertices. The $2\mu + 1$  old control points defining the $\mu$ old faces that share the old vertex intervene in the computational of the new position of the control point. The weight of each old control point is shown in Fig. \ref{refrules} f).
  \end{itemize}
\end{itemize}

\section{Convergence study}

\begin{figure} [t!] 
 \centering
 \subfigure[Control net]{\includegraphics[scale=.83]{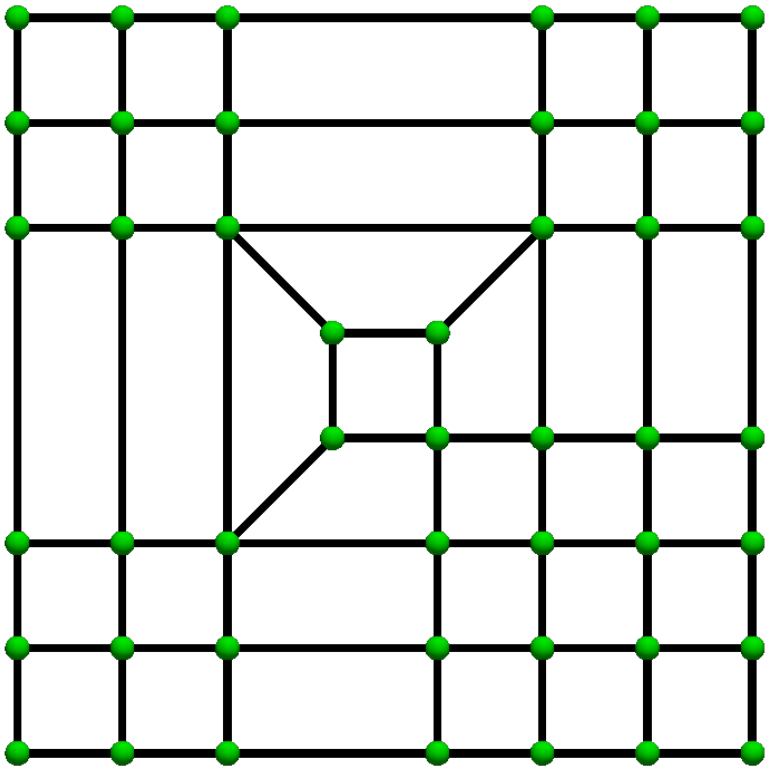}} \hspace*{+5.0mm}
 \subfigure[G-spline surface and B\'ezier mesh]{\includegraphics[scale=.82]{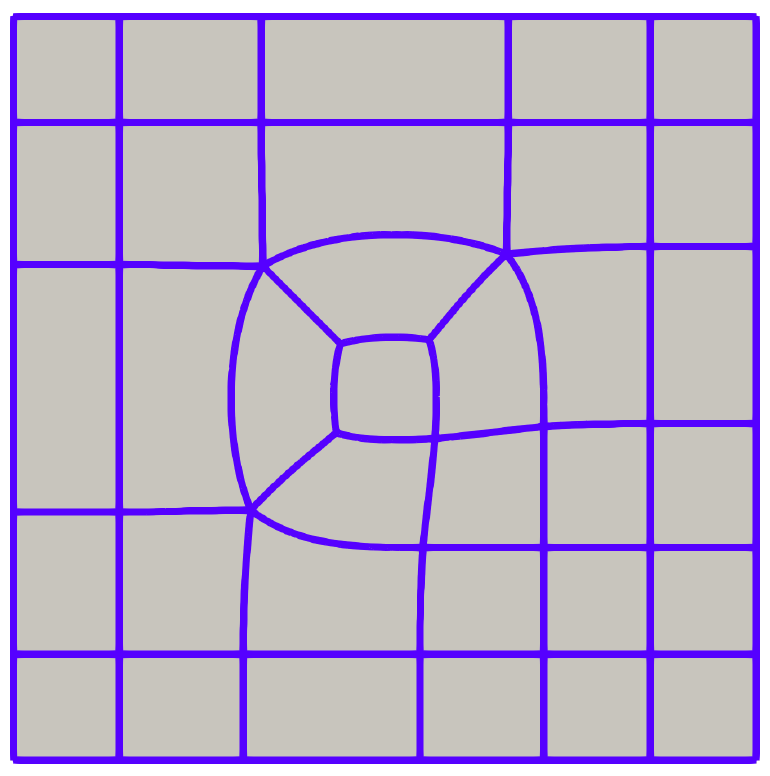}} \\
\caption{(Color online) Unit square used in the convergence study.}
\label{beziermeshcs}
\end{figure}

	\begin{figure} [t!] 
\centering
\includegraphics[width=12cm]{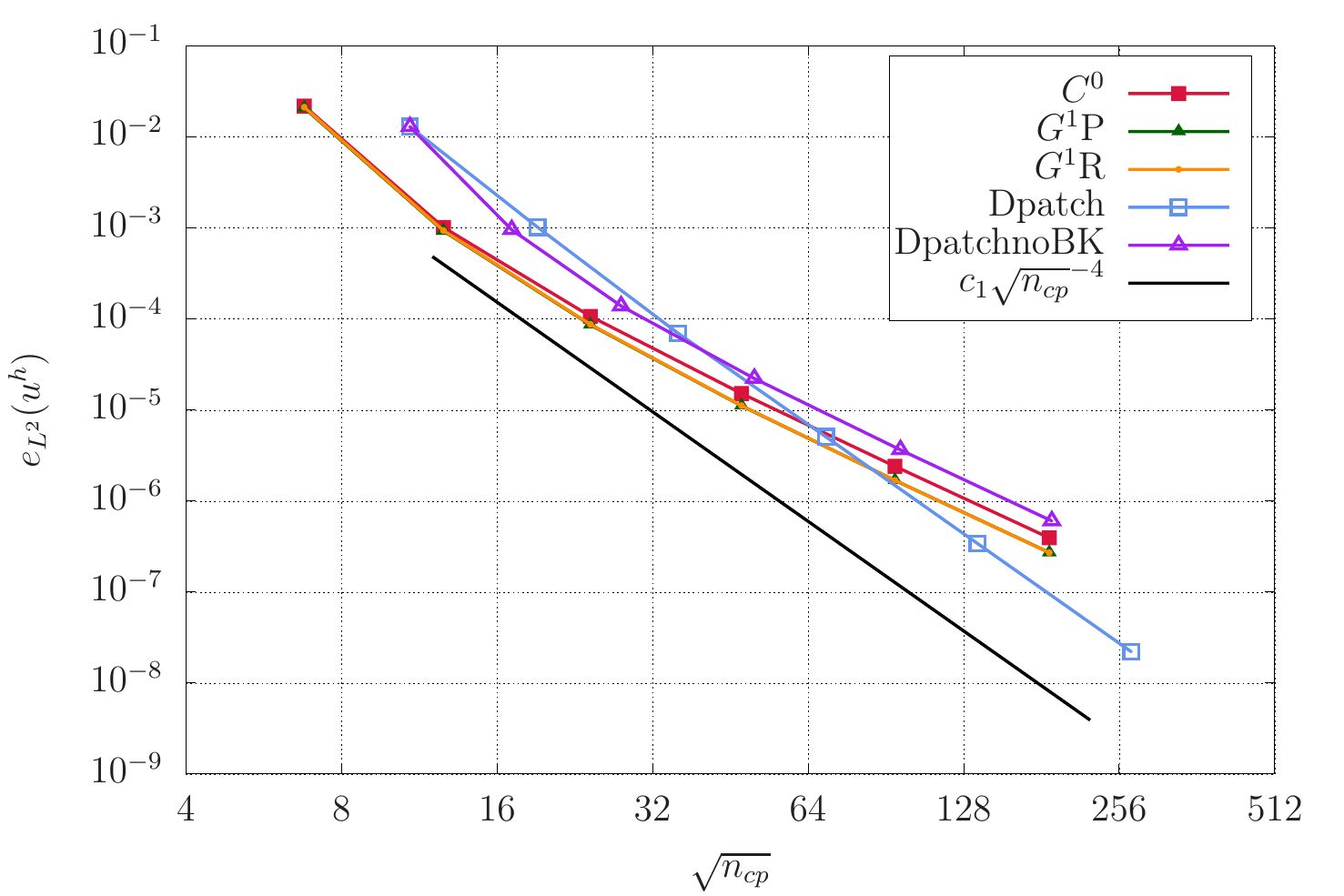}
\caption{(Color online) Convergence study of a second-order linear elliptic boundary-value problem. Convergence in $L^2$ norm for different EP constructions. } 
\label{l2norm}
\end{figure}

In this section, we study the convergence of G-splines under uniform global refinement. We solve the Poisson equation on a unit square ($\Omega = [0,1]^2$) with homogeneous Dirichlet boundary conditions, viz.,
\begin{align}
  \Delta u  &= - 2 \pi^2 \text{sin}(\pi x)\text{sin}(\pi y), \quad (x,y) \in \Omega  \text{,} \\ 
  u  &= 0, \quad (x,y) \in \partial \Omega  \text{.}  
\end{align}
The exact solution of the boundary-value problem defined above is
\begin{equation}
  u = \text{sin}(\pi x)\text{sin}(\pi y)  \text{.} 
\end{equation}
We study the convergence of the numerical solution in $L^2$ norm, $L^{\infty}$ norm, and $H^1$ norm. In order to do so, we define the relative errors of the numerical solution in $L^2$ norm, $L^{\infty}$  norm, and $H^1$ norm as
%
%
\begin{align}
   e_{L^2}(u^h) &=  \frac{ \sqrt{\int_{\Omega} \left( u^h - u \right)^2  \mathrm d\Omega }}{ \sqrt{ \int_{\Omega}  u^2 \, \mathrm d\Omega} }   \text{,} \\
   e_{L^{\infty}}(u^h) &=  \frac{ \max \vert u^h - u \vert }{ \max \vert u \vert }   \text{,} \\
   e_{H^1}(u^h) &=  \frac{ \sqrt{ \int_{\Omega} ( u^h - u )^2 \, \mathrm d\Omega + \int_{\Omega} ( \frac{\partial u^h}{\partial x}  - \frac{\partial u}{\partial x} )^2 \,  \mathrm d\Omega + \int_{\Omega} ( \frac{\partial u^h}{\partial y} - \frac{\partial u}{\partial y} )^2 \,  \mathrm d\Omega }}{ \sqrt{ \int_{\Omega}  u^2 \, \mathrm d\Omega  +  \int_{\Omega}  (\frac{\partial u}{\partial x})^2 \, \mathrm d\Omega +  \int_{\Omega}  (\frac{\partial u}{\partial y})^2 \, \mathrm d\Omega } }\text{,}
\end{align}
respectively. The five EP constructions used in this section are the following

	\begin{figure} [t!] 
\centering
\includegraphics[width=12cm]{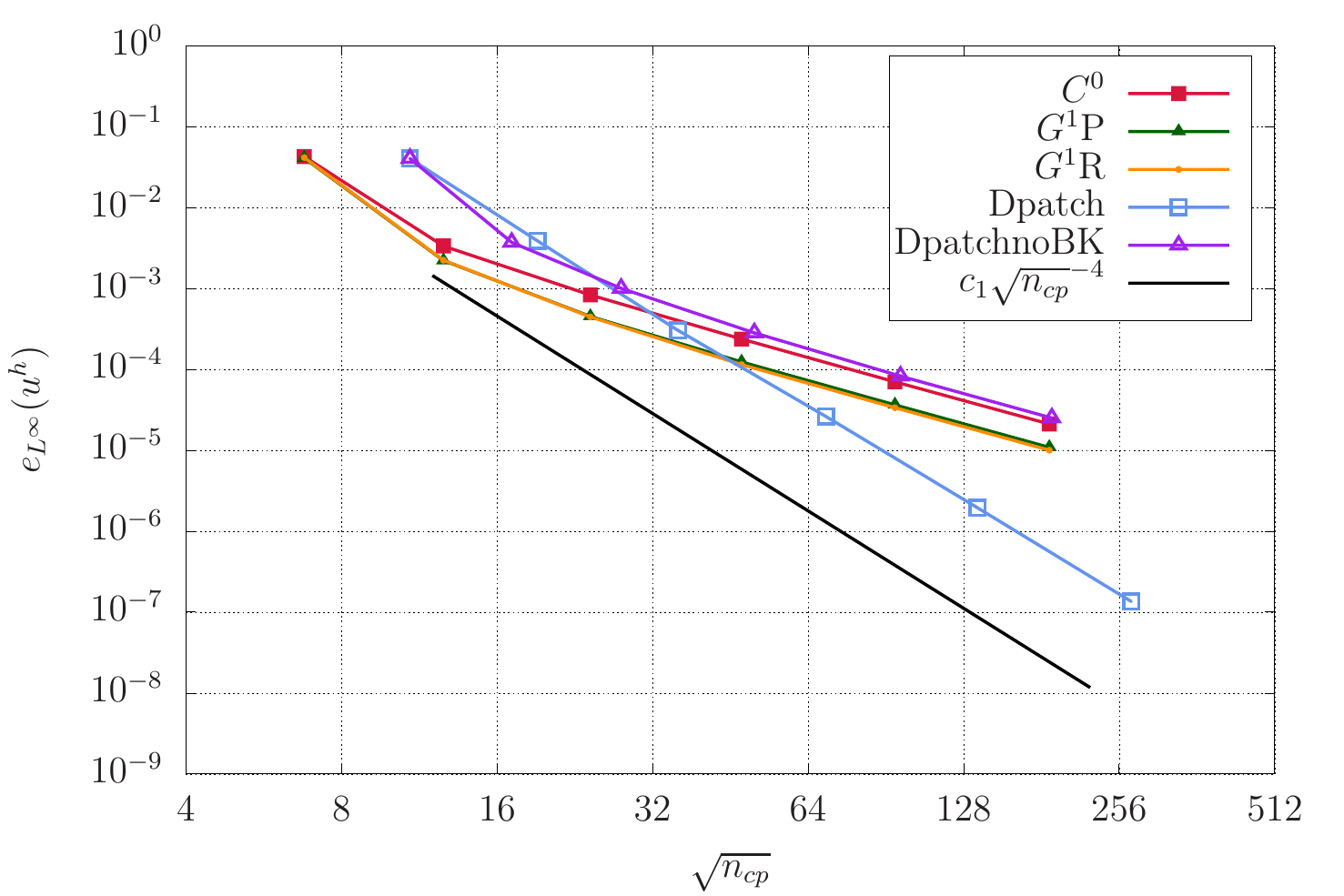}
\caption{(Color online) Convergence study of a second-order linear elliptic boundary-value problem. Convergence in $L^{\infty}$ norm for different EP constructions. } 
\label{linfinitynorm}
\end{figure}

	\begin{figure} [t!] 
\centering
\includegraphics[width=12cm]{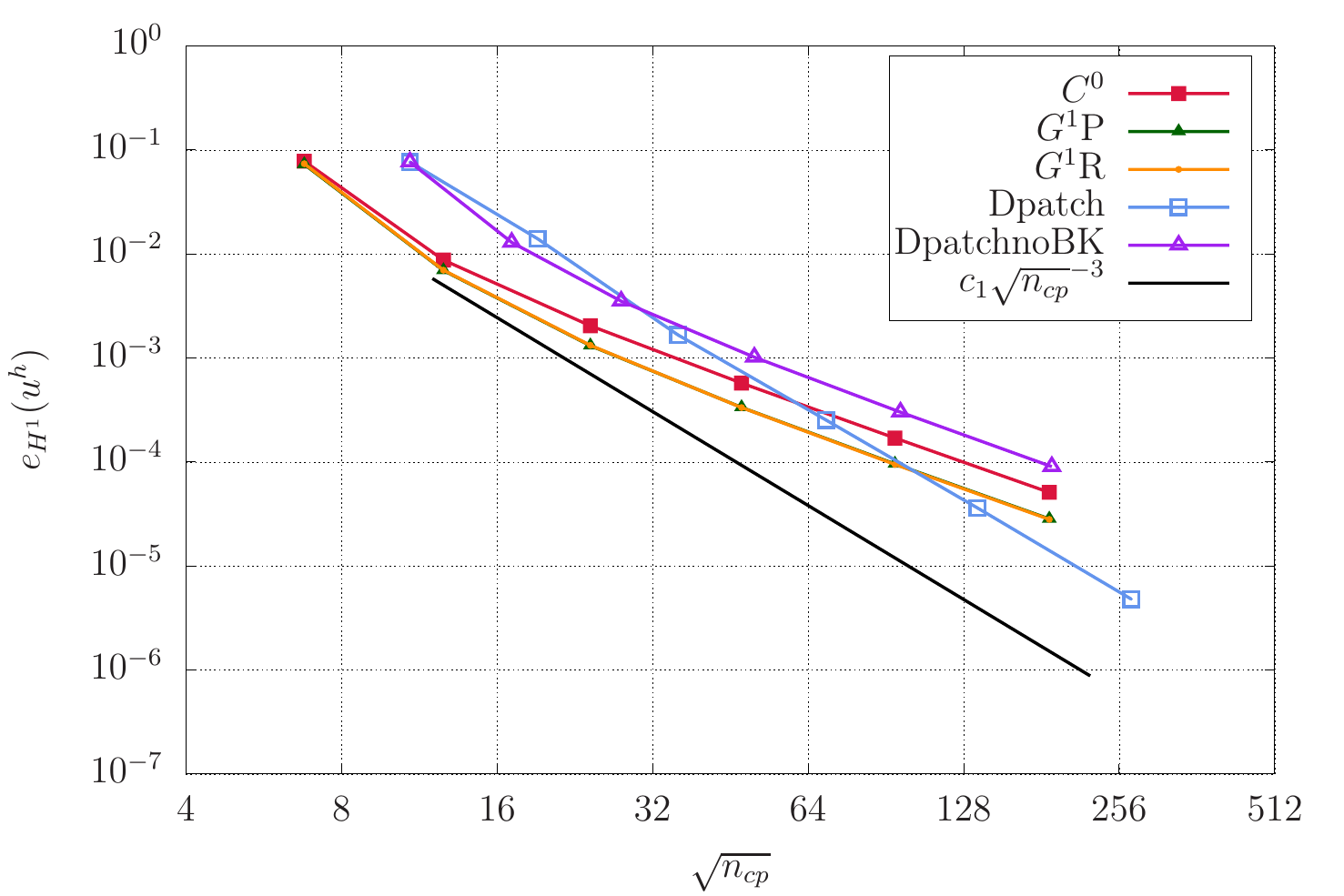}
\caption{(Color online) Convergence study of a second-order linear elliptic boundary-value problem. Convergence in $H^1$ norm for different EP constructions. } 
\label{h1norm}
\end{figure}

\begin{itemize}

\item $C^0$: This is the EP construction explained in Section 2.3.1.

\item $G^1$P: This is the EP construction explained in Section 2.3.2 whose basis functions are polynomial functions.

\item $G^1$R: This is the EP construction explained in Section 2.3.2 whose basis functions are rational functions.

\item Dpatch: This is the EP construction based on the D-patch framework \cite{reif1997refineable} explained in \cite{wei2022analysis} which involves book-keeping.

\item DpatchnoBK: This is the EP construction based on the D-patch framework \cite{reif1997refineable} explained in \cite{wei2022analysis} without using booking-keeping, that is, maintaining one ring of irregular elements throughout all the refinement levels.

\end{itemize}

Since the constructions Dpatch and DpatchnoBK cannot handle boundary EPs, we use a control net with interior EPs, but without boundary EPs. The control net and the B\'ezier mesh for the construction $G^1$P are shown in Fig. \ref{beziermeshcs}. After level 0 is built, we perform five levels of global uniform refinement to compare the accuracy of the five EP constructions considered in this section.

The convergence of the numerical solution in $L^2$, $L^{\infty}$, and $H^1$  norms with the five EP constructions are plotted in Figs. \ref{l2norm}, \ref{linfinitynorm}, and \ref{h1norm}, respectively. As shown in Figs. \ref{l2norm}, \ref{linfinitynorm}, and \ref{h1norm}, the accuracy of the constructions $G^1$P and $G^1$R is indistinguishable at the scale of the plots. The accuracy of the constructions $G^1$P and $G^1$R  is superior to the accuracy of the constructions $C^0$ and DpatchnoBK for all mesh resolutions. For meshes with coarse and intermediate resolution, the accuracy of the constructions $G^1$P and $G^1$R is superior to the construction Dpatch. However, for fine meshes, the accuracy of the construction Dpatch is superior to the accuracy of the constructions $G^1$P and $G^1$R  since the construction Dpatch has optimal asymptotic convergence. As explained in Section 2.5, having the option of performing several refinement levels with book-keeping is not common in real-world engineering applications. In addition, the errors are already negligible when the construction Dpatch becomes more accurate than the constructions $G^1$P and $G^1$R, namely, the errors are already smaller than 0.0003\%, 0.02\%, and 0.01\% in $L^2$, $L^{\infty}$, and $H^1$ norms, respectively. Thus, these results suggest that the constructions $G^1$P and $G^1$R are preferable for real-world industrial applications. The development of an EP construction that is at least as accurate as the constructions $G^1$P and $G^1$R for coarse meshes while also having optimal asymptotic convergence rates is a worthwhile direction of future work.

\section{Surface quality}

\begin{figure} [t!] 
 \centering
 \subfigure[]{\includegraphics[scale=.10]{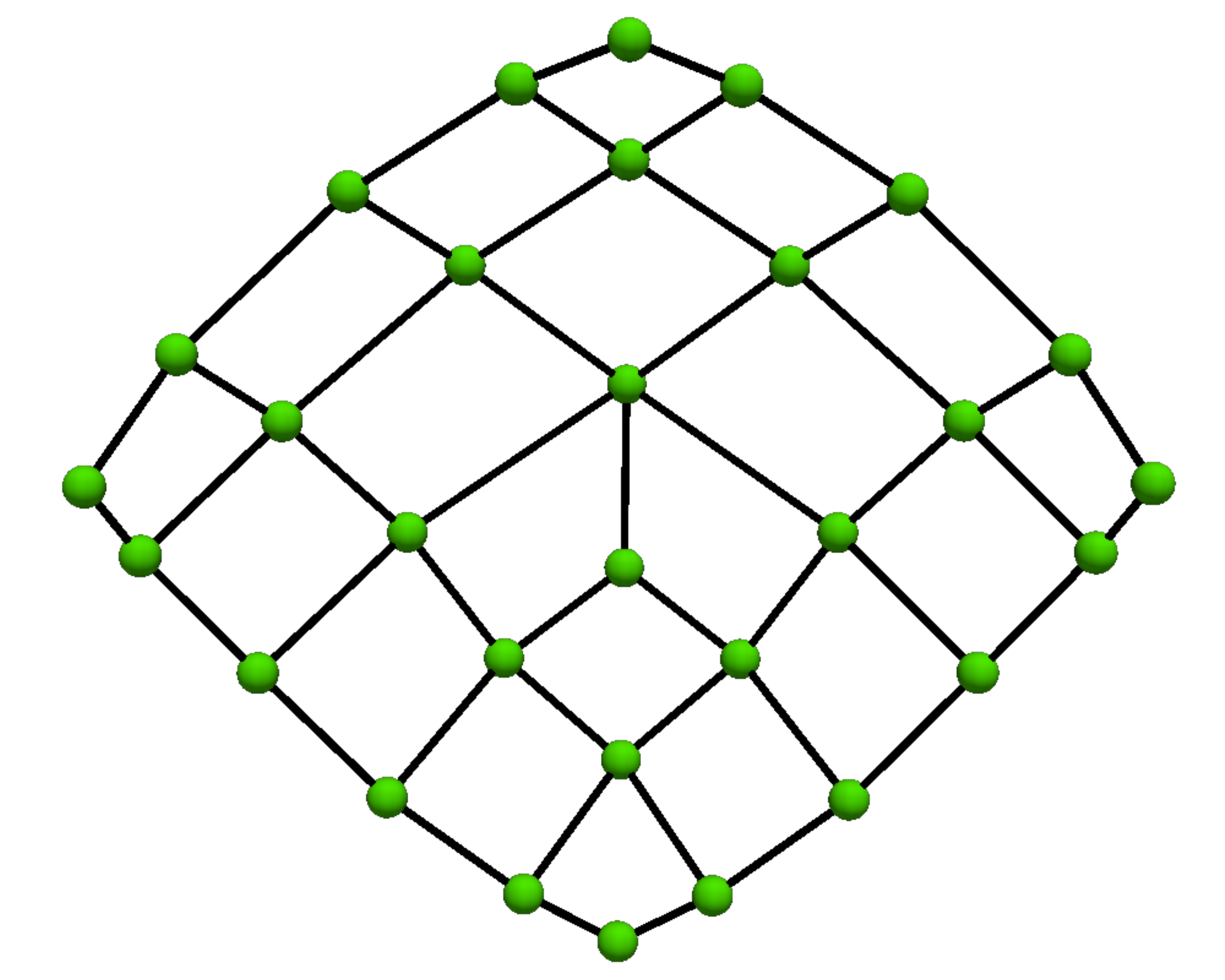}} 
 \subfigure[]{\includegraphics[scale=.10]{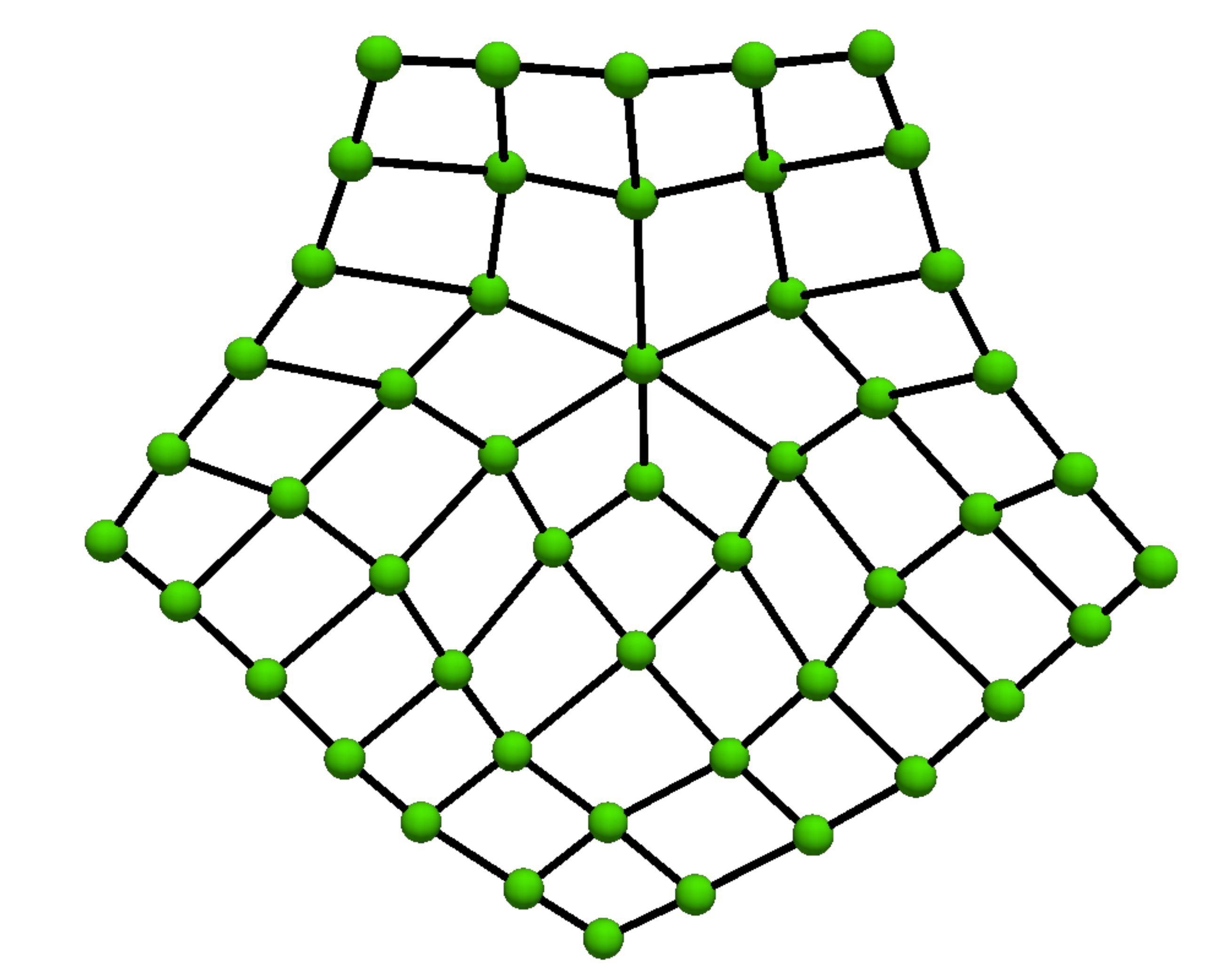}}
 \subfigure[]{\includegraphics[scale=.10]{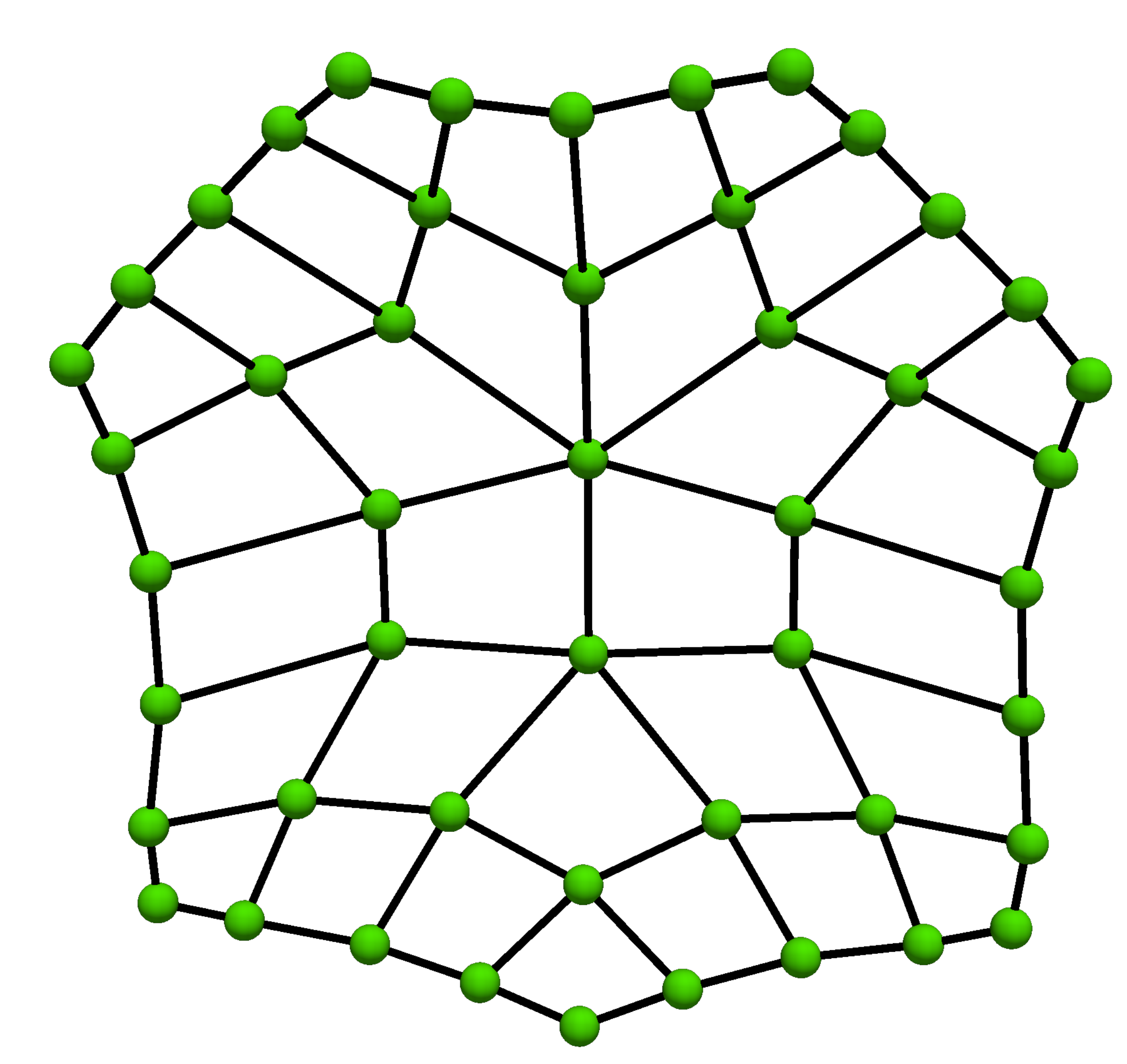}}
 \subfigure[]{\includegraphics[scale=.10]{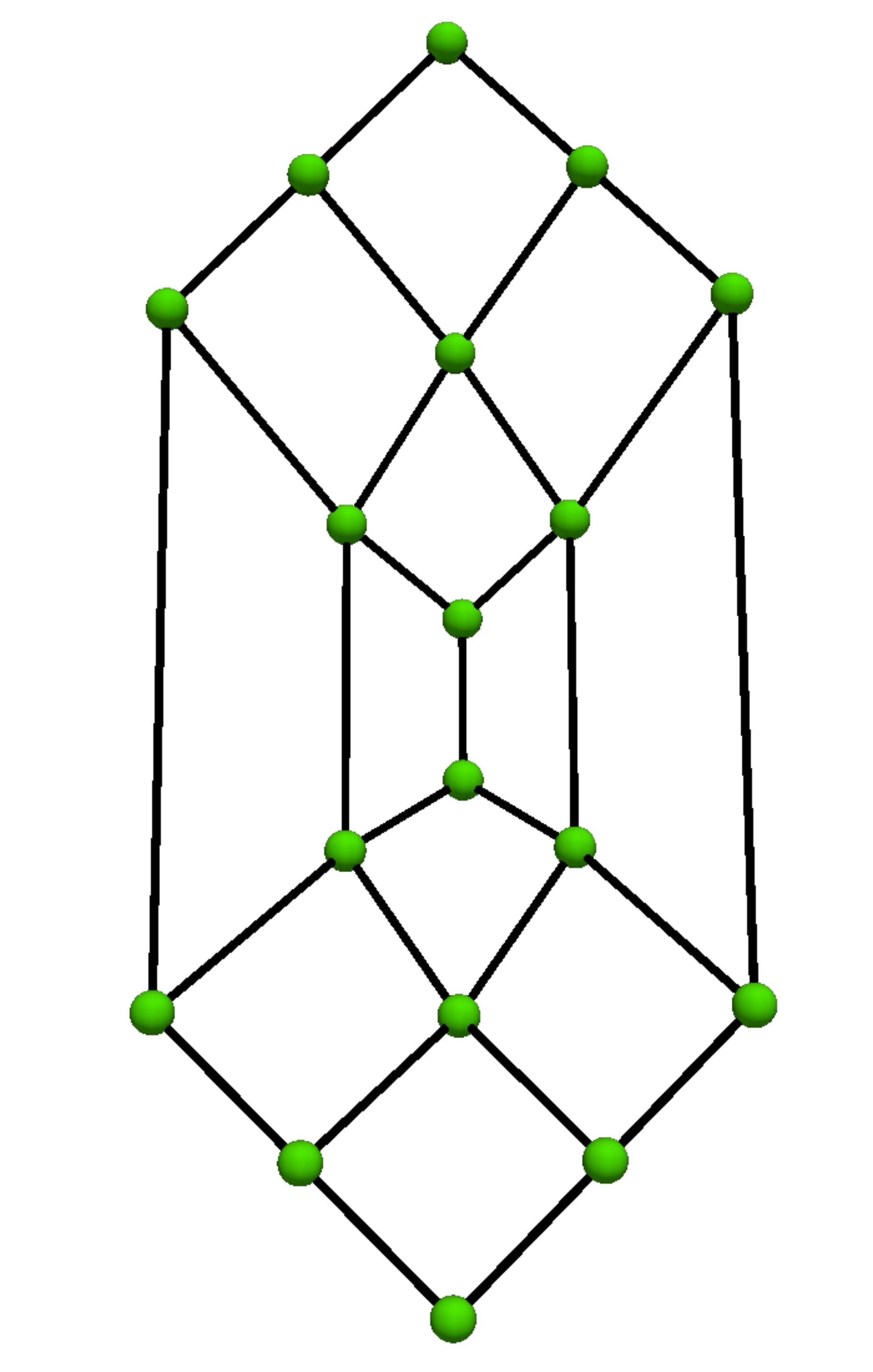}}
 \subfigure[]{\includegraphics[scale=.13]{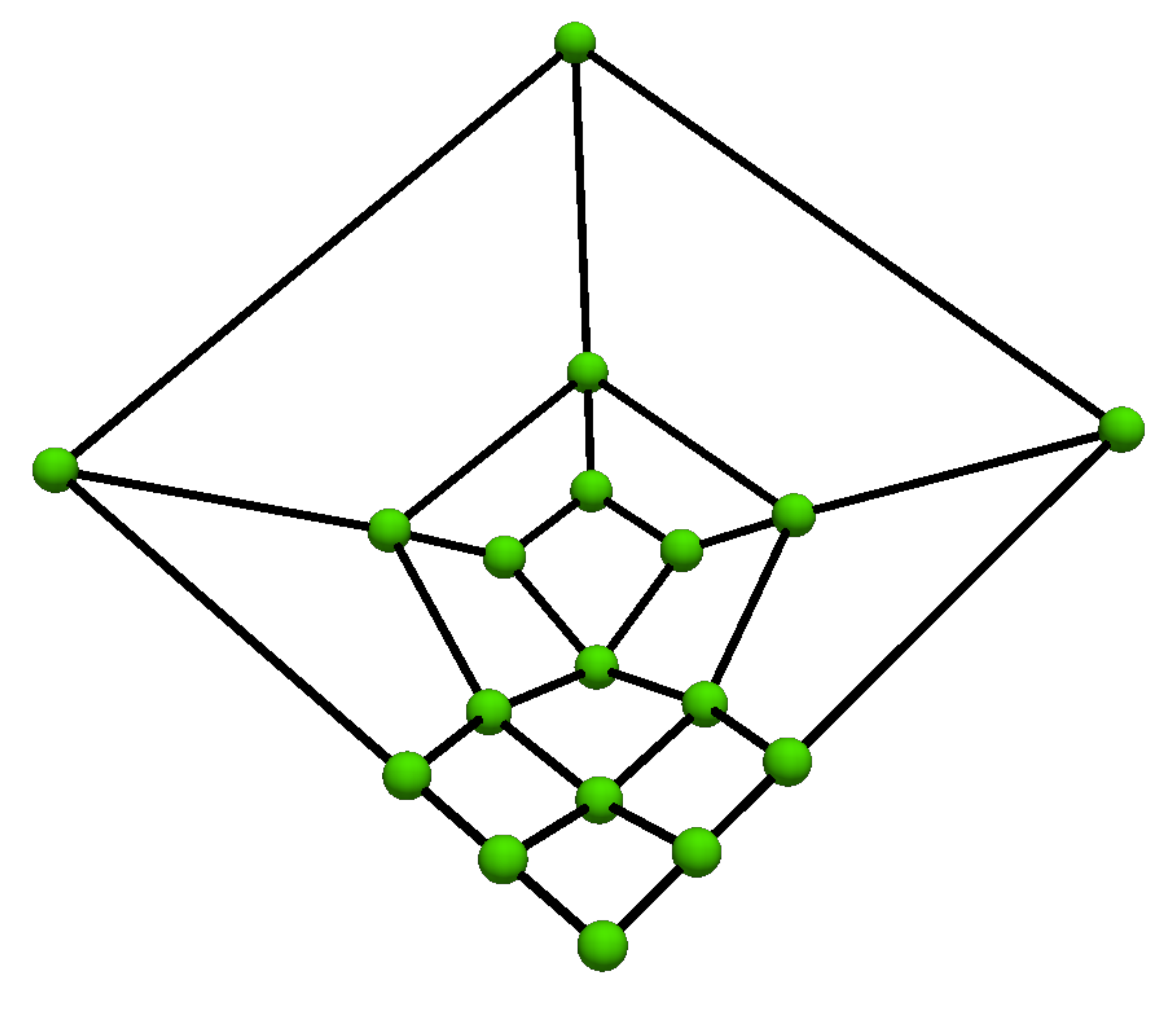}}
 \subfigure[]{\includegraphics[scale=.11]{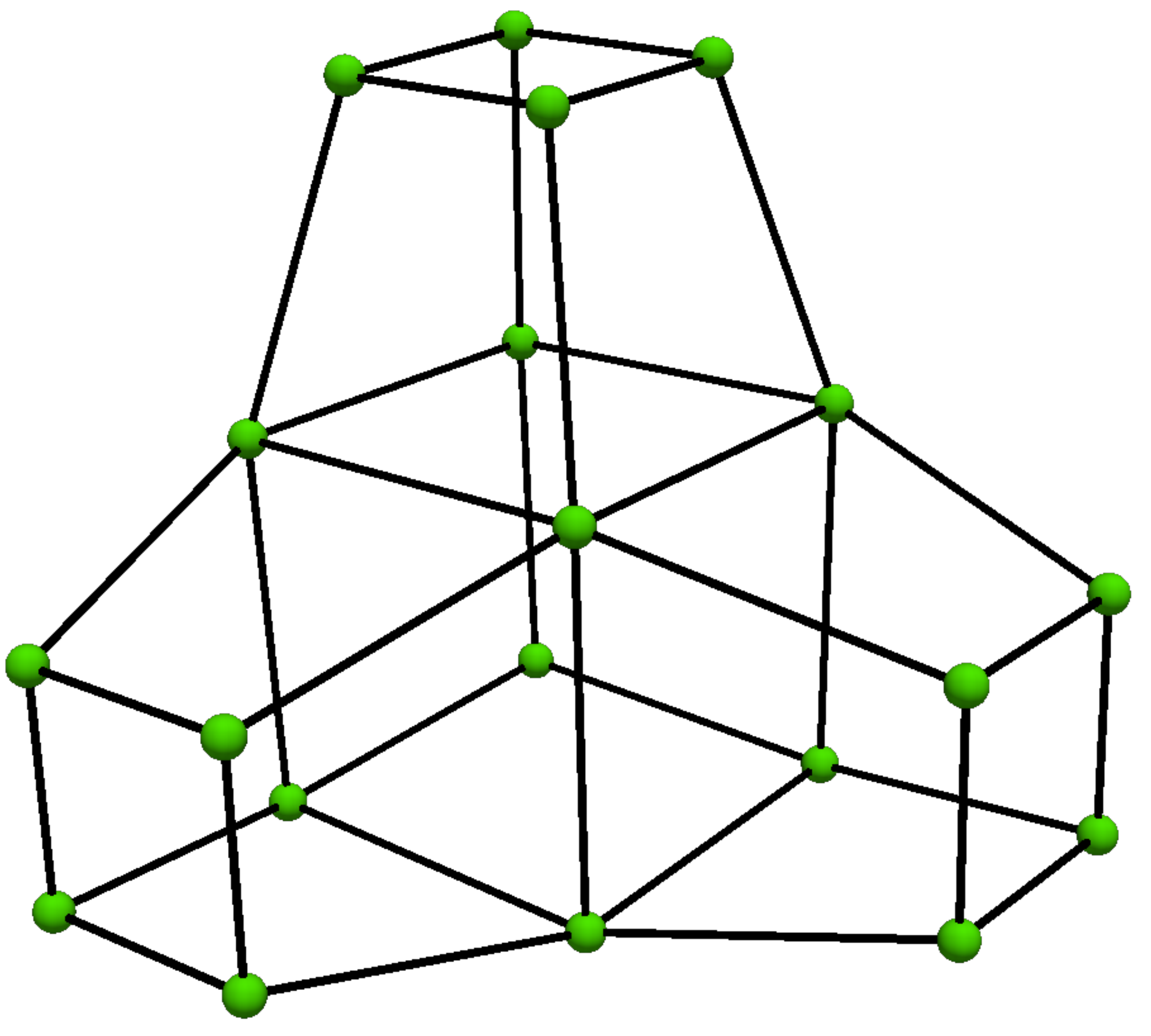}} \\
\caption{Control nets used to study the surface quality. (a) EPs with valences 3 and 5 on the same face. (b) EPs with valences 3 and 6 on the same face. (c) EPs with valences 5 and 6 on the same face. (d) Two EPs with valence 3 on the same face. (e) Three EPs with valence 3 on the same face. (f) Control net having four EPs with valence 3 in some faces and other faces that combine EPs with valences 3, 5, and 6.}
\label{sqcontrolnets}
\end{figure}

\begin{figure} [t!] 
 \centering
 \subfigure[]{\includegraphics[scale=.09]{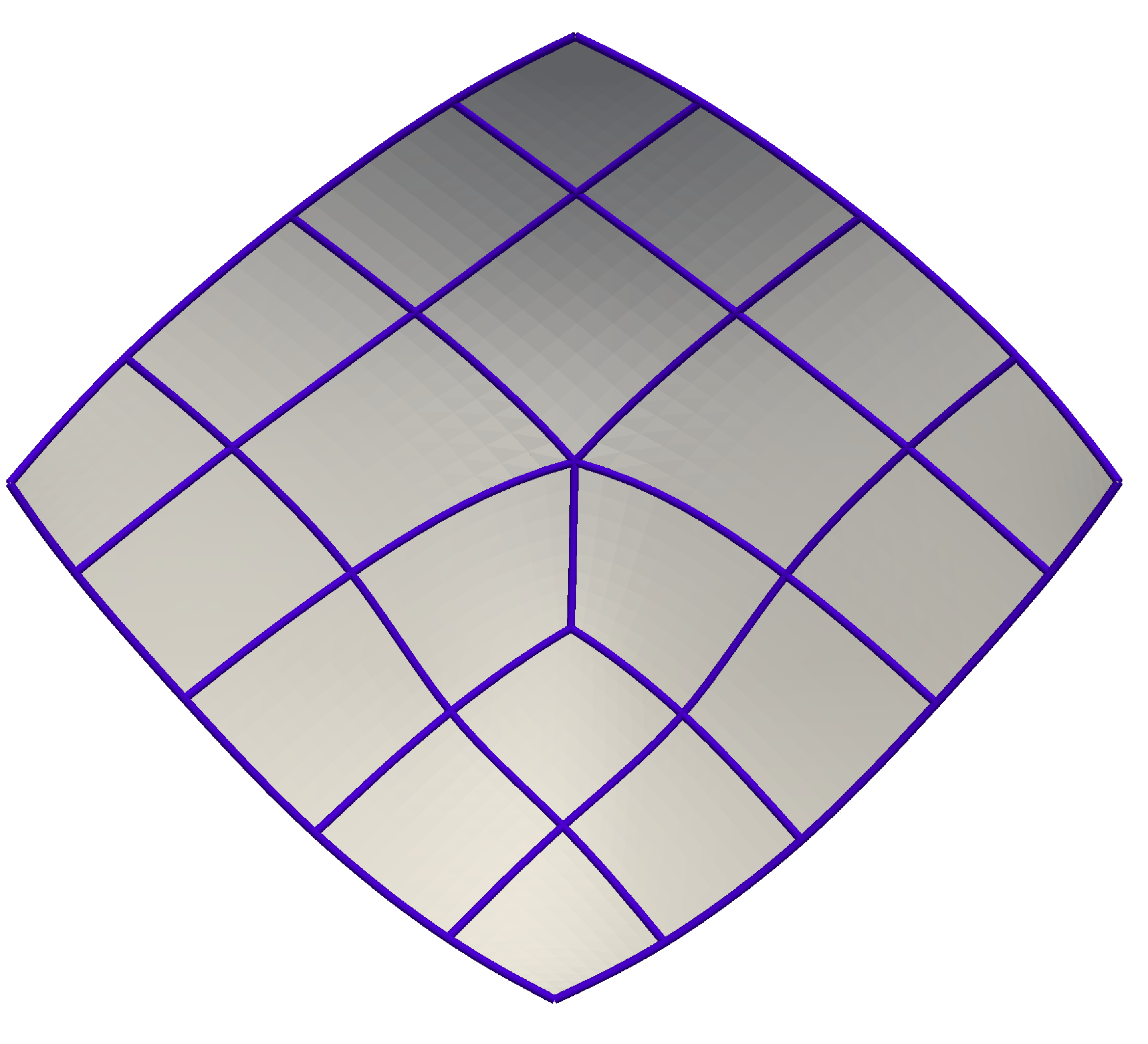}} 
 \subfigure[]{\includegraphics[scale=.09]{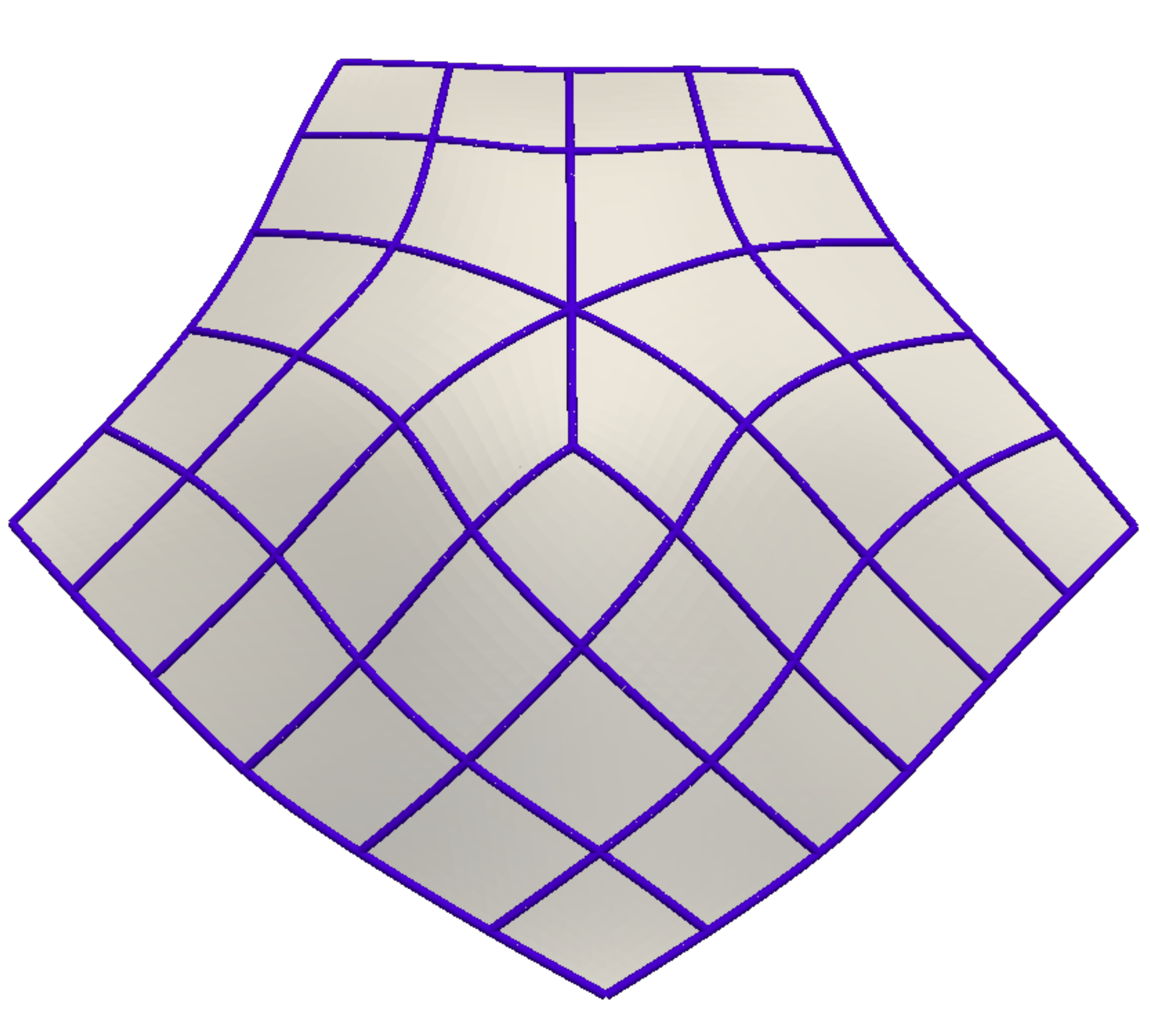}}
 \subfigure[]{\includegraphics[scale=.09]{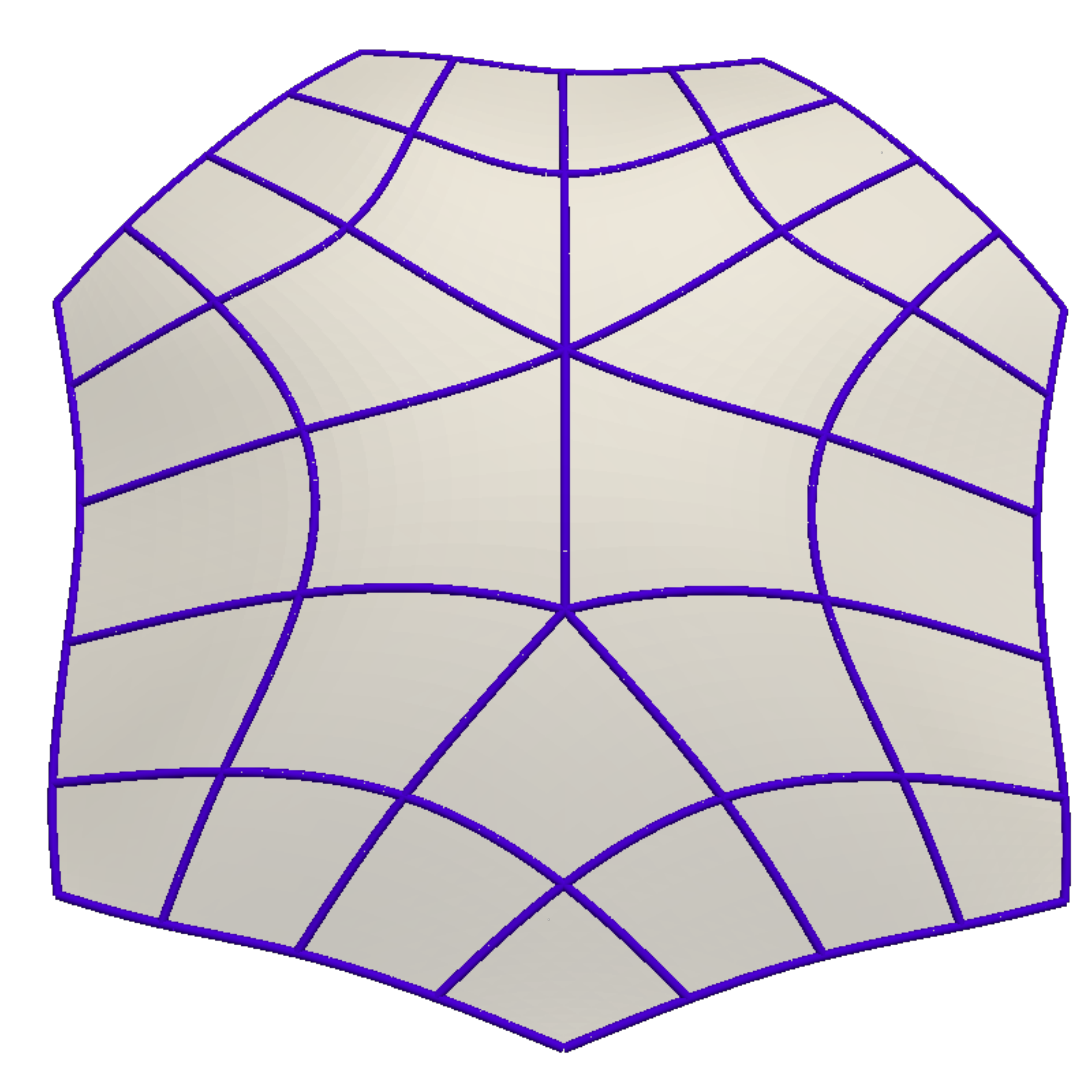}}
 \subfigure[]{\includegraphics[scale=.06]{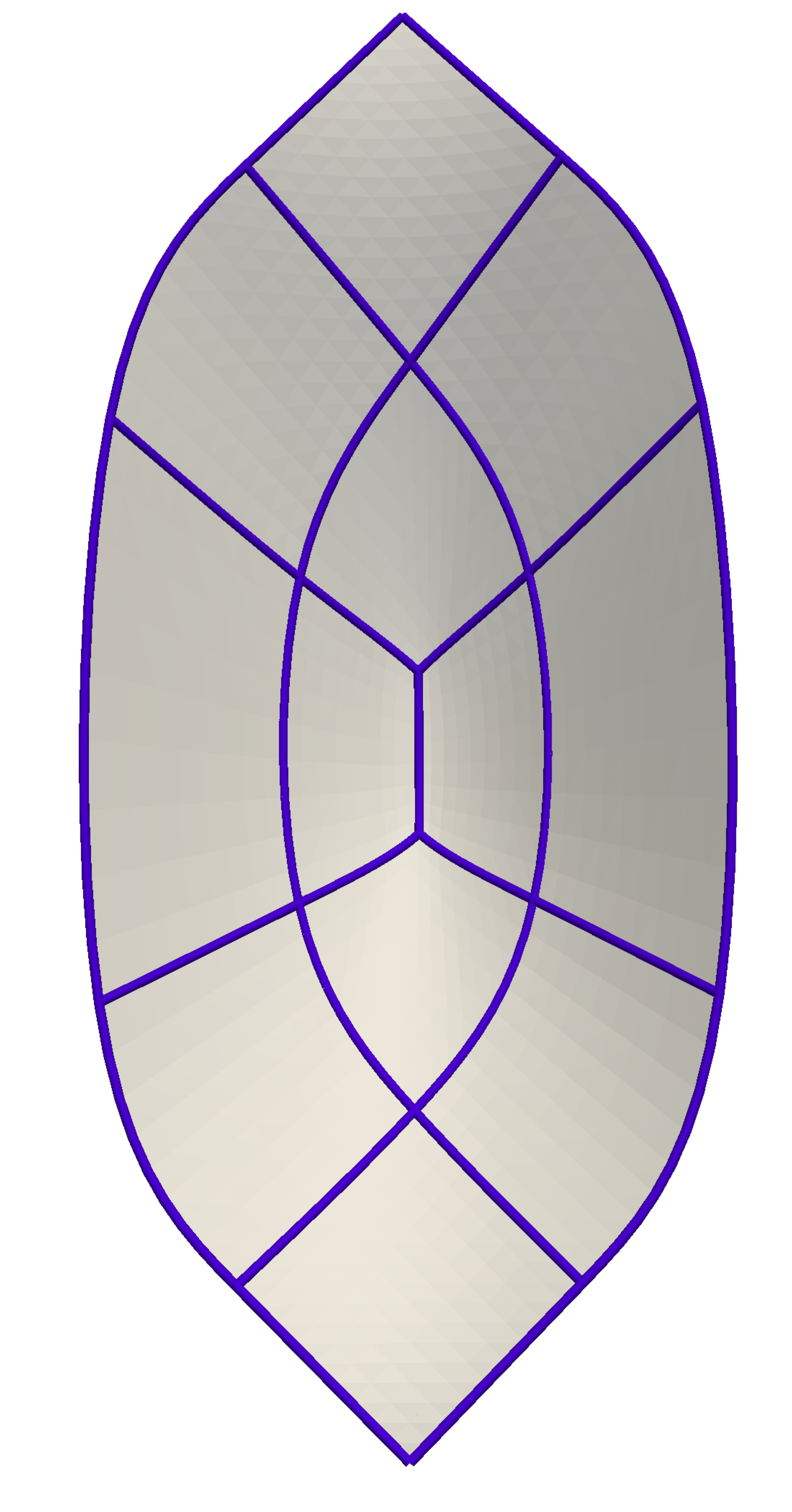}}
 \subfigure[]{\includegraphics[scale=.09]{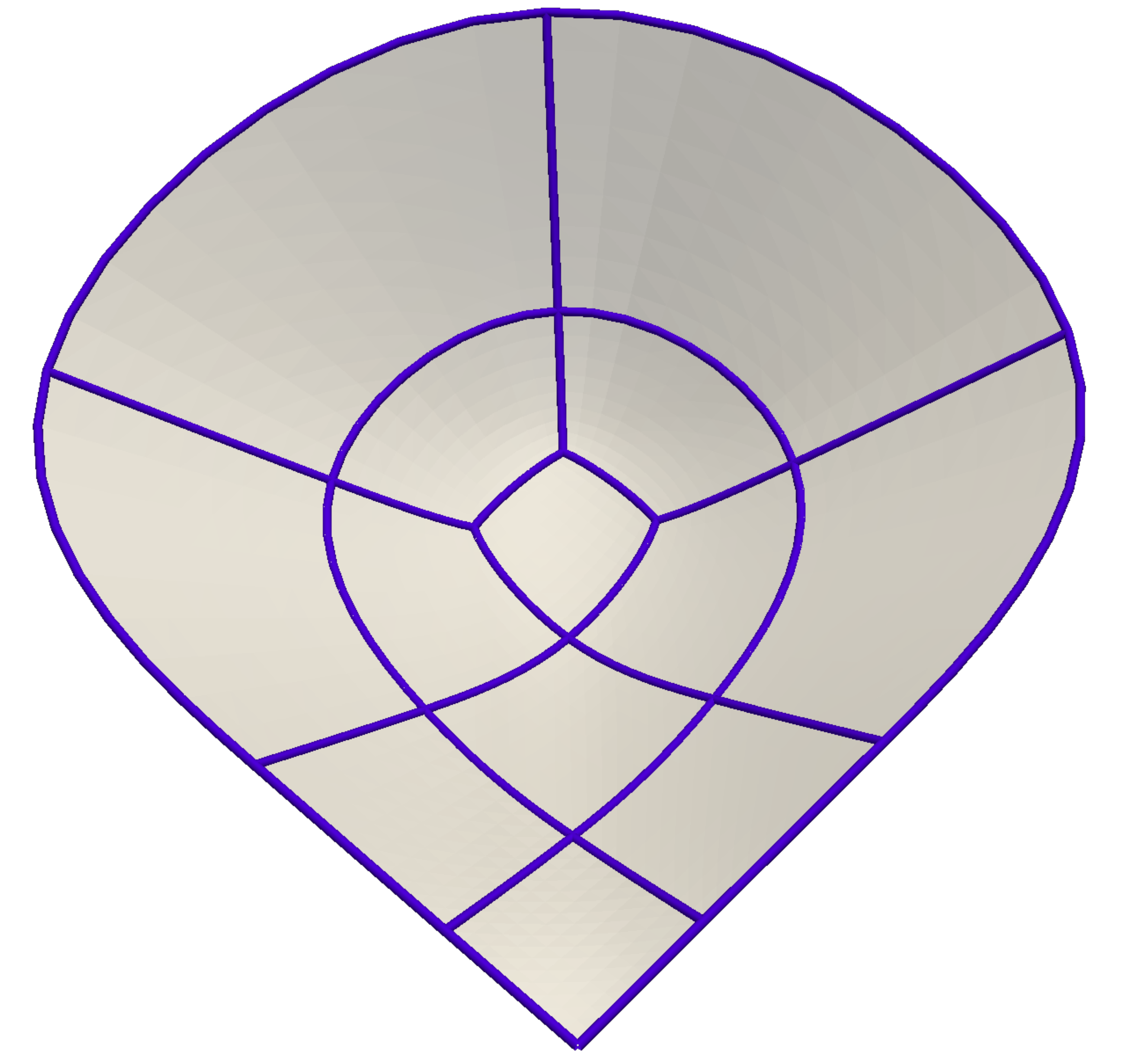}}
 \subfigure[]{\includegraphics[scale=.10]{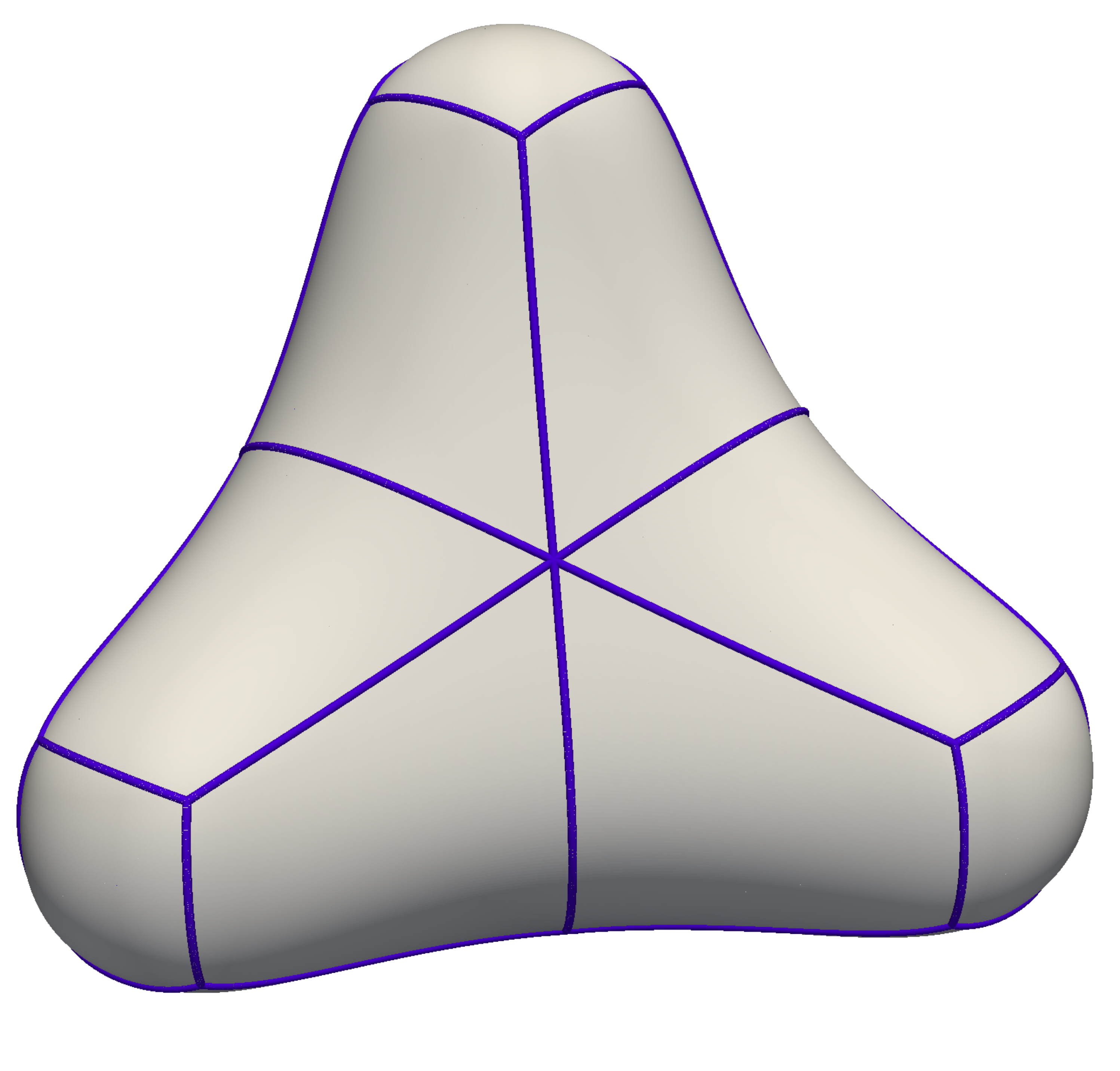}} \\
\caption{G-spline surfaces and B\'ezier meshes associated with the control nets given in Fig. \ref{sqcontrolnets}. (a) Surface having EPs with valences 3 and 5 on the same face. (b) Surface having EPs with valences 3 and 6 on the same face. (c) Surface having EPs with valences 5 and 6 on the same face. (d) Surface having two EPs with valence 3 on the same face. (e) Surface having three EPs with valence 3 on the same face. (f) Surface having four EPs with valence 3 in some faces and other faces that combine EPs with valences 3, 5, and 6.}
\label{sqbeziermeshes}
\end{figure}


EP constructions that enforce the tangent plane to be continuous have a tendency to result in surfaces with limited surface quality. The mission of the fairing equations that are used in EP constructions is to maintain a good surface quality near EPs. In order to avoid unaesthetic surfaces in CAD programs, the surface quality of an EP construction is usually evaluated by plotting either the distribution of the highlight lines \cite{beier1994highlight} or the mean and Gauss curvatures \cite{karvciauskas2015improved, karvciauskas2015can, karvciauskas2016generalizing}.

 \begin{table}[t!]
   \caption{Minimum value of the thickness that results in invalid area element for each geometry and each EP construction.} \label{tablesq}
   \bigskip
     \centering
     \renewcommand{\arraystretch}{1.20}
     \begin{tabular}{c@{\hspace{12.0mm}}  c@{\hspace{12.0mm}}  c@{\hspace{12.0mm}}  c@{\hspace{12.0mm}} c@{\hspace{12.0mm}}}
\hline
  & $C^0$ & Dpatch & $G^1$P & $G^1$R  \\
\hline
Val35   & 1.98  & 0.30 & 2.09 & 2.18   \\          
Val36   & 3.41  & 0.36 & 3.70  & 3.83  \\     
Val56   & 2.70  & 0.76 & 2.62  & 2.62   \\          
Val33   & 0.72  & 0.16 & 0.73  & 0.76    \\    
Val333   & 0.88  & 0.11 & 0.91  & 0.93  \\     
Val3456   & 0.34   & 0.06 & 0.29  & 0.29  \\ 
Inner Part of B-pillar  & 5.10  & 1.10 & 5.20  & 5.20   \\        
Outer Part of B-pillar   & 2.40  & 0.40 & 2.60 & 2.60   \\    
Stiffener of B-pillar  & 3.60  & 0.80 & 3.50   & 3.50  \\         
\hline      
     \end{tabular}     
   \end{table}



Shell formulations with material nonlinearities require to perform numerical integration in the through-thickness direction. When splines with EPs are used in FEA, limited surface quality near EPs can cause invalid area elements at quadrature points off of the midsurface when performing the through-thickness numerical integration. If an invalid area element at a quadrature point is found, the FEA simulation cannot be carried out. Thus, in FEA, the surface quality of an EP construction can be evaluated by obtaining the minimum thickness value that results in an invalid area element at a quadrature point. All lengths are given in millimeters in this section.


In this section, we compare the surface quality of the EP constructions $C^0$, $G^1$P, and $G^1$R, and D-patch. In order to compute the area element at quadrature points, we begin by computing the non-unit tangent vectors to the midsurface, viz.,
\begin{equation}
 \mathbf{a}_{1}=  \frac{\partial \mathbf{x}^e }{\partial \xi}, \quad  \mathbf{a}_{2}=  \frac{\partial \mathbf{x}^e }{\partial \eta} \text{.}
\end{equation}
We compute the unit normal vector to the midsurface as follows
\begin{equation}
 \mathbf{a}_3 = \frac{\mathbf{a}_1\times\mathbf{a}_2}{||\mathbf{a}_1\times\mathbf{a}_2||},
\end{equation}
We compute the covariant metric coefficients of the midsurface and the covariant curvature coefficients of the midsurface as
\begin{align}
 a_{\alpha\beta} & =\mathbf{a}_{\alpha}\cdot\mathbf{a}_{\beta} , \quad \alpha, \beta \in \{1,2\}  \text{,} \\
 b_{\alpha1} & = \frac{\partial \mathbf{a}_{\alpha}}{\partial\xi} \cdot\mathbf{a}_{3}, \quad  b_{\alpha2}  = \frac{\partial \mathbf{a}_{\alpha}}{\partial\eta} \cdot\mathbf{a}_{3}, \quad \alpha \in \{1,2\}  \text{,}
\end{align}
respectively. Any point in the shell that is off of the midsurface can be described by the following position vector
\begin{equation}
 \mathbf{r}^e =  \mathbf{x}^e + \zeta \mathbf{a}_3  \text{,}
\end{equation}
where $\zeta \in [-t/2,t/2]$ is the thickness coordinate and $t$ is the thickness value. We can now compute the covariant base vectors of the shell and the metric coefficients of the shell as
\begin{align}
  \mathbf{g}_{1} & =  \frac{\partial \mathbf{r}^e }{\partial \xi}, \quad \mathbf{g}_{2}  =  \frac{\partial \mathbf{r}^e }{\partial \eta}  \text{,} \\
  g_{\alpha\beta} & =\mathbf{g}_{\alpha}\cdot\mathbf{g}_{\beta} = a_{\alpha\beta} - 2 \zeta b_{\alpha\beta} + \zeta^2 \mathbf{a}_{3,\alpha} \cdot \mathbf{a}_{3,\beta}, \quad \alpha, \beta \in \{1,2\}  \text{,}
\end{align}
, respectively. As explained in \cite{bischoff2004models, Kiendl2015}, the quadratic term is neglected to obtain a linear strain distribution through the thickness. Thus,
\begin{equation}
g_{\alpha\beta} =\mathbf{g}_{\alpha}\cdot\mathbf{g}_{\beta} = a_{\alpha\beta} - 2 \zeta b_{\alpha\beta}, \quad \alpha, \beta \in \{1,2\}  \text{.}
\end{equation}
The area element at any point in the shell can be obtained as
\begin{equation}
 g =  \sqrt{| g_{\alpha\beta} |} =  \sqrt{g_{11}g_{22} - g_{12}g_{21}}  \text{.}
\end{equation}

If $| g_{\alpha\beta} |$ is negative at any quadrature point, the area element is invalid and the FEA simulation cannot be carried out. As quadrature rule in the two surface directions, we use a Gauss-Legendre quadrature rule with $p+1$ quadrature points in each direction. As quadrature rule in the through-thickness direction, we use a Gauss-Lobatto quadrature rule with 5 quadrature points. For a given thickness, Gauss-Lobatto quadrature rules are the most likely to have a quadrature point with invalid area element since these quadrature rules result in quadrature points at the top and bottom surfaces of the shell.

\begin{figure} [t!] 
 \centering
 \subfigure[]{\includegraphics[scale=.105]{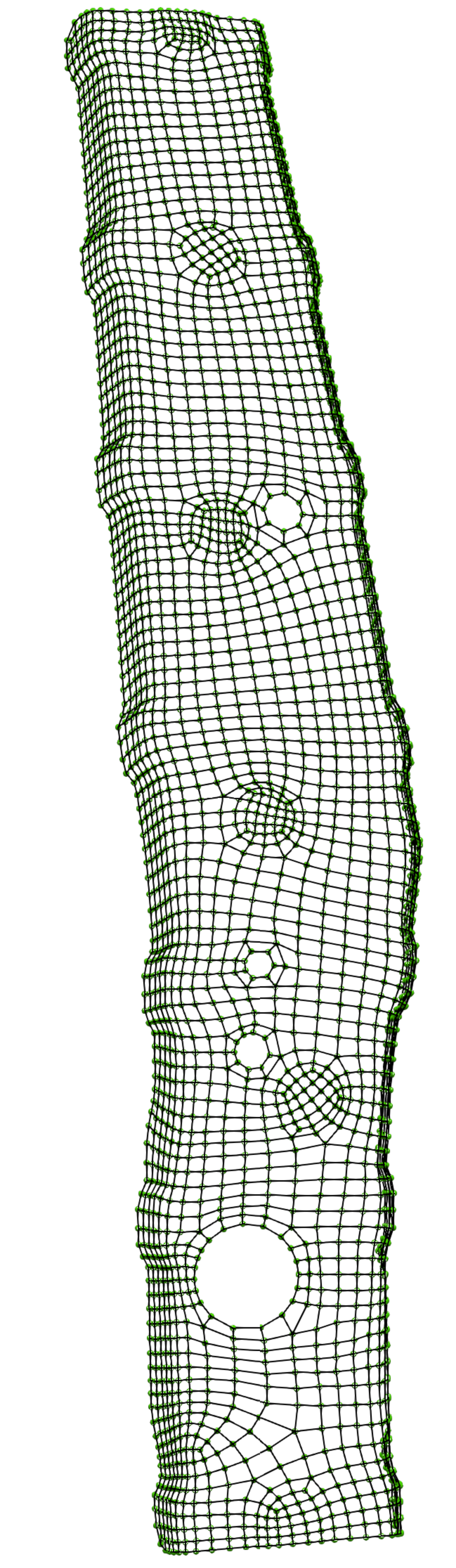}} 
 \subfigure[]{\includegraphics[scale=.21]{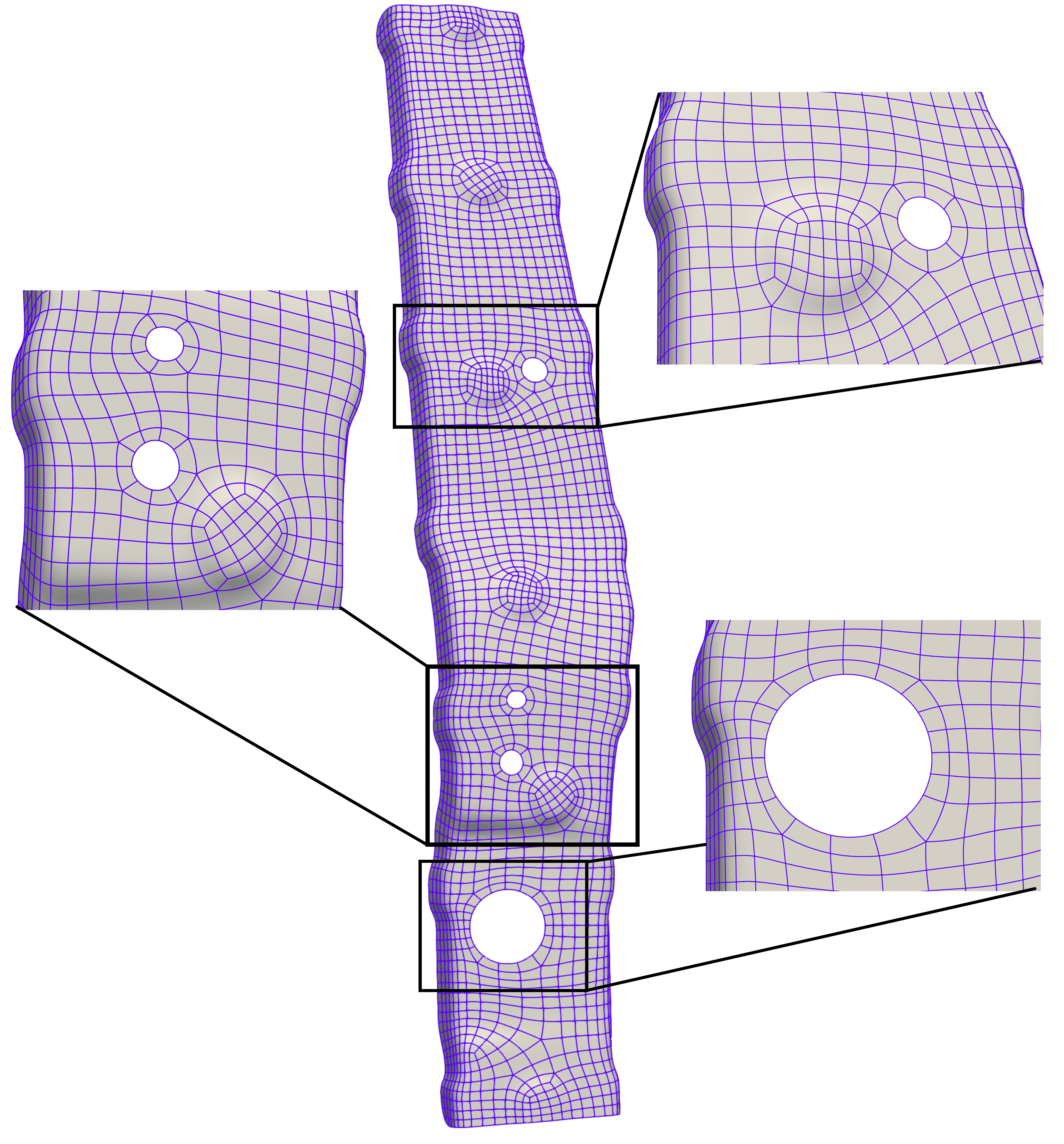}}\\
\caption{Stiffener of an automotive B-pillar. (a) Control net with 2,785 control points. (b) G-spline surface and B\'ezier mesh. The B\'ezier mesh has 2,633 elements with an average element size of 5 mm.}
\label{stiffener}
\end{figure}

\begin{figure} [t!] 
 \centering
 \subfigure[]{\includegraphics[scale=.105]{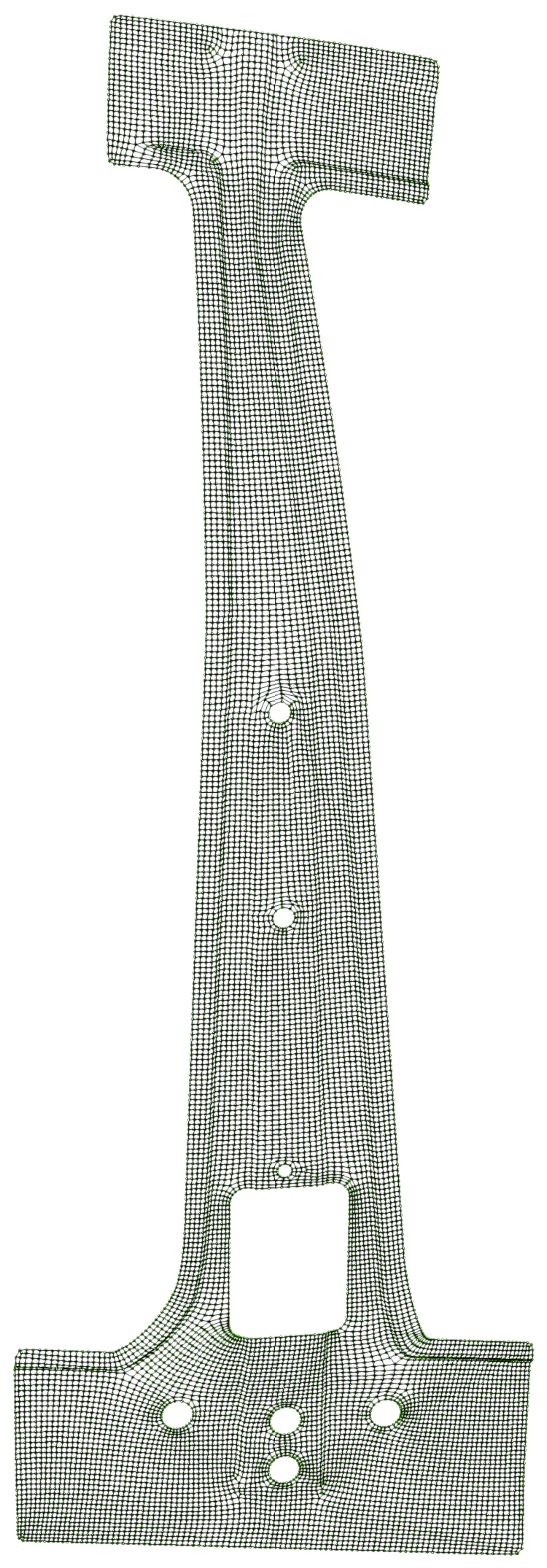}} 
 \subfigure[]{\includegraphics[scale=.21]{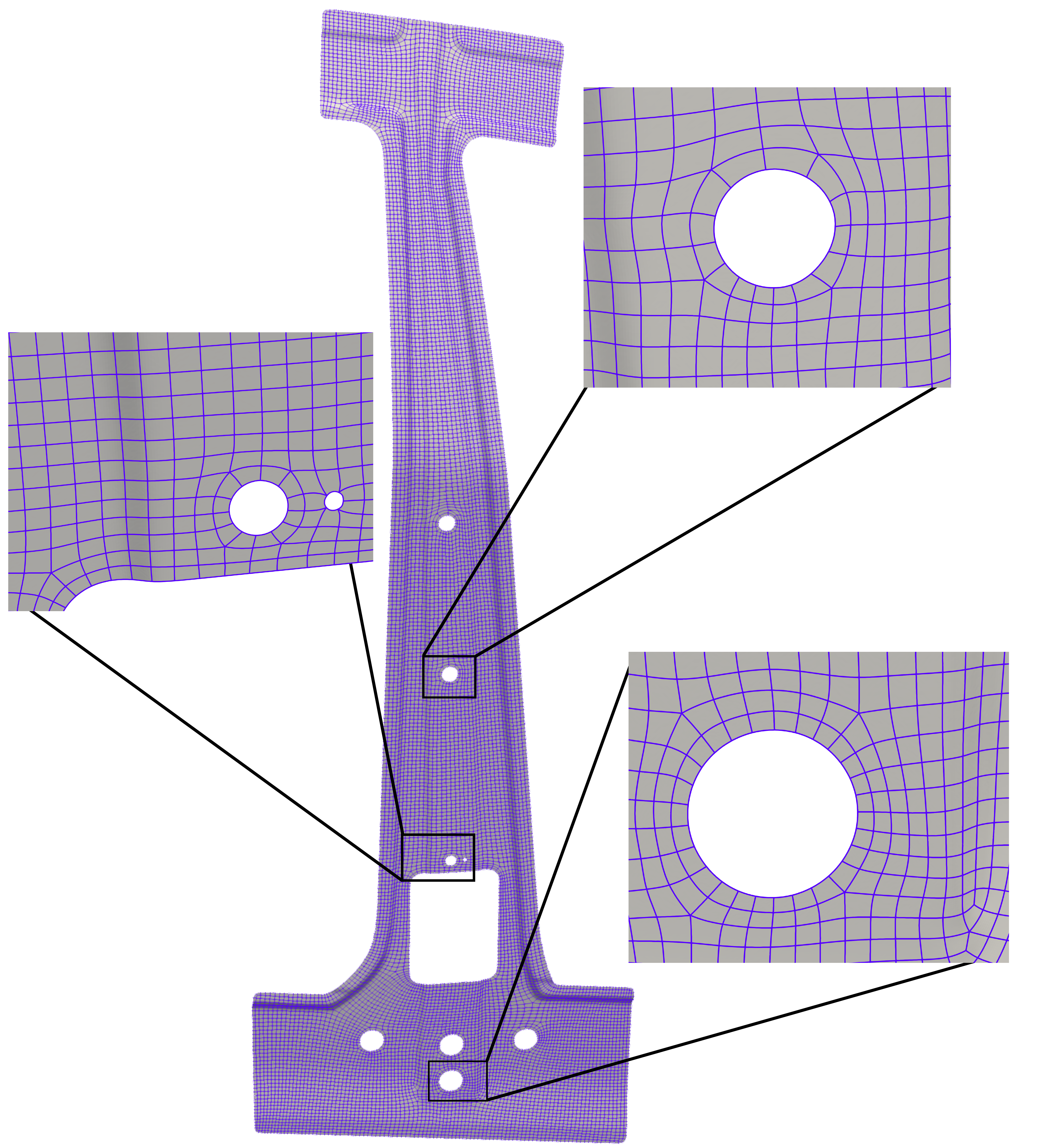}}\\
\caption{Inner part of an automotive B-pillar. (a) Control net with 13,575 control points. (b) G-spline surface and B\'ezier mesh. The B\'ezier mesh has 13,037 elements with an average element size of 5 mm. }
\label{inner}
\end{figure}

\begin{figure} [t!] 
 \centering
 \subfigure[]{\includegraphics[scale=.11]{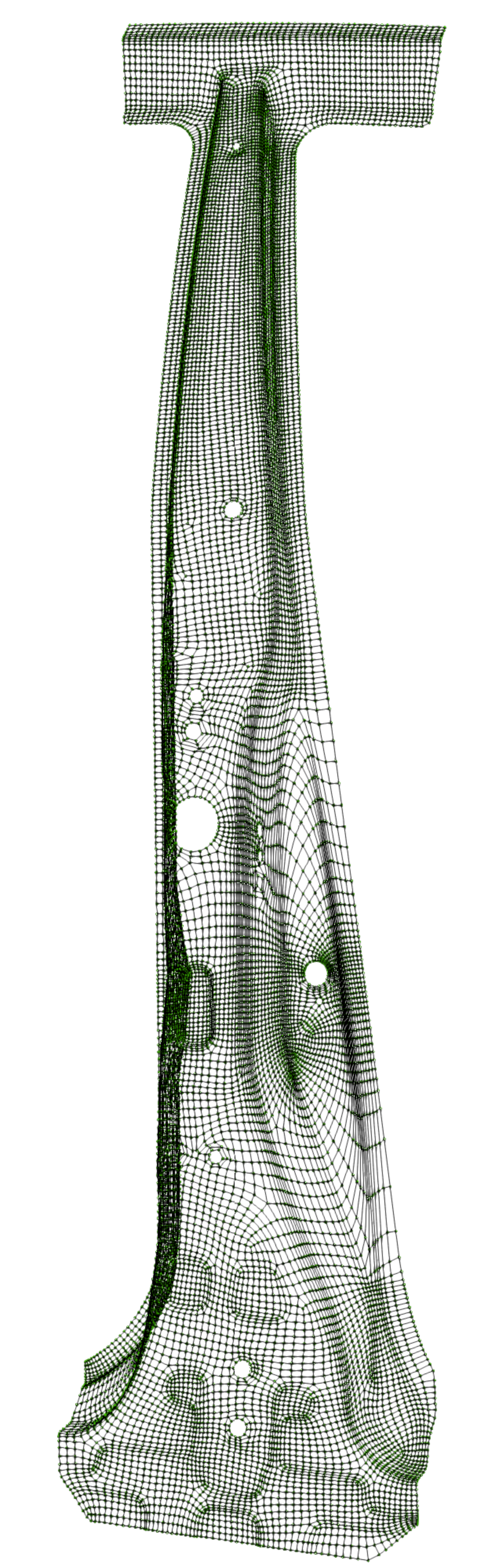}} 
 \subfigure[]{\includegraphics[scale=.21]{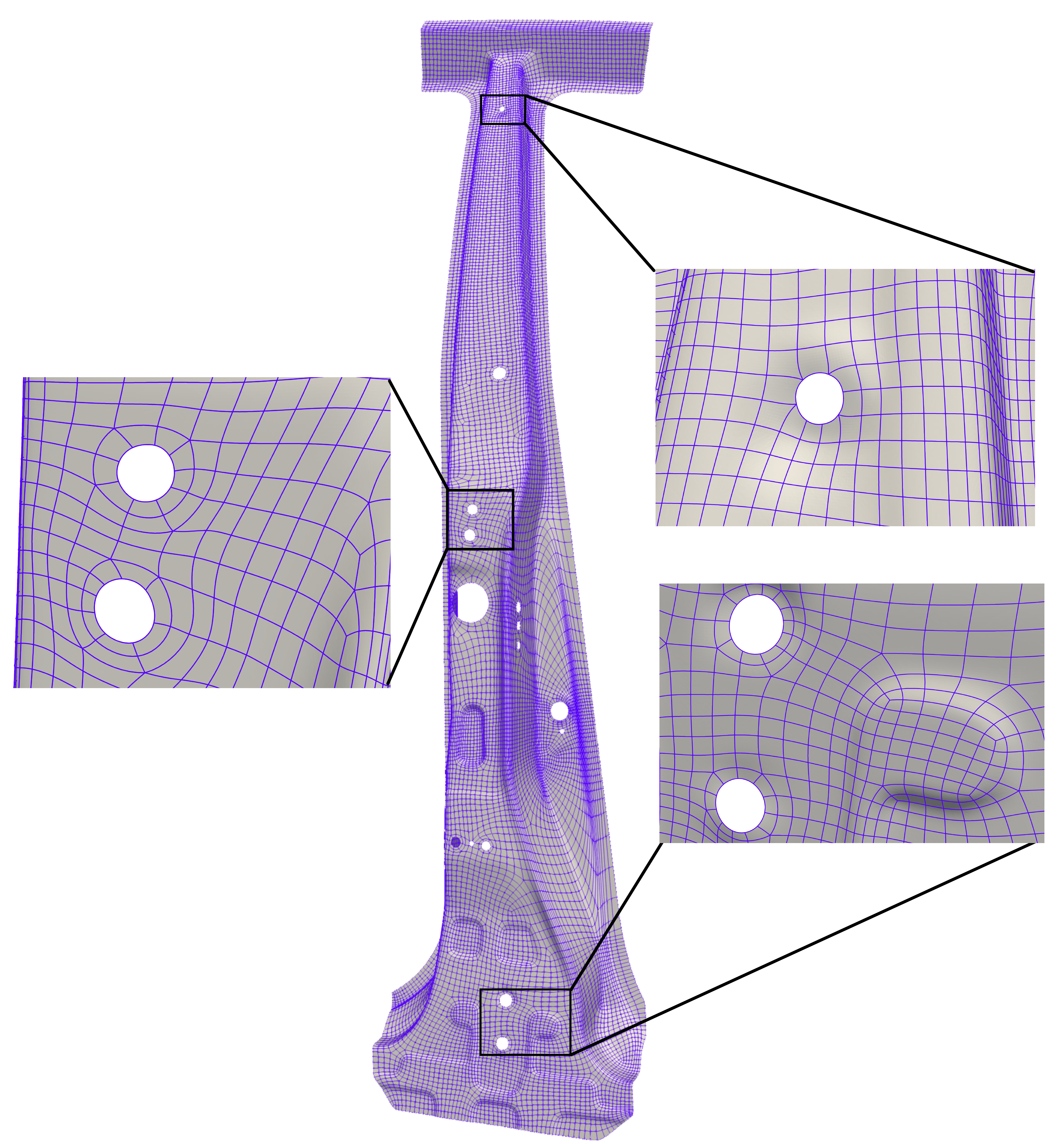}}\\
\caption{Outer part of an automotive B-pillar. (a) Control net with 16,108 control points. (b) G-spline surface and B\'ezier mesh. The B\'ezier mesh has 15,685 elements with an average element size of 5 mm. }
\label{outer}
\end{figure}

The construction $C^0$ is known to result in good surface quality in element interiors. The constructions $G^1$P, $G^1$R, and D-patch all involve additional steps on top of the steps needed for the construction $C^0$. Thus, we expect that these additional steps done to increase the surface continuity across element boundaries can also maintain a good surface quality in element interiors.

When representing manifold surfaces with arbitrary topological genus using splines with extraordinary points, it is rare to encounter EPs other than interior EPs with valences 3, 5, and 6 and boundary EPs with valence 3. Thus, we will focus on these types of EPs in our surface-quality studies. We constructed six control nets involving multiple interior EPs per face with different combinations of valences as specified in Figs. \ref{sqcontrolnets} and \ref{sqbeziermeshes}. Table \ref{tablesq} shows the constructions $G^1$P and $G^1$R preserve the good surface quality of the construction $C^0$ in element interiors since these three EP constructions have similar minimum thickness with invalid area element for the six control nets considered. However, as shown in Table \ref{tablesq}, the minimum thickness with invalid area element using the construction Dpatch is several times smaller than with the other three EP constructions. This evinces the poor surface quality of the D-patch framework.

We now consider the stiffener of a B-pillar, the inner part of a B-pillar, and the outer part of a B-pillar. In \cite{wei2022analysis}, we designed from scratch the outer part of a B-pillar and a side outer panel using the T-spline tools of Autodesk Fusion360. Then, we exported their control nets and use them as input in our in-house code. We could obtain control nets using the same type of workflow employing the T-spline tools of Autodesk Fusion360 again or the SubD tools of Rhinoceros 3D. This type of workflow is the closest to having the same geometric representation for the design and the analysis of thin-walled structures that current commercial software tools allow. However, we decided to take a different path here to illustrate an alternative workflow that can be used whenever a standard CAD file that represents the thin-walled structure using trimmed NURBS already exists. The stiffener, the inner part, and the outer part of a B-pillar were created in the commercial CAD software CATIA using trimmed NURBS. The stiffener, the inner part, and the outer part have 349, 311, and 845 trimmed NURBS patches, respectively. We then used the retopology tools of the computer animation software 3ds Max (namely, the Reform option of the Retopology modifier) to obtain a control net for each of the three parts. In the computer-animation industry, these retopology tools are used to transfer either a trimmed NURBS CAD file or a very fine triangular mesh into a SubD, but in our case we are only interested in obtaining the control net and export it from Autodesk 3ds Max as an .obj file. These control nets are shown in Figs. \ref{stiffener}, \ref{inner}, and \ref{outer} together with the G-spline surfaces and B\'ezier meshes obtained using these control nets as input in our in-house code that builds a G-spline surface for any given control net. The stiffener, the inner part, and the outer part have  4, 9, and 15 holes, respectively. In all three parts, the average element size is 5 mm, which is the most common element size used in crash simulations by the automotive industry. The G-spline representation of the stiffener contains 27 interior EPs with valence 3 and 39 interior EPs with valence 5. The G-spline representation of the inner part contains 18 interior EPs with valence 3, 46 interior EPs with valence 5, and 4 boundary EPs with valence 3. The G-spline representation of the outer part contains 67 interior EPs with valence 3, 101 interior EPs with valence 5, and 14 boundary EPs with valence 3. All three control nets have multiple EPs per face. This demonstrates the need for having EP constructions suitable for IGA that can handle multiple EPs per face as well as interior and boundary EPs. Since the construction D-patch cannot handle boundary EPs, we manually changed the connectivity of the control nets to remove the boundary EPs so that we can include comparisons among the four EP constructions for these three complex structural parts. As shown in Table \ref{tablesq}, the constructions $G^1$P, $G^1$R, and $C^0$ have similar minimum thickness with invalid area element, but the construction D-patch has a minimum thickness with invalid area element that is several times smaller. Using the construction D-patch, the element in which the minimum thickness with invalid area element appears is an irregular element of an interior EP with valence 3 for all three geometries, that is, it is in a region that is not affected by the manual changes that we introduced to remove the boundary EPs when using the construction D-patch. The actual thickness of the inner part is 1 mm and the actual thickness of the outer part and the stiffener is 1.6 mm. Thus, when using the construction D-patch, the shell discretizations of the outer part and the stiffener are invalid and FEA simulations cannot be carried out.

\section{Eigenvalue tests}

\begin{table}[t!]
   \caption{Eigenvalues of the stiffener obtained with G-splines and with bilinear quadrilaterals using the consistent mass matrix.} \label{stiffcon}
   \bigskip
     \centering
     \renewcommand{\arraystretch}{1.20}
     \begin{tabular}{c@{\hspace{12.0mm}}  c@{\hspace{12.0mm}}  c@{\hspace{12.0mm}}  c@{\hspace{12.0mm}}}
\hline
 Mode & IGA K-L & IGA R-M & FEM \\
\hline
7   & $6.831958 \times 10^{-1}$ & $6.802610 \times 10^{-1}$   & $6.729194\times 10^{-1}$   \\          
8   & $1.643057 \times 10^{0}$ & $1.625134\times 10^{0}$ & $1.669634\times 10^{0}$  \\     
9   & $3.837714 \times 10^{0}$  & $3.808307\times 10^{0}$  & $3.806768\times 10^{0}$  \\          
10   & $1.008562 \times 10^{1}$ & $9.985609\times 10^{0}$  & $1.023542 \times 10^{1}$ \\    
11   & $1.655282 \times 10^{1}$ & $1.638740\times 10^{1}$ & $1.652496\times 10^{1}$  \\     
12   & $2.737640 \times 10^{1}$  & $2.715919\times 10^{1}$ & $2.780284\times 10^{1}$   \\         
\hline      
     \end{tabular}     
   \end{table}

\begin{table}[t!]
   \caption{Eigenvalues of the stiffener obtained with G-splines and with bilinear quadrilaterals using the lumped mass matrix.} \label{stifflum}
   \bigskip
     \centering
     \renewcommand{\arraystretch}{1.20}
     \begin{tabular}{c@{\hspace{12.0mm}}  c@{\hspace{12.0mm}}  c@{\hspace{12.0mm}}  c@{\hspace{12.0mm}}}
\hline
 Mode & IGA K-L & IGA R-M & FEM \\
\hline
7   & $6.820109\times 10^{-1}$ & $6.791284\times 10^{-1}$   & $6.690467\times 10^{-1}$   \\          
8   & $1.655572\times 10^{0}$  & $1.637535\times 10^{0}$ & $1.667858\times 10^{0}$     \\     
9   & $3.829959\times 10^{0}$ & $3.800838\times 10^{0}$  & $3.784465\times 10^{0}$   \\          
10   & $1.013302\times 10^{1}$ & $1.003303\times 10^{1}$  & $1.020809\times 10^{1}$  \\    
11   & $1.651432\times 10^{1}$ & $1.635013\times 10^{1}$  & $1.642579\times 10^{1}$   \\     
12   & $2.734857\times 10^{1}$  & $2.713475\times 10^{1}$ & $2.764796\times 10^{1}$   \\         
\hline      
     \end{tabular}     
   \end{table}

G-spline surfaces, as any other type of splines that admits B\'ezier extraction, can be imported in the commercial FEA software LS-DYNA. In this section, we solve eigenvalue problems with G-splines in LS-DYNA and perform comparisons with conventional finite elements. The geometries considered are the stiffener of a B-pillar, the inner part of a B-pillar, and the outer part of a B-pillar shown in Figs. Figs. \ref{stiffener}, \ref{inner}, and \ref{outer}, respectively. In this section, we use millimeters, milliseconds, and kilograms as length, time, and mass units, respectively, which is the unit system more frequently used in the automotive industry. The value of the density, the Young modulus, and the Poisson ratio are $\rho = 7.8 \times 10^{-6} $, $E = 2.0 \times 10^{2}$, $\nu = 0.3$, respectively. 



\begin{table}[t!]
   \caption{Eigenvalues of the inner part obtained with G-splines and with bilinear quadrilaterals using the consistent mass matrix.} \label{innercon}
   \bigskip
     \centering
     \renewcommand{\arraystretch}{1.20}
     \begin{tabular}{c@{\hspace{12.0mm}}  c@{\hspace{12.0mm}}  c@{\hspace{12.0mm}}  c@{\hspace{12.0mm}}}
\hline
 Mode & IGA K-L & IGA R-M & FEM \\
\hline
7   & $1.707981\times 10^{-3}$ & $1.711364\times 10^{-3}$  & $1.672901\times 10^{-3}$   \\          
8   & $1.086697\times 10^{-2}$  & $1.089283\times 10^{-2}$ & $1.054207\times 10^{-2}$   \\     
9   & $1.545168\times 10^{-2}$  & $1.546120\times 10^{-2}$  & $1.509589\times 10^{-2}$  \\          
10   & $4.470250\times 10^{-2}$ & $4.483026\times 10^{-2}$  & $4.387936\times 10^{-2}$ \\    
11   & $7.676406\times 10^{-2}$ & $7.688878\times 10^{-2}$ & $7.640507\times 10^{-2}$ \\     
12   & $9.107535\times 10^{-2}$  & $9.122090\times 10^{-2}$ & $9.131929\times 10^{-2}$  \\         
\hline      
     \end{tabular}     
   \end{table}
   
\begin{table}[t!]
   \caption{Eigenvalues of the inner part obtained with G-splines and with bilinear quadrilaterals using the lumped mass matrix.} \label{innerlum}
   \bigskip
     \centering
     \renewcommand{\arraystretch}{1.20}
     \begin{tabular}{c@{\hspace{12.0mm}}  c@{\hspace{12.0mm}}  c@{\hspace{12.0mm}}  c@{\hspace{12.0mm}}}
\hline
 Mode & IGA K-L & IGA R-M & FEM \\
\hline
7   & $1.723343\times 10^{-3}$ & $1.726760\times 10^{-3}$  & $1.672105\times 10^{-3}$   \\          
8   & $1.097192\times 10^{-2}$  & $1.099804\times 10^{-2}$ & $1.054039\times 10^{-2}$   \\     
9   & $1.558502\times 10^{-2}$  & $1.559465\times 10^{-2}$ & $1.508588\times 10^{-2}$  \\          
10   & $4.509849\times 10^{-2}$ & $4.522745\times 10^{-2}$  & $4.385520\times 10^{-2}$ \\    
11   & $7.737074\times 10^{-2}$ & $7.749579\times 10^{-2}$ & $7.632417\times 10^{-2}$ \\     
12   & $9.180853\times 10^{-2}$  & $9.195564\times 10^{-2}$ & $9.123395\times 10^{-2}$ \\         
\hline      
     \end{tabular}     
   \end{table}

   \begin{table}[t!]
   \caption{Eigenvalues of the outer part obtained with G-splines and with bilinear quadrilaterals using the consistent mass matrix.} \label{outercon}
   \bigskip
     \centering
     \renewcommand{\arraystretch}{1.20}
     \begin{tabular}{c@{\hspace{12.0mm}}  c@{\hspace{12.0mm}}  c@{\hspace{12.0mm}}  c@{\hspace{12.0mm}}}
\hline
 Mode & IGA K-L & IGA R-M & FEM \\
\hline
7   & $2.111531\times 10^{-2}$    & $2.107359\times 10^{-2}$   & $2.075109\times 10^{-2}$  \\          
8   & $1.439462\times 10^{-1}$    & $1.437286\times 10^{-1}$   & $1.419425\times 10^{-1}$    \\     
9   & $4.161993\times 10^{-1}$    & $4.157969\times 10^{-1}$   & $4.164362\times 10^{-1}$     \\          
10  & $5.400375\times 10^{-1}$   & $5.393084\times 10^{-1}$    & $5.413323\times 10^{-1}$    \\    
11  & $1.093012\times 10^{0}$    & $1.090889\times 10^{0}$     & $1.095769\times 10^{0}$      \\     
12  & $1.332509\times 10^{0}$    & $1.330410\times 10^{0}$     & $1.336470\times 10^{0}$       \\         
\hline      
     \end{tabular}     
   \end{table}   
   
 \begin{table}[t!]
   \caption{Eigenvalues of the outer part obtained with G-splines and with bilinear quadrilaterals using the lumped mass matrix.} \label{outerlum}
   \bigskip
     \centering
     \renewcommand{\arraystretch}{1.20}
     \begin{tabular}{c@{\hspace{12.0mm}}  c@{\hspace{12.0mm}}  c@{\hspace{12.0mm}}  c@{\hspace{12.0mm}}}
\hline
 Mode & IGA K-L & IGA R-M & FEM \\
\hline
7   & $2.128929\times 10^{-2}$               & $2.124744\times 10^{-2}$   & $2.073354\times 10^{-2}$         \\          
8   & $1.450363\times 10^{-1}$          & $1.448187\times 10^{-1}$        & $1.417880\times 10^{-1}$         \\     
9   & $4.196939\times 10^{-1}$          & $4.192901\times 10^{-1}$        & $4.161236\times 10^{-1}$           \\          
10   & $5.430577\times 10^{-1}$          & $5.423398\times 10^{-1}$            & $5.400843\times 10^{-1}$        \\    
11   & $1.098894\times 10^{0}$      & $1.096792\times 10^{0}$             & $1.093141\times 10^{0}$          \\     
12   & $1.340526\times 10^{0}$         & $1.338441\times 10^{0}$          & $1.333670\times 10^{0}$           \\         
\hline      
     \end{tabular}     
   \end{table}

An element size of 5 mm is used for both IGA and FEM simulations. For the FEM simulations, we used bilinear quadrilateral elements and the shell formulation ELFORM 16 of LS-DYNA. For the IGA simulations, we used both the construction $G^1$P and the construction $G^1$R and the shell formulations ELFORM 3 and ELFORM 2 of LS-DYNA. ELFORM 3 is a Reissner-Mindlin (R-M) shell formulation and ELFORM 2 is a Kirchhoff-Love (K-L) shell formulation. For all the eigenvalues included in this section, the differences between the construction $G^1$P and the construction $G^1$R are negligible (smaller than 0.02\% in all cases). Thus, for brevity, in the tables of this section, we only include the results obtained with the construction $G^1$P.

   \begin{figure} [] 
 \centering
 \subfigure[]{\includegraphics[scale=.34]{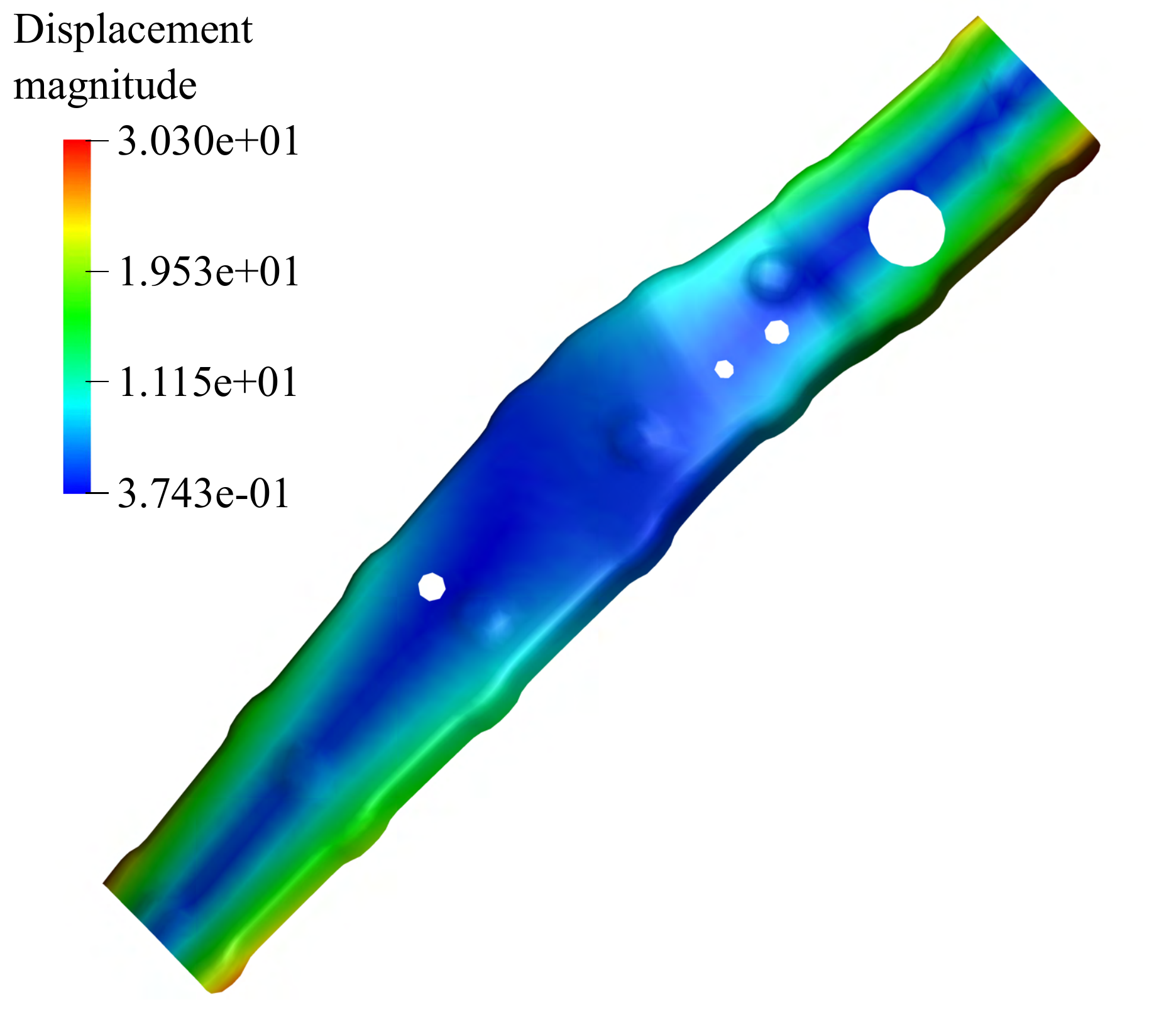}} 
 \subfigure[]{\includegraphics[scale=.34]{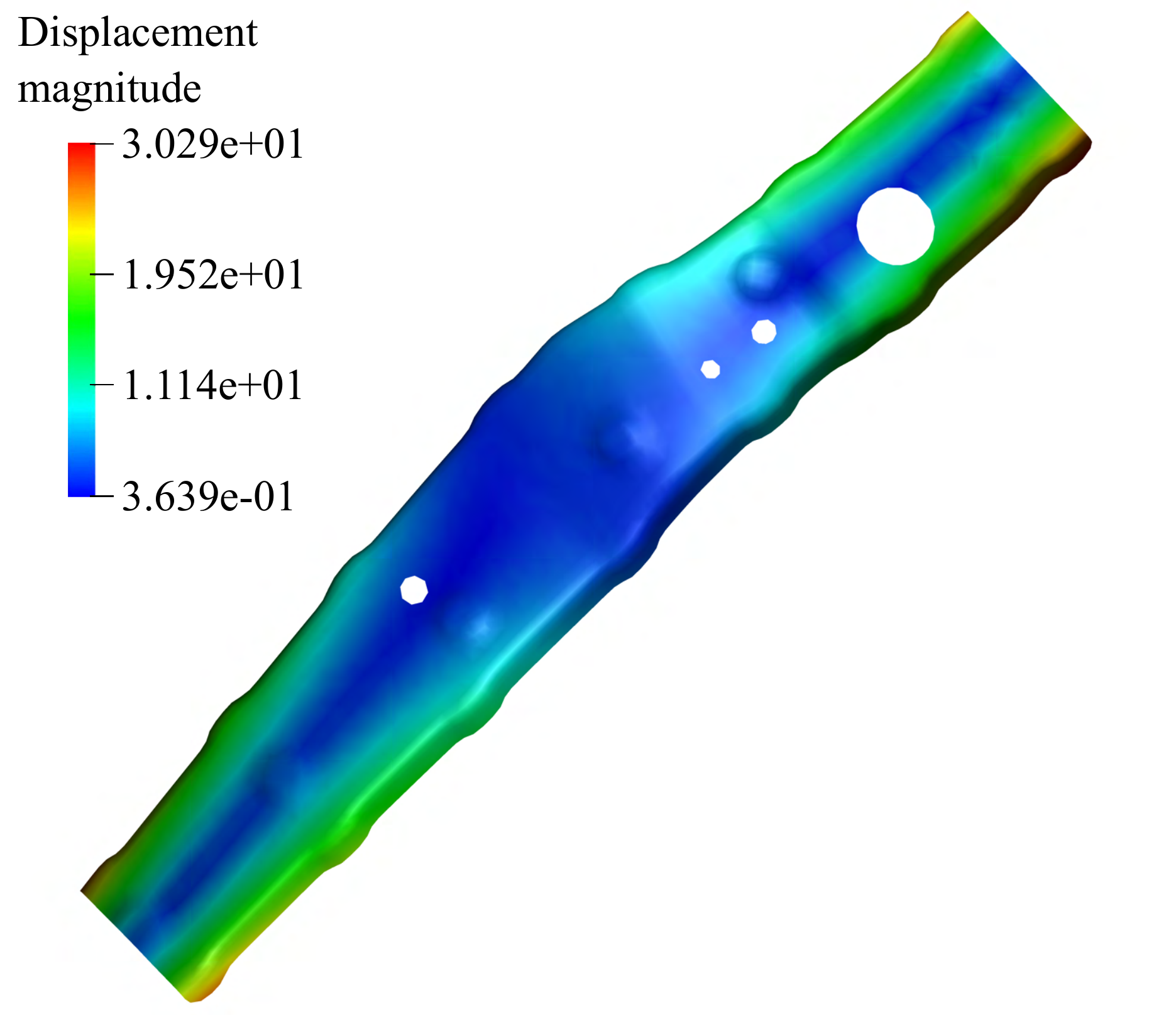}}
 \subfigure[]{\includegraphics[scale=.34]{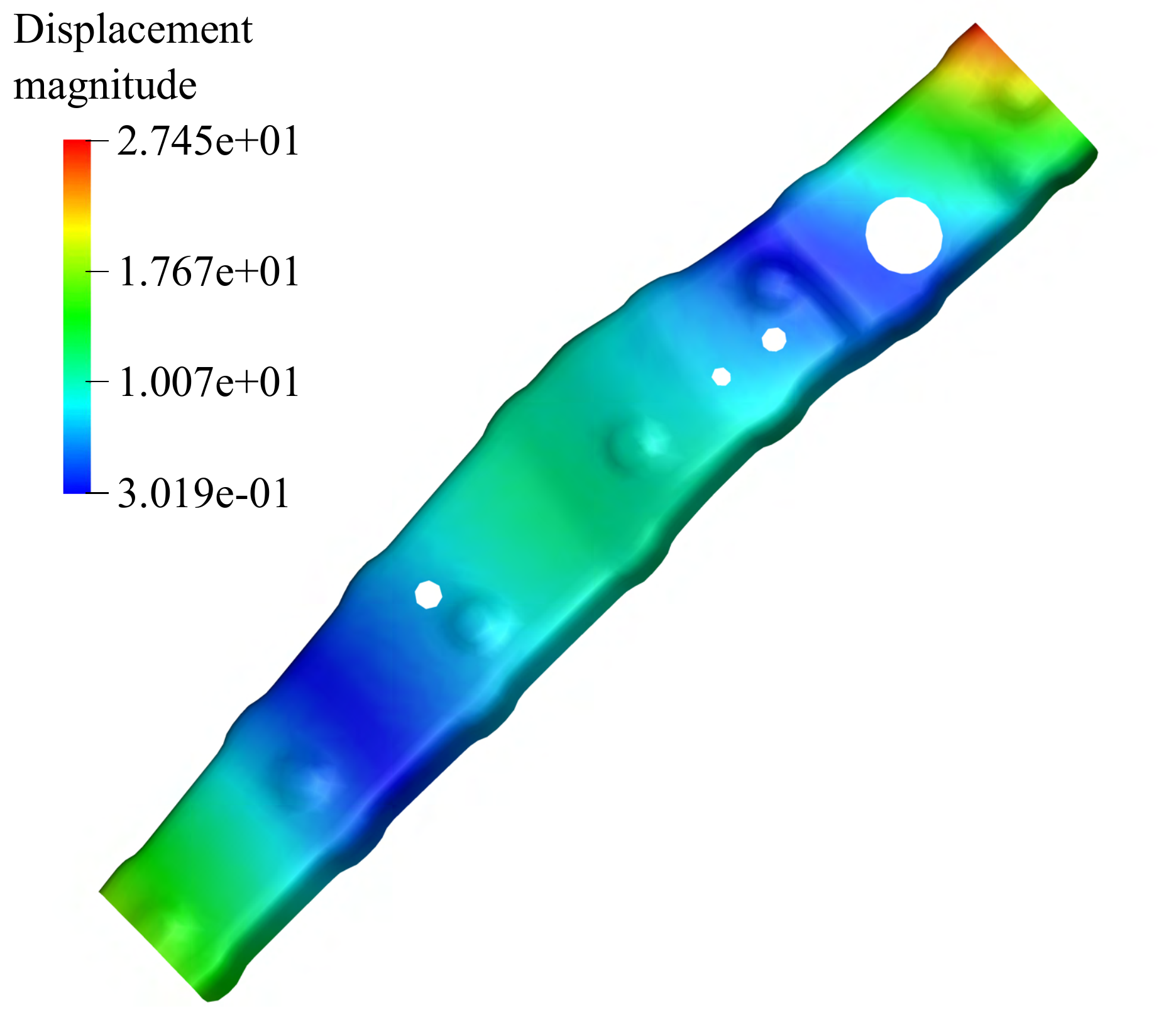}} 
 \subfigure[]{\includegraphics[scale=.34]{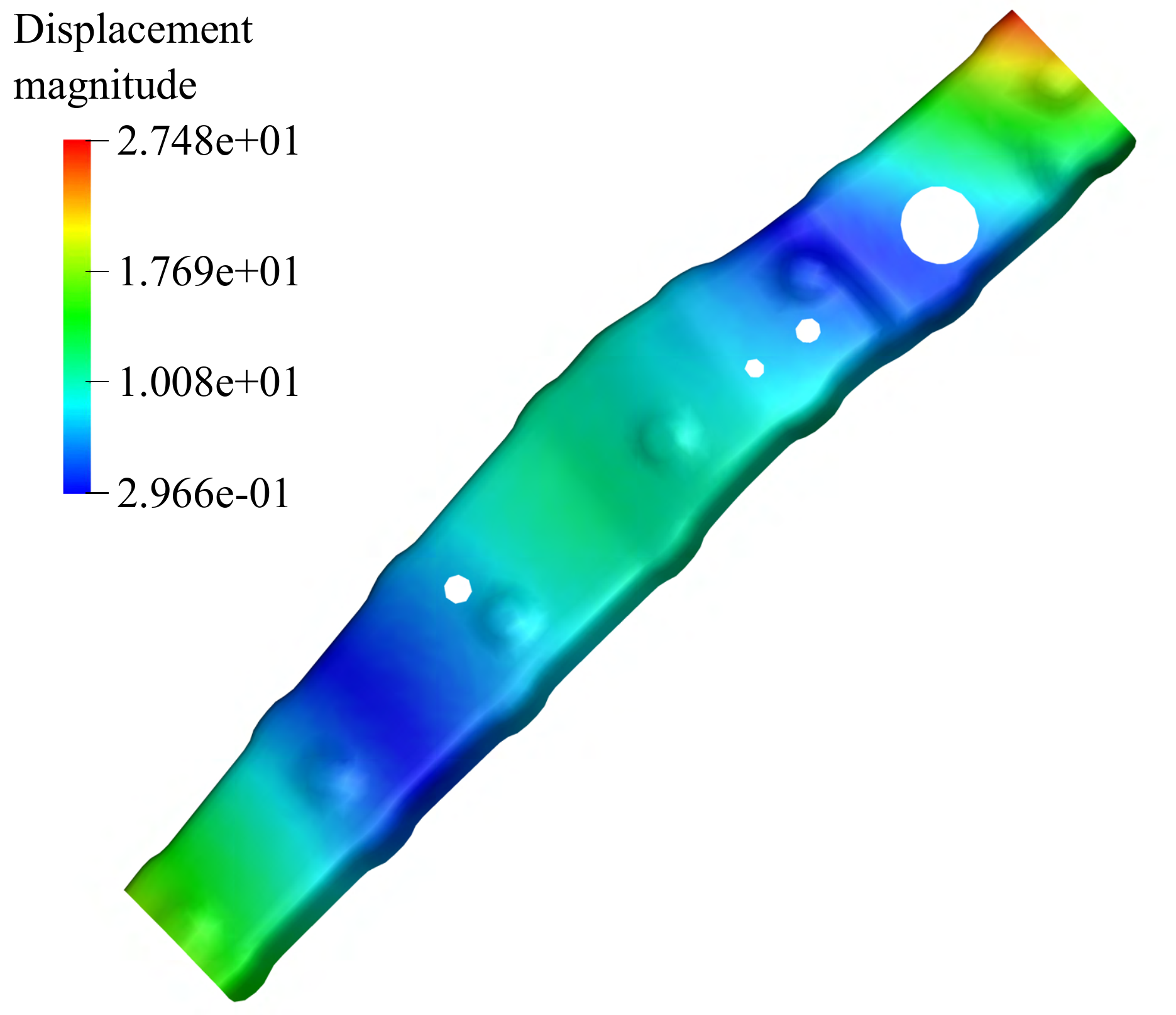}}
 \subfigure[]{\includegraphics[scale=.34]{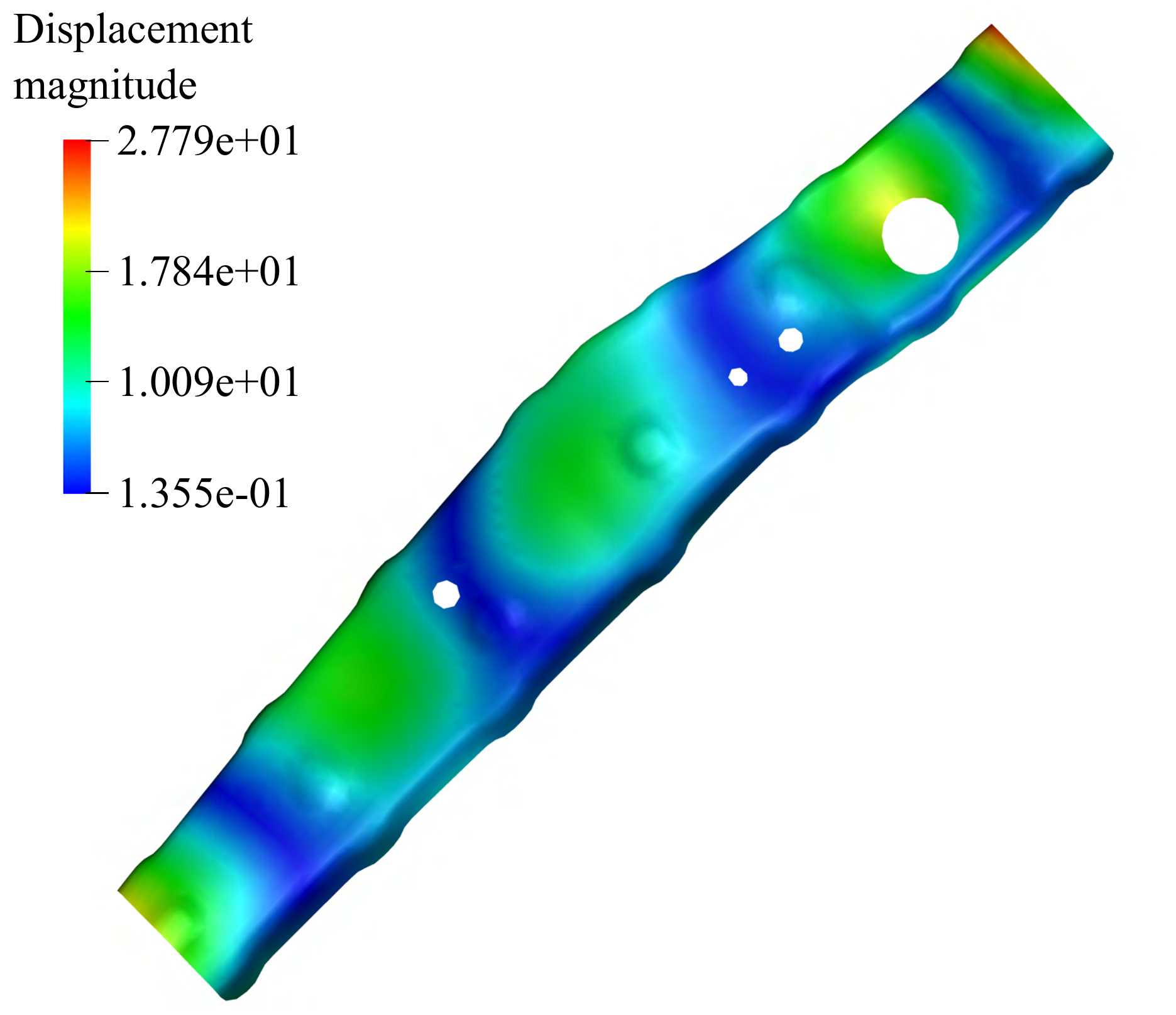}} 
 \subfigure[]{\includegraphics[scale=.34]{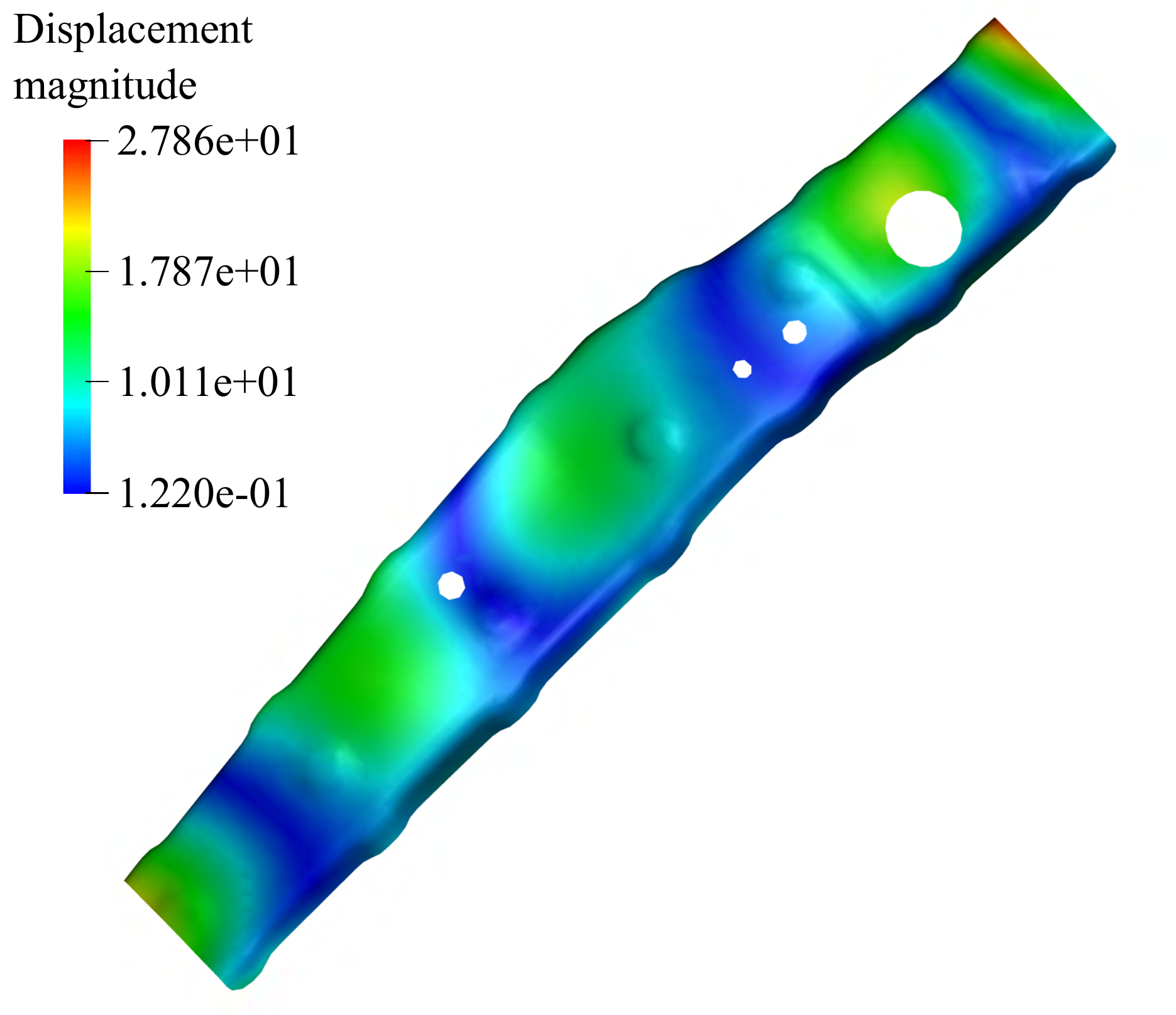}}\\
\caption{Stiffener of a B-pillar. (a), (c), and (e) illustrate the seventh, eighth, and twelfth mode shapes, respectively, using G-splines and Reissner-Mindlin shells. (b), (d), and (f) illustrate the seventh, eighth, and twelfth mode shapes, respectively, using conventional finite elements.}
\label{modeshapesstiff}
\end{figure}

   \begin{figure} [] 
 \centering
 \subfigure[]{\includegraphics[scale=.35]{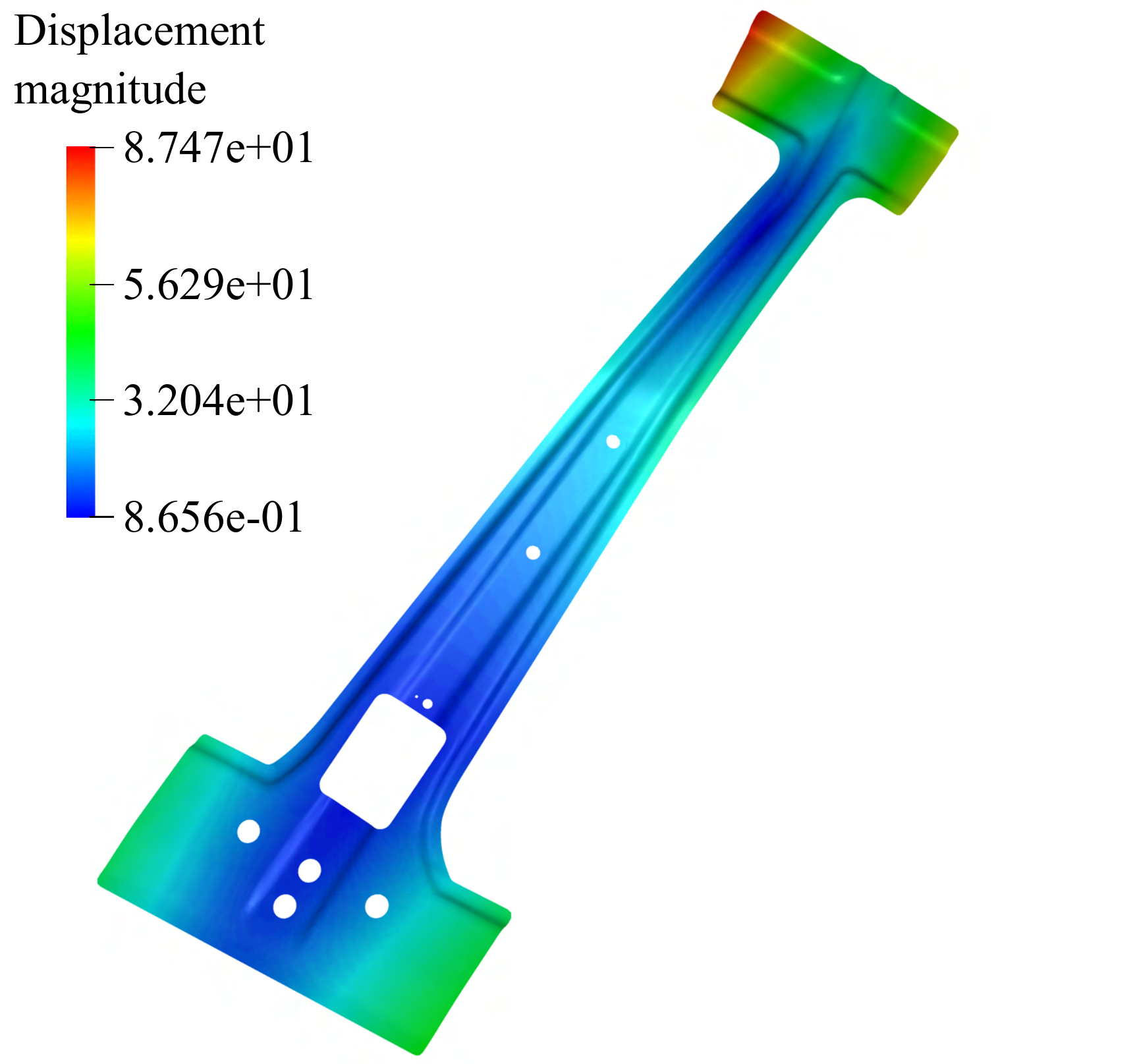}} 
 \subfigure[]{\includegraphics[scale=.35]{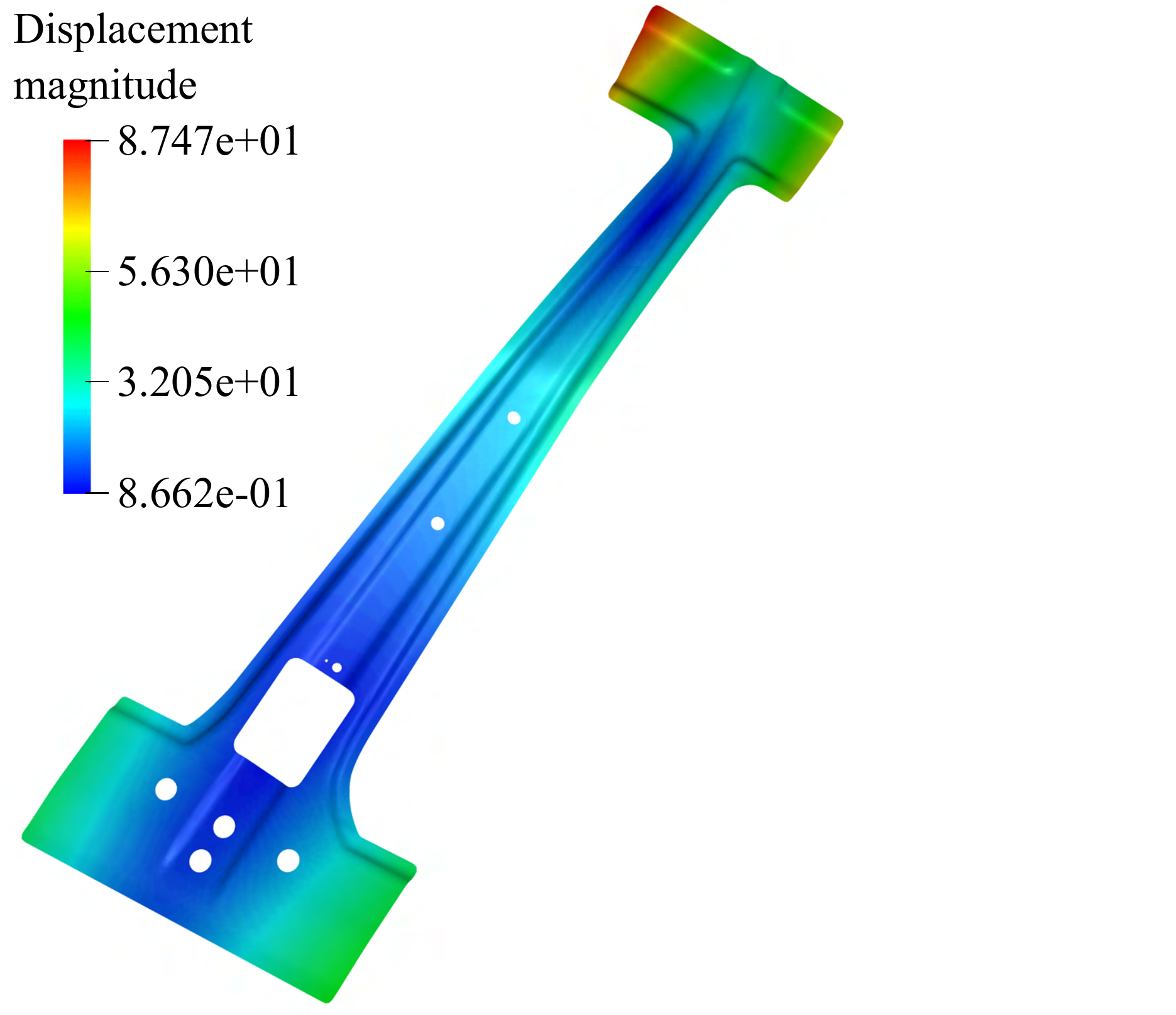}}
 \subfigure[]{\includegraphics[scale=.35]{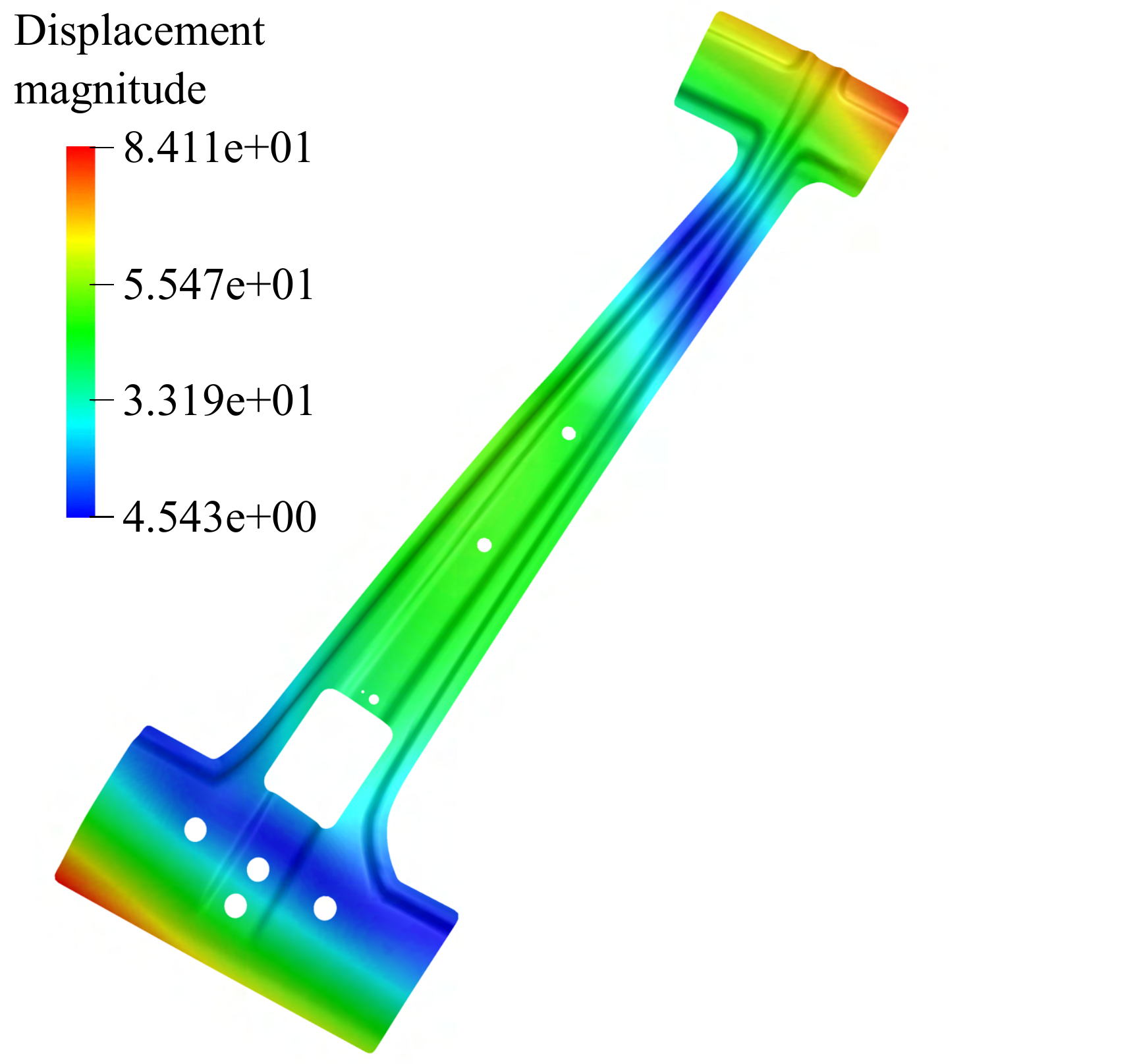}} 
 \subfigure[]{\includegraphics[scale=.35]{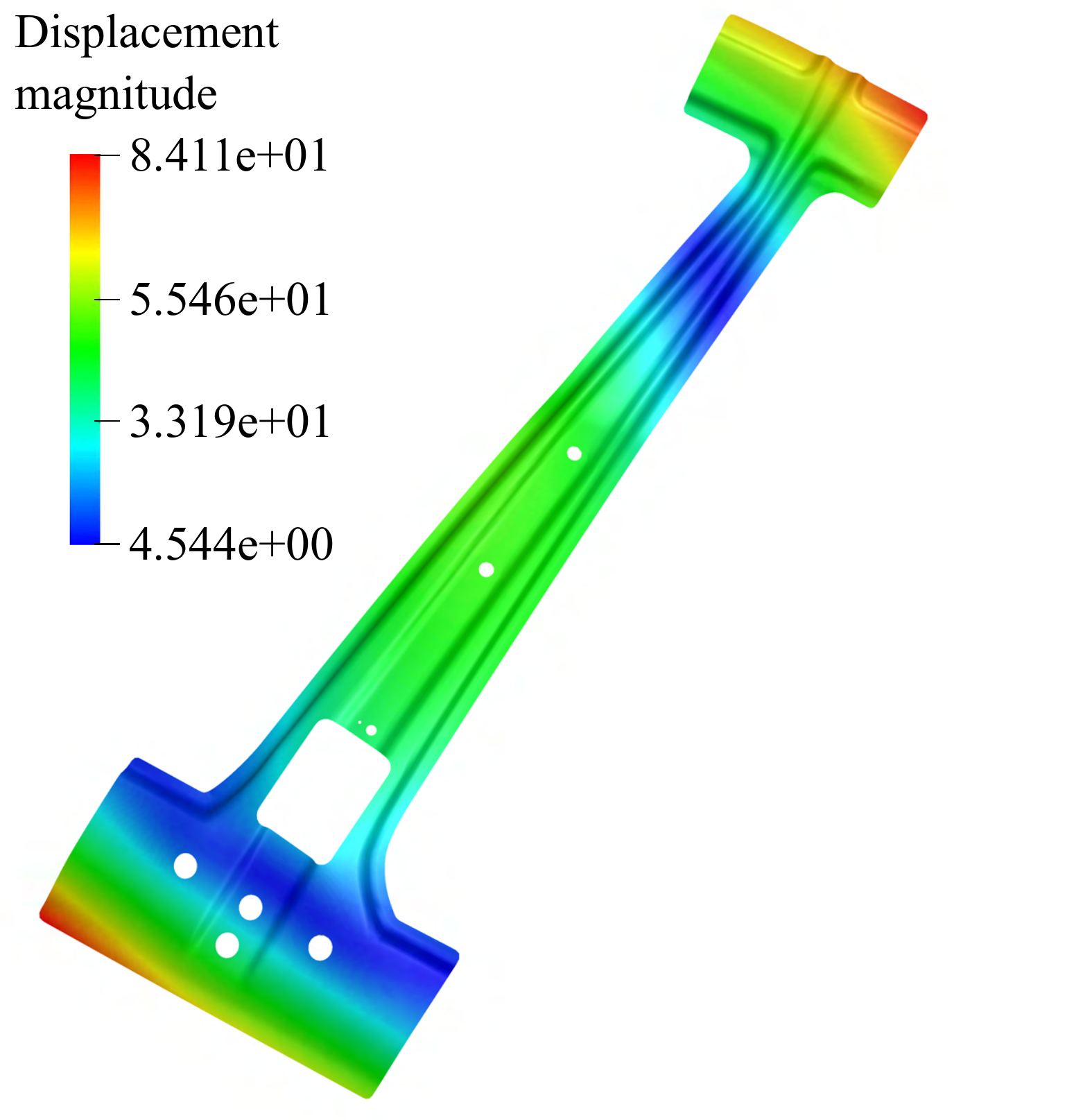}}
 \subfigure[]{\includegraphics[scale=.35]{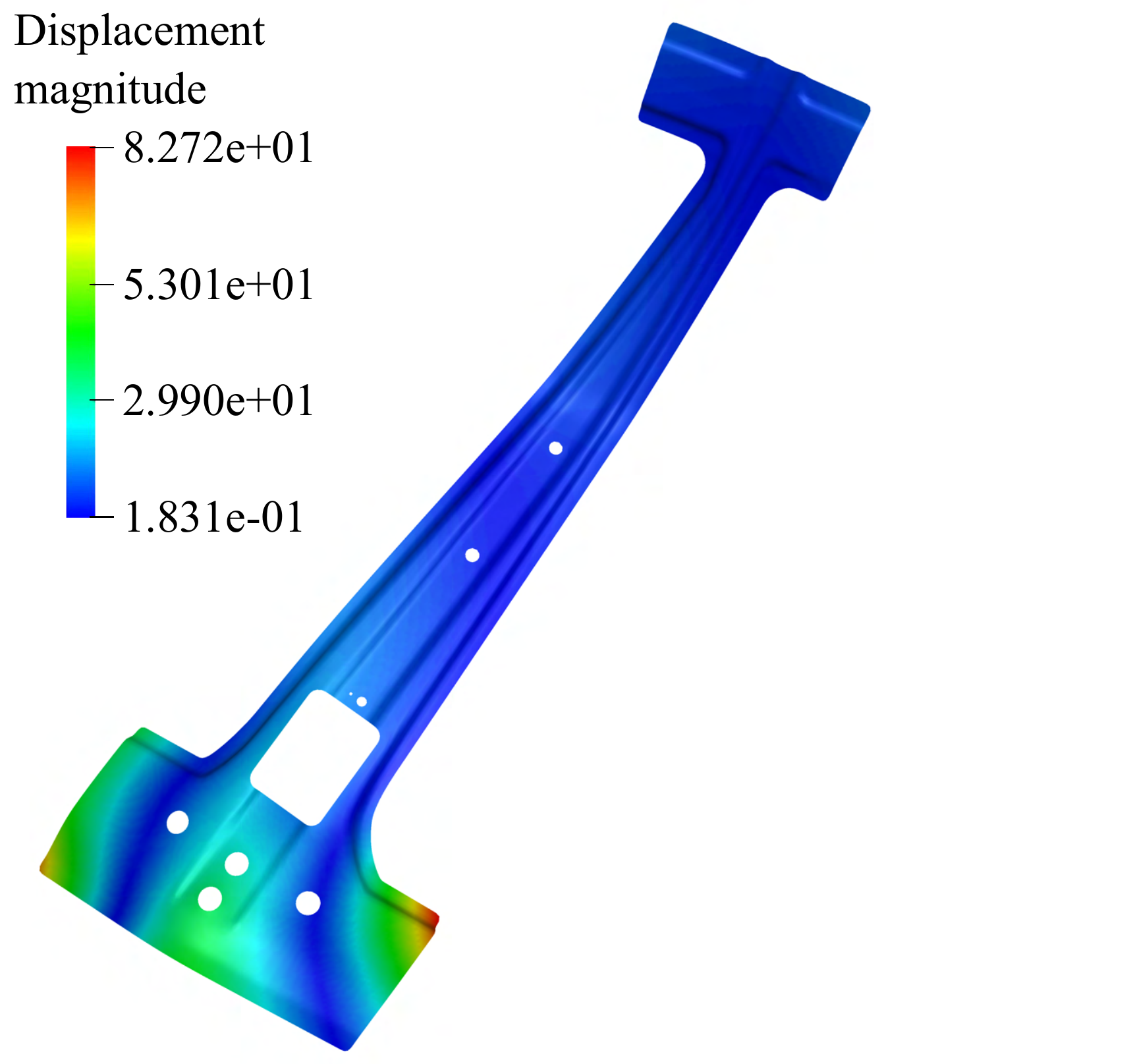}} 
 \subfigure[]{\includegraphics[scale=.35]{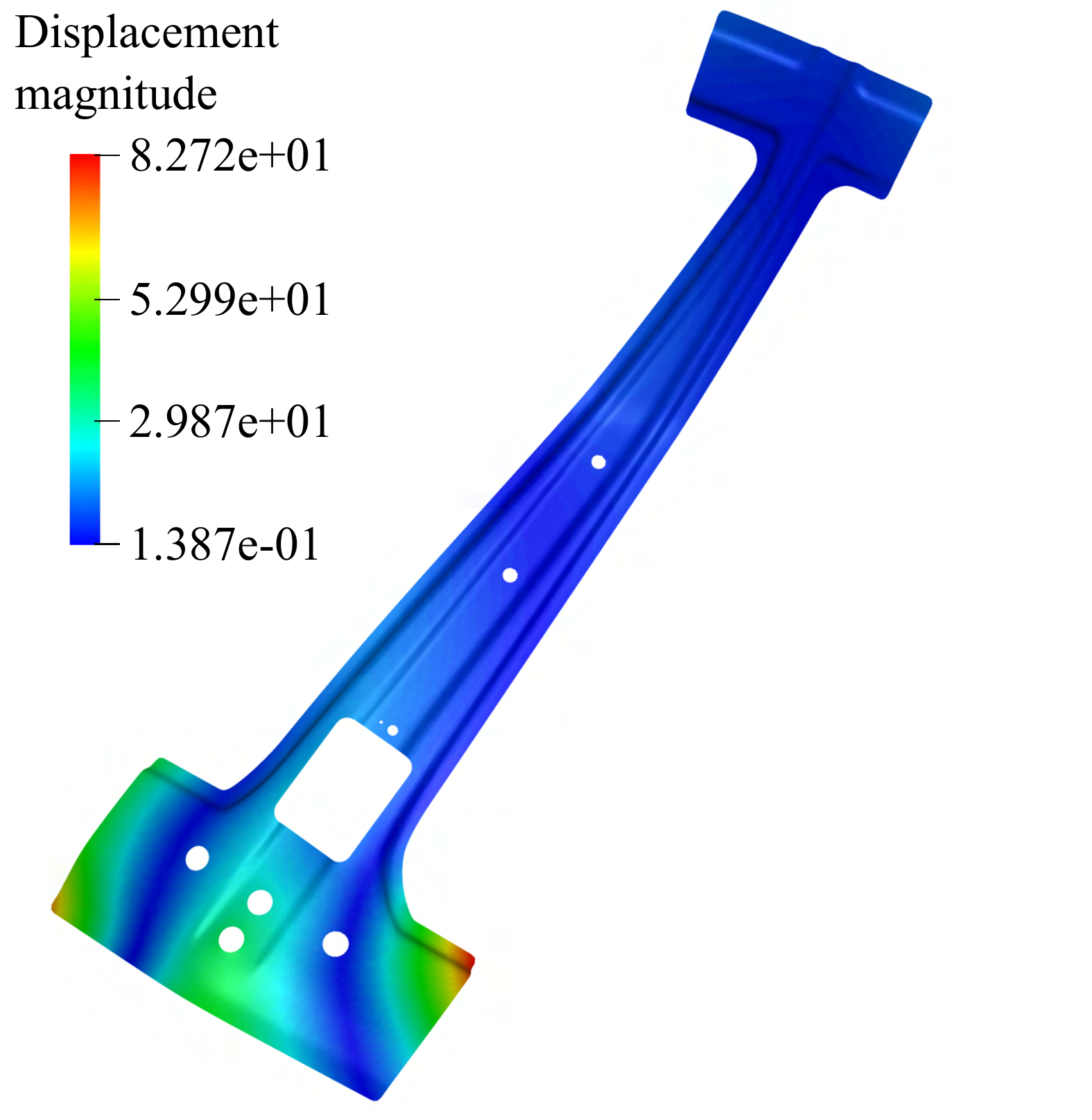}}\\
\caption{Inner part of a B-pillar. (a), (c), and (e) illustrate the seventh, eighth, and twelfth mode shapes, respectively, using G-splines and Reissner-Mindlin shells. (b), (d), and (f) illustrate the seventh, eighth, and twelfth mode shapes, respectively, using G-splines and Kirchhoff-Love shells.}
\label{modeshapesINNwithIGA}
\end{figure}

Since no Dirichlet boundary conditions are applied, the first six mode shapes and eigenvalues are expected to be related to rigid-body motions and we verified that is the case for all the IGA and FEM simulations that we run. Tables \ref{stiffcon} and \ref{stifflum} include the seventh, eighth, ninth, tenth, eleventh, twelfth eigenvalues of the stiffener obtained with FEM, IGA with R-M shells, and IGA with K-L shells using the consistent mass matrix and the lumped mass matrix, respectively. Tables \ref{innercon} and \ref{innerlum} and Tables \ref{outercon} and \ref{outerlum} do the same for the inner part and the outer part, respectively. As shown in Tables \ref{stiffcon}, \ref{stifflum}, \ref{innercon}, \ref{innerlum}, \ref{outercon}, and \ref{outerlum}, excellent agreement among FEM, IGA with R-M shells, and IGA with K-L shells is obtained in all cases.

\begin{figure} [] 
 \centering
 \subfigure[]{\includegraphics[scale=.35]{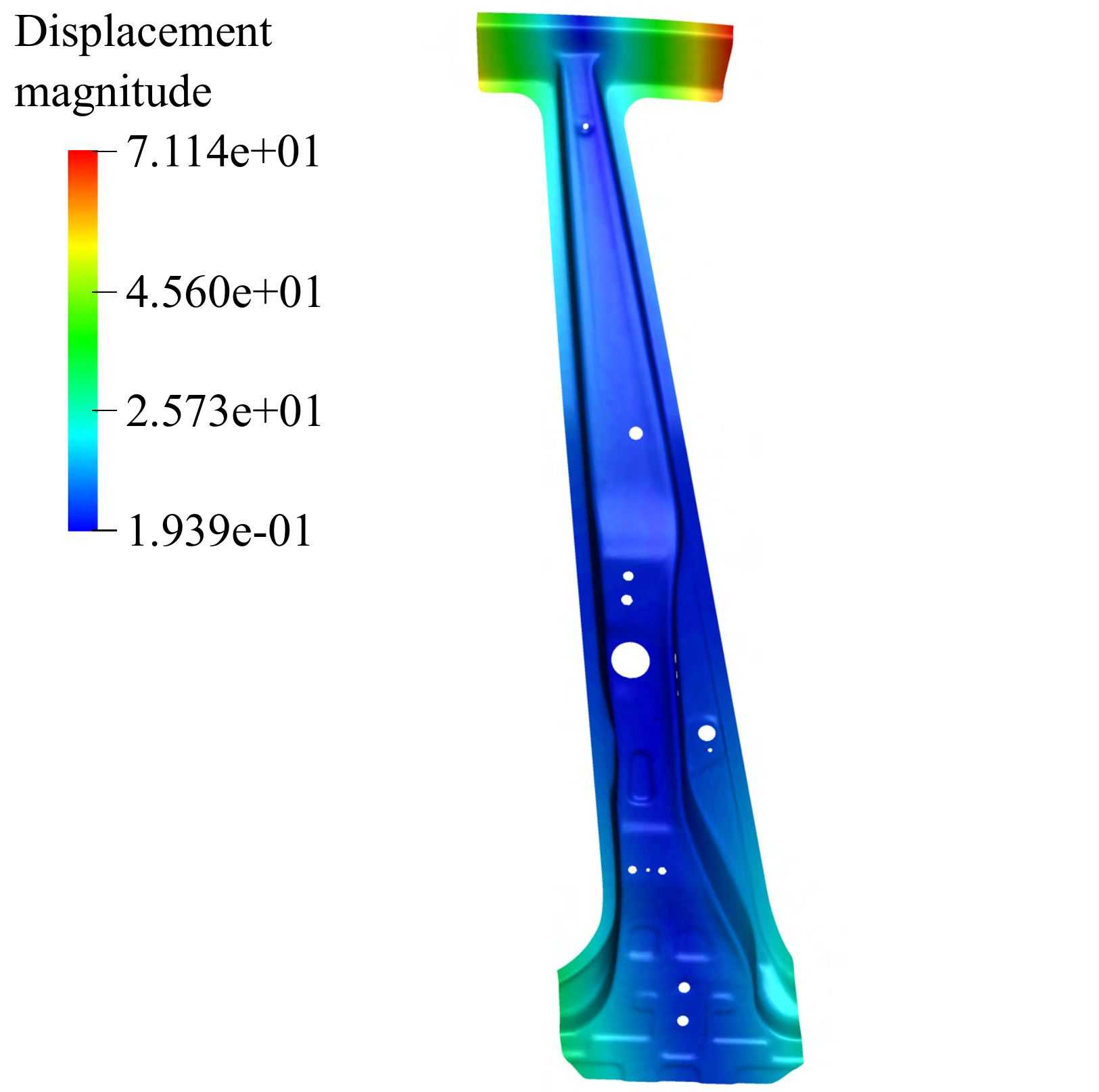}} 
 \subfigure[]{\includegraphics[scale=.35]{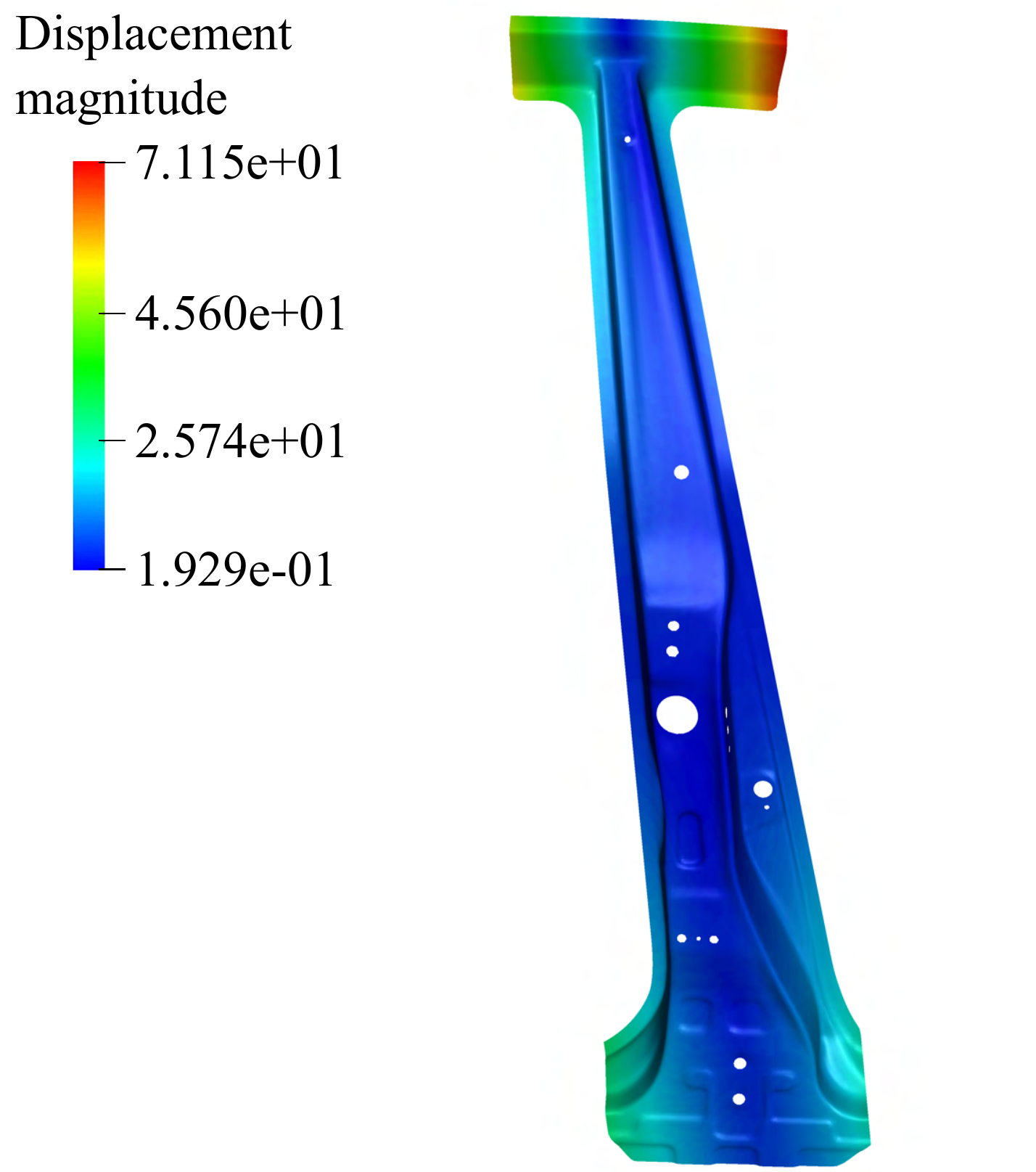}}
 \subfigure[]{\includegraphics[scale=.35]{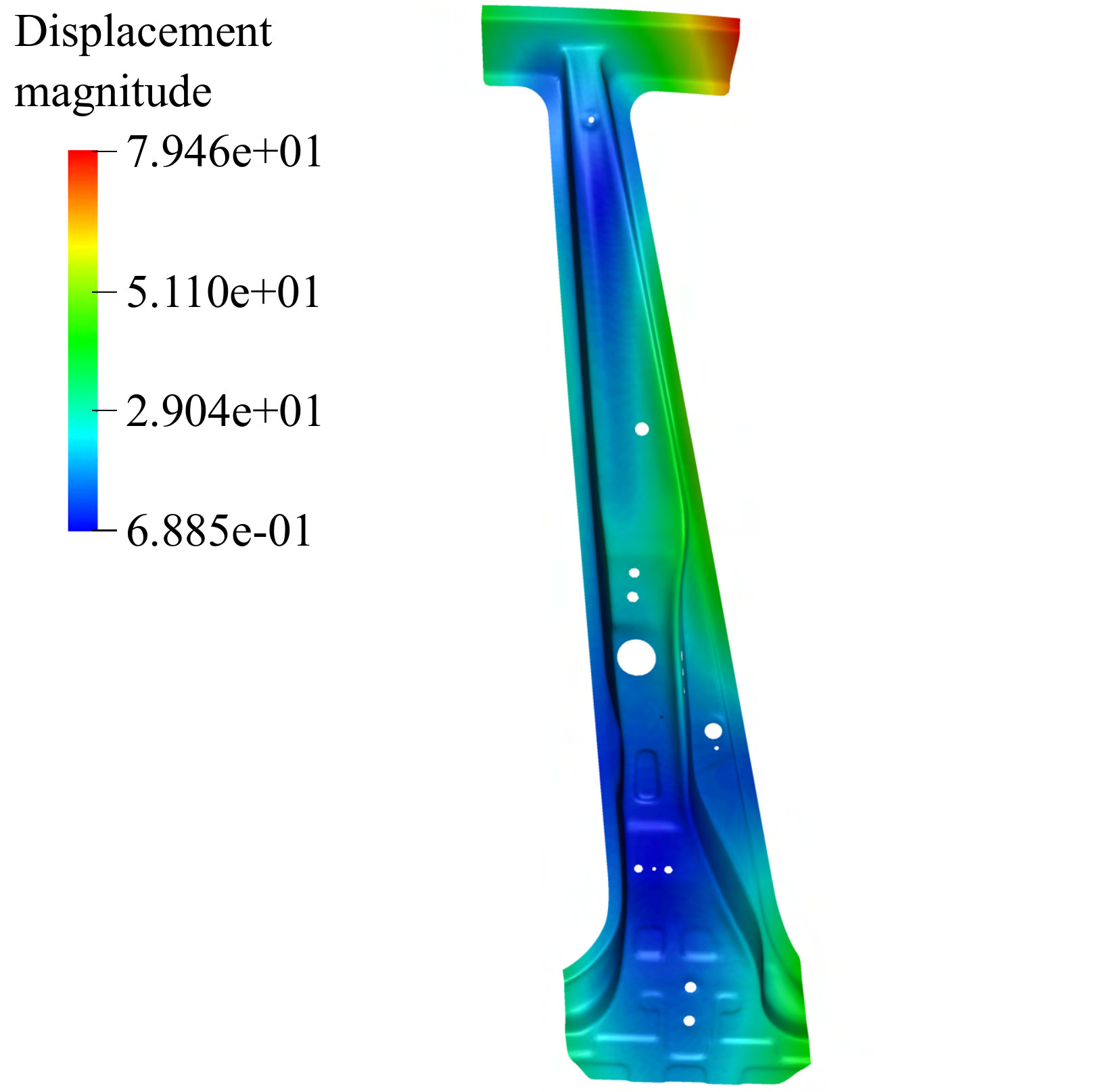}} 
 \subfigure[]{\includegraphics[scale=.35]{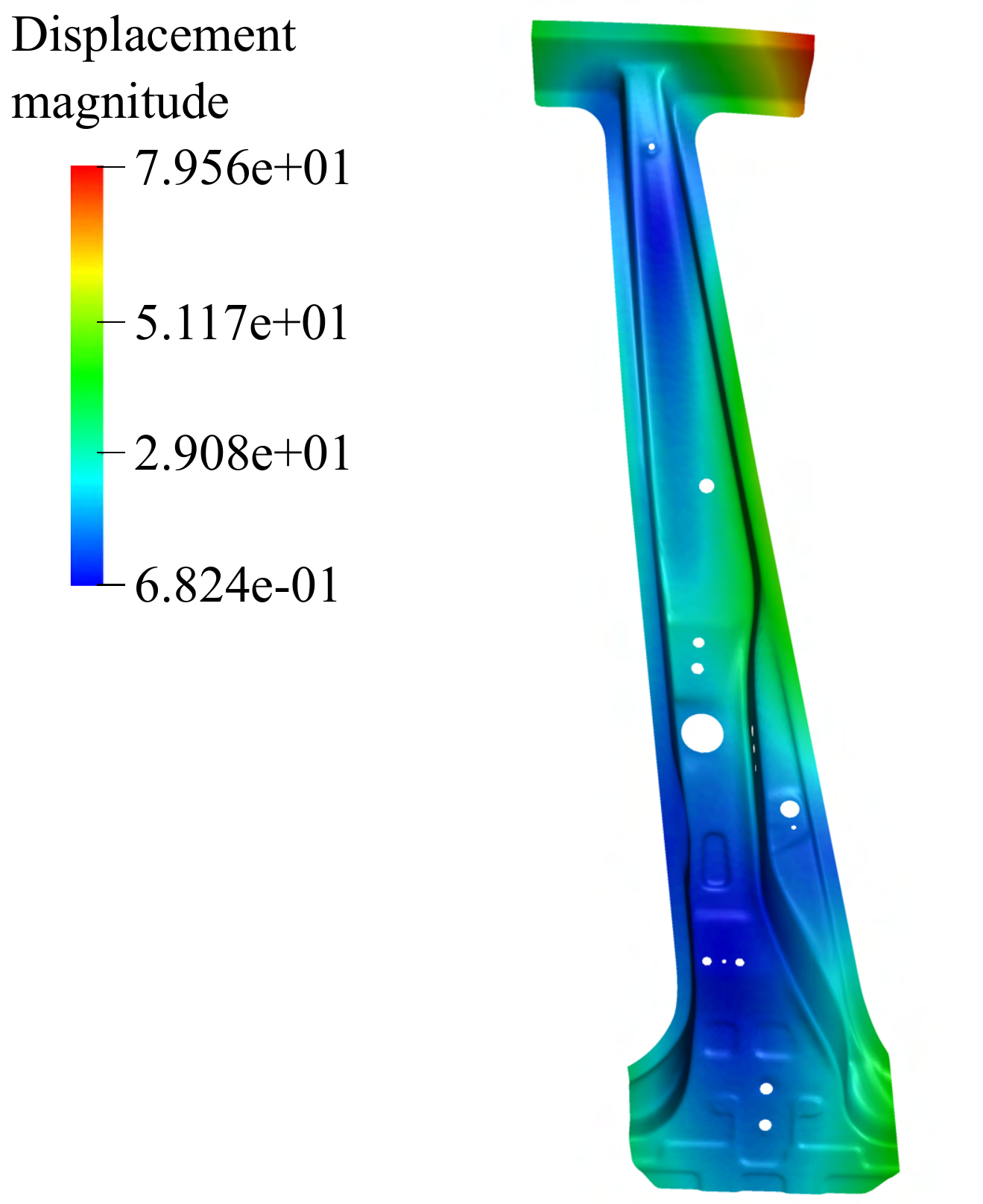}}
 \subfigure[]{\includegraphics[scale=.35]{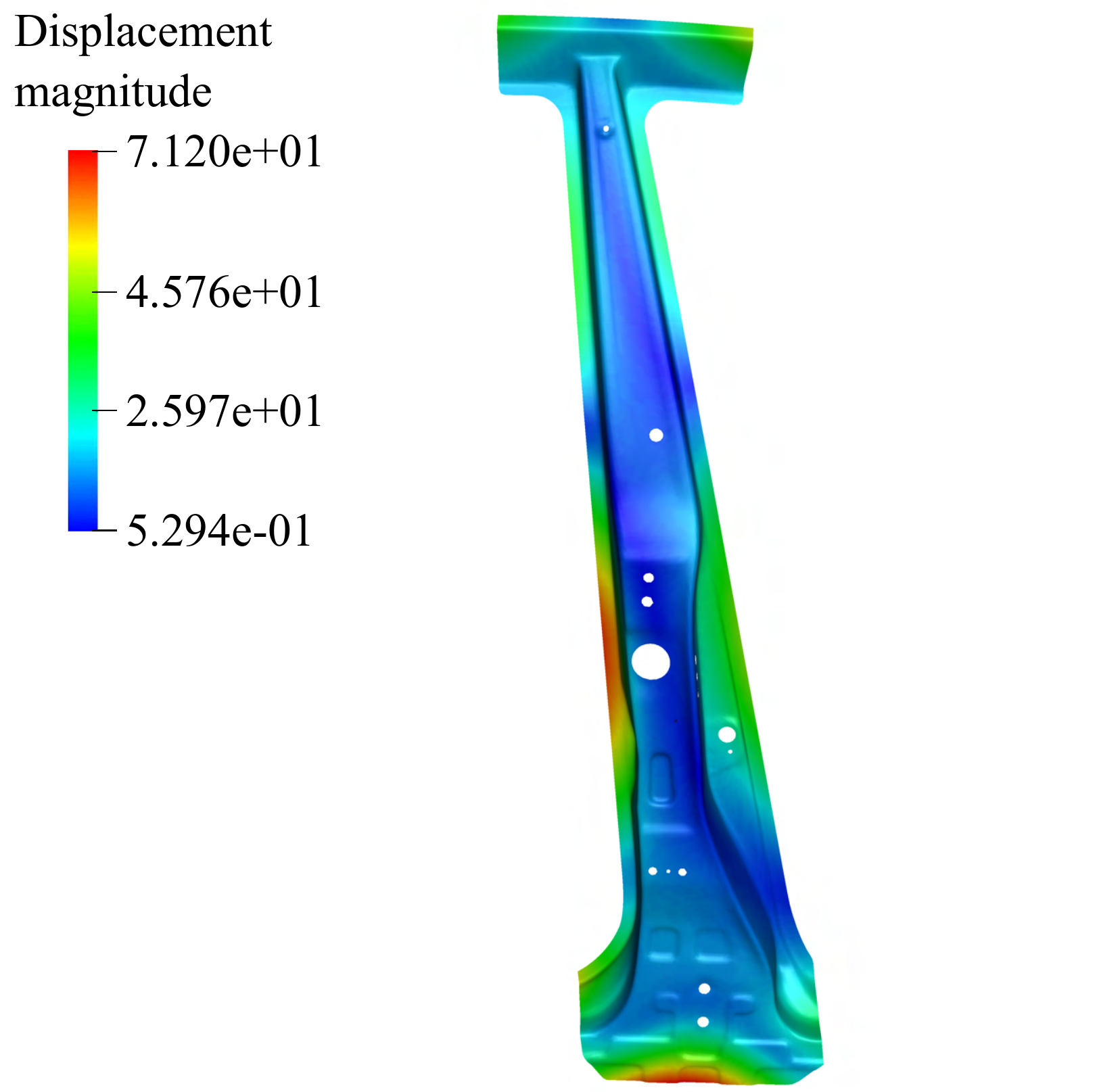}} 
 \subfigure[]{\includegraphics[scale=.35]{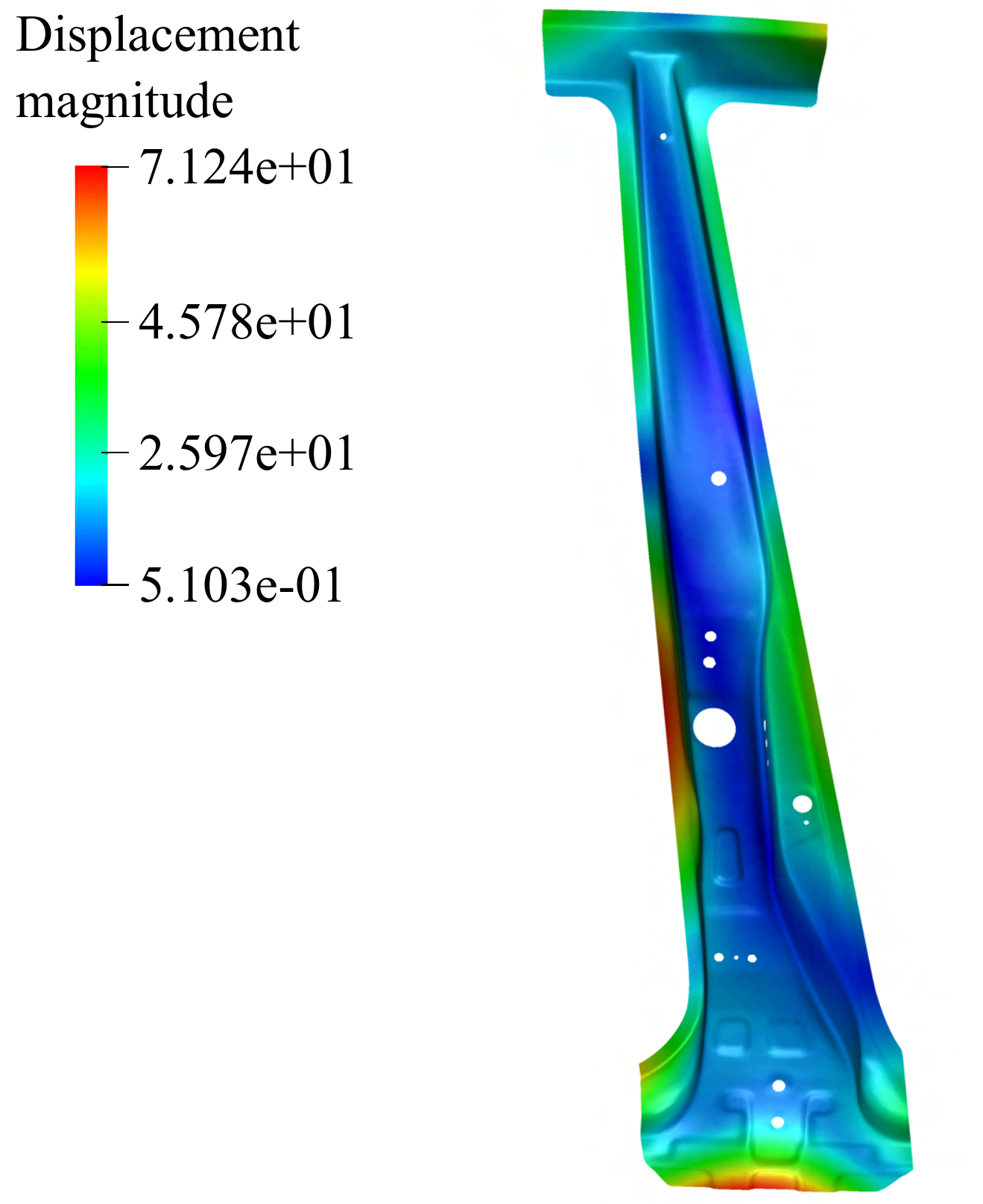}}\\
\caption{Outer part of a B-pillar. (a), (c), and (e) illustrate the seventh, eighth, and twelfth mode shapes, respectively, using G-splines and Kirchhoff-Love shells. (b), (d), and (f) illustrate the seventh, eighth, and twelfth mode shapes, respectively, using conventional finite elements.}
\label{modeshapesOuterwithIGA}
\end{figure}

Fig. \ref{modeshapesstiff} plots the seventh, eighth, and twelfth mode shapes of the stiffener obtained with FEM and IGA with R-M shells using the lumped mass matrix. Fig. \ref{modeshapesINNwithIGA} plots the seventh, eighth, and twelfth mode shapes of the inner part obtained with IGA with R-M shells and IGA with K-L shells using the lumped mass matrix. Fig. \ref{modeshapesOuterwithIGA} plots the seventh, eighth, and twelfth mode shapes of the outer part obtained with FEM and IGA with K-L shells using the lumped mass matrix. All the mode shapes in Figs. \ref{modeshapesstiff}, \ref{modeshapesINNwithIGA}, and \ref{modeshapesOuterwithIGA}  are in excellent agreement.

Finally,  Table \ref{lasteigen} shows the last eigenvalue obtained with FEM, IGA with R-M shells, and IGA with K-L shells using the lumped mass matix for the three geometries considered in this section. As explained in \cite{Hughes2012}, the value of the last eigenvalue limits the maximum stable time-step size in explicit dynamics (the smaller the last eigenvalue, the larger the time-step size can be). As shown in Table \ref{lasteigen}, IGA with R-M shells results in either the same value or smaller value than FEM for the last eigenvalue using the same element size. As expected, IGA with K-L shells results in significantly smaller values for the last eigenvalue using the same element size since this shell formulation does not have rotational degrees of freedom.

  \begin{table}[h!]
   \caption{Last eigenvalues for different geometries obtained with G-splines and with bilinear quadrilaterals using the lumped mass matrix.} \label{lasteigen}
   \bigskip
     \centering
     \renewcommand{\arraystretch}{1.20}
     \begin{tabular}{c@{\hspace{9.0mm}}  c@{\hspace{9.0mm}}  c@{\hspace{9.0mm}}  c@{\hspace{9.0mm}}}
\hline
  & IGA K-L & IGA R-M & FEM \\
\hline
Inner Part of B-pillar    & $9.664515\times 10^{6}$          & $1.183618\times 10^{11}$   & $1.183520\times 10^{11}$       \\          
Outer Part of B-pillar    & $1.042185\times 10^{8}$       & $8.127251\times 10^{10}$         & $1.340322\times 10^{11}$      \\     
Stiffener of B-pillar     & $1.398117\times 10^{7}$     & $4.616443\times 10^{10}$        & $4.617944\times 10^{10}$         \\                
\hline      
     \end{tabular}     
   \end{table}


\section{Conclusions}

We introduced two EP constructions based on imposing $G^1$ constraints which can handle control nets with \textit{any} unstructured quadrilateral layout. Among other reasons, developing EP constructions without any restriction in the distribution of EPs throughout the control net is required to capture small features (e.g., holes) whose size is similar to the element size. The studies of convergence, surface quality, and eigenvalue problems show that there are no relevant differences between the performance of the two proposed EP constructions. Thus, it is up to the user to decide to choose either an EP construction with polynomial basis functions in irregular elements, but with increased support of the basis functions (construction $G^1$P) or an EP construction with rational basis functions in irregular elements, but the support of the basis functions does not increase after enforcing the $G^1$ constraints (construction $G^1$R). The studies of convergence and surface quality also suggest that G-splines are more suitable for real-world engineering applications involving thin-walled structures than EP constructions based on the D-patch framework. Since only vertex-based control points are used and these control points behave as geometric shape handles, the two proposed versions of G-splines can be used for both the design and the analysis of complex thin-walled structures. We have represented the stiffener, the inner part, and the outer part of a B-pillar using G-spline surfaces. We have solved eigenvalue problems on these geometries using both G-splines and bilinear quadrilaterals in the commercial software LS-DYNA. Excellent agreement was found between G-splines and conventional finite elements.

\section*{Acknowledgements}

H. Casquero and M. S. Faruque were partially supported by the National Science Foundation (NSF) of USA
(grant number CMMI-2138187), Honda Motor Co., Japan, and Ansys Inc, USA. X. Wei is supported by the National
Natural Science Foundation of China (grant number 12202269).





\bibliographystyle{elsarticle-num} 
\bibliography{./Bibliography}


\end{document}